%% file: pqc-inverse.tex
\documentclass[3p, number,sort&compress,times,fleqn]{elsarticle}
\usepackage{amsmath}
\usepackage{amssymb}
\usepackage{mathtools}
\usepackage{subfig}
\usepackage[colorlinks = true,
linkcolor = blue,
urlcolor  = blue,
citecolor = blue,
anchorcolor = blue]{hyperref}
\usepackage[utf8]{inputenc}
\usepackage{caption}
\usepackage{mathdots}
\usepackage{graphicx}
\usepackage{tikz}
\usepackage{quantikz}
\usepackage{braket}
\usepackage{tikz-cd}
\usepackage{xcolor}
\usepackage{booktabs}
\usepackage{todonotes}
\usepackage{algorithm}
\usepackage{algpseudocode}
\usepackage{lineno}

\definecolor{purplefill}{RGB}{238,237,254}
\definecolor{purplestroke}{RGB}{83,74,183}
\definecolor{purpletext}{RGB}{60,52,137}

\definecolor{tealfill}{RGB}{225,245,238}
\definecolor{tealstroke}{RGB}{15,110,86}
\definecolor{tealtext}{RGB}{8,80,65}

\definecolor{grayfill}{RGB}{241,241,241}
\definecolor{graystroke}{RGB}{95,95,95}
\definecolor{graytext}{RGB}{68,68,68}

\definecolor{bluefill}{RGB}{225,240,250}
\definecolor{bluestroke}{RGB}{30,130,190}
\definecolor{bluetext}{RGB}{15,80,130}

\definecolor{amberfill}{RGB}{250,238,218}     
\definecolor{amberstroke}{RGB}{133,79,11}     
\definecolor{ambertext}{RGB}{99,56,6}         

\definecolor{new-blue}{RGB}{0,110,254}
\definecolor{dark-blue-custom}{RGB}{0,73,137}
\definecolor{accentblue}{RGB}{90,140,200}
\definecolor{accentgreen}{RGB}{70,140,90}

\usetikzlibrary{arrows.meta,positioning,matrix,positioning,shapes.geometric,calc,fit,backgrounds}

\usepackage{csml}

\begin{document}
	
	\begin{frontmatter}
		
		\title{A quantum-assisted latent surrogate model for Bayesian inversion of heterogeneous material properties on NISQ hardware}
		
		\author[1]{Dewangga Alfarisy}
		\author[2]{Catherine F. Higham}
		\author[1]{Eky Febrianto\corref{cor1}}
		\ead{eky.febrianto@glasgow.ac.uk}
		
		\cortext[cor1]{Corresponding author}
		
		\address[1]{Glasgow Computational Engineering Centre, James Watt School of Engineering, University of Glasgow, University Avenue, Glasgow, G12 8QQ, UK}
		\address[2]{School of Computing Science, University of Glasgow,
			University Avenue, Glasgow, G12 8QQ, UK}
		
		\begin{abstract}
			
			Bayesian inversion is widely used in engineering applications, including
			material characterisation, structural damage identification, and medical
			diagnostics such as tumour detection. These inverse problems are typically
			ill-posed, because localised variations in material properties must be inferred
			from noisy observations, and they are computationally intensive,
			since they often require many evaluations of a high-fidelity forward solver such
			as the finite element method. Quantum computing has recently attracted
			considerable academic and industrial interest, partly because some near-term
			algorithms are feasible on noisy intermediate-scale quantum (NISQ) hardware. In
			particular, parameterised quantum circuits (PQCs) combined with classical
			optimisation enable a hybrid quantum-classical surrogate model. A central 
			obstacle to applying such circuits to inversion is that the discretised parameter 
			and response fields are far too high-dimensional to be embedded in a near-term 
			quantum register. In this work we propose a quantum-assisted latent surrogate model (QLSM) for the Bayesian inversion of inhomogeneous material properties, in which latent 
			compression bridges these high-dimensional fields and the small
			registers accessible to near-term circuits. The low intrinsic dimensional structure of
			the spatially varying conductivity and potential fields is exploited by reducing
			the conductivity through principal component analysis (PCA) and the response
			through a convolutional autoencoder. A parameterised quantum circuit is then
			trained to map the latent conductivity to the latent response, and stochastic
			variational inference is performed entirely in latent space, so that posterior
			inference never invokes a full finite element solve. The methodology is
			demonstrated on one- and two-dimensional elliptic inverse problems, recovering
			localised inclusions, structured inclusions, and unstructured fields under both
			informative and non-informative priors. We further evaluate the model
			under realistic hardware noise, using a trapped-ion noise-aware simulator
			(IonQ) and a physical superconducting device (Rigetti Cepheus-1-108Q), and
			use a lightweight learned denoising stage that recovers the noiseless
			accuracy to within a small residual.
			
		\end{abstract}
		
		
	\end{frontmatter}
	
	\input{introduction}
	
	\input{forward-map}

	\input{quantum-latent}

	\input{stochastic-vi}
\input{examples}
\input{conclusions}
	
	\section*{Acknowledgments}
	DA is supported by a Graduate School Scholarship from the University of Glasgow. CFH was supported by EPSRC grants EP/T021020/1 and EP/T00097X/1 and by a Fellowship from the Centre for Data Science and AI at the University of Glasgow. EF is supported through a Reward for Excellence Award from the Early Career Development Programme at the College of Science and Engineering, University of Glasgow.
	\bibliographystyle{elsarticle-num-names}
	\bibliography{pqc-inverse.bib}
	
\end{document}

%% file: introduction.tex
\section{Introduction}
\label{sec:introduction}

Bayesian inversion provides a principled framework for inferring spatially varying material properties from indirect, and noisy observations. It has been applied across a wide range of engineering settings, including subsurface flow~\cite{Iglesias2014, Zhang2026}, thermal and electrical characterisation of materials~\cite{alberts2026bayesian,Taghizadeh2020}, the identification of elastic properties~\cite{Koutsourelakis2012, Vigliotti2018, Kucerova2023, Scholz2025WNVI, Jafari2025, zheng2025automatic}, and the detection of structural damage~\cite{Muralidhar2023}. For physical fields governed by partial differential equations, the inverse problem is typically ill-posed, because localised variations in the material parameter field must be recovered from only limited measurements of the response. The Bayesian formulation addresses this ill-posedness by combining a prior on the parameter field with a likelihood derived from a forward model and data, yielding a posterior distribution that quantifies both the reconstruction and its uncertainty~\cite{Stuart2010, Kaipio2005, Dashti2017}. The dominant computational cost of this framework lies in the forward model itself. Each evaluation of the posterior requires a high-fidelity solve, most commonly by the finite element method (FEM), and inference through Markov chain Monte Carlo or stochastic variational inference can demand many thousands of such evaluations~\cite{Cui2014, Hoffman2013SVI, Girolami2021,  povala2022variational}. The forward solver therefore remains the principal computational bottleneck, and this motivates the use of fast approximate surrogates in place of repeated full-order solves.

A range of surrogate strategies has been developed to reduce this cost by learning an approximate parameter-to-observation map from a training set of high-fidelity solutions. Projection-based reduced-order models compress the response onto a low-dimensional subspace through proper orthogonal decomposition and then solve a reduced Galerkin problem in this space~\cite{Benner2015,Quarteroni2016}. Stochastic spectral methods instead represent the parameter field through Karhunen--Lo\`{e}ve expansions or polynomial chaos, precomputing spectral coefficients in order to accelerate repeated evaluations~\cite{GhanemSpanos1991, XiuKarniadakis2002, Sun2023}. More recently, data-driven neural surrogates have learned nonlinear parameter-to-response mappings directly from data, among them convolutional autoencoders~\cite{LeeCarlberg2020, Deshpande2025}, neural operators~\cite{Lu2021, Li2021FNO}, and physics-informed networks~\cite{Raissi2019}, and several of these have been integrated into Bayesian  inversion workflows with reported cost reductions of one to two orders of magnitude~\cite{Yang2021,Wu2023}. A theme common to all of these approaches is dimensionality reduction. The spatially varying fields of interest are nominally high-dimensional, yet they possess a low intrinsic dimension that a compact latent representation can capture, and it is within this reduced space that surrogate construction becomes tractable.

Alongside these classical developments, quantum computing has emerged as a candidate paradigm for accelerating computationally demanding tasks, and the noisy intermediate-scale quantum era has made a class of hybrid algorithms feasible on present-day hardware~\cite{Preskill2018NISQ, Bharti2022NISQreview}. Central to this class are parameterised quantum circuits (PQC), which act as trainable function approximators when combined with classical optimisation in a hybrid quantum-classical learning loop~\cite{Cerezo2021VQA, Benedetti2019PQCML, MitaraiNegoroKitagawaFujii2018QCL}. The expressive power of such circuits can be enhanced through data re-uploading, in which the input is encoded repeatedly across successive circuit layers in order to enrich the family of representable functions~\cite{PerezSalinas2020Quantum, SchuldSwekeMeyer2021}, and related constructions have been explored for the solution of differential equations~\cite{Kyriienko2021DiffEq, Lubasch2020Nonlinear, febrianto2026quantum} and for surrogate modelling in computational engineering~\cite{Sedykh2024QCPINN, alavi2025hybrid, Sukulthanasorn2026}. The practical use of these circuits is nonetheless constrained in two important respects. Their trainability degrades as circuit width and depth grow, through phenomena such as barren plateaus~\cite{McClean2018BarrenPlateaus, Larocca2025barren}, and the hardware available in the near term offers only a small number of qubits with limited coherence times. These constraints sit awkwardly with the discretised parameter and response fields that arise in PDE-based inversion, which routinely involve thousands to millions of degrees of freedom and cannot be embedded directly into a near-term quantum register. This incompatibility is the central obstacle that any quantum-assisted inversion framework must confront, and the natural way past it is the low intrinsic dimension that classical surrogates already exploit.

\begin{figure}[]
	\centering
	\includegraphics[width=0.9\textwidth]{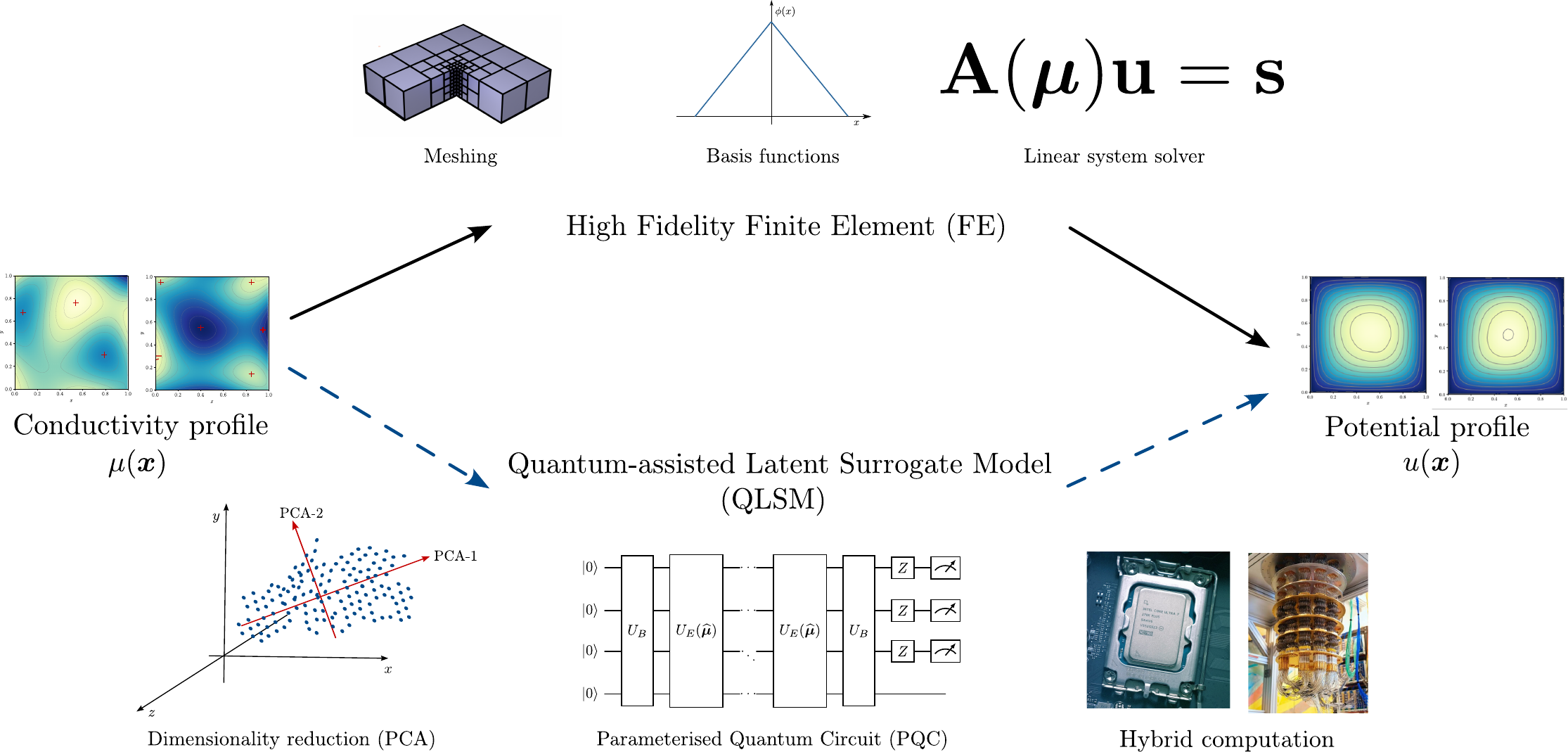}
	\caption{Diagram of quantum-assisted latent surrogate model (QLSM) trained on high fidelity finite element (FE) simulations. QLSM utilises dimensionality reduction into the latent space of the conductivity population and autoencoder to obtain the map between potential profile data and their respective latent representation. The nonlinear forward map between these latent spaces is constructed using parameterised quantum circuit (PQC). The forward and inverse computational pipeline is simulated through hybrid classical-quantum computing.}
	\label{fig:intro}
\end{figure}

In this work we propose a quantum-assisted latent surrogate model (QLSM) for forward and inverse problems involving inhomogeneous material properties, in which latent compression provides the bridge between the high-dimensional PDE fields and the small registers accessible to near-term quantum circuits. The high-dimensional conductivity field is first projected onto a low-dimensional space by principal component analysis, while the potential response field is compressed through a convolutional autoencoder. A parameterised quantum circuit (PQC) is then trained to map the latent conductivity to the latent potential response. Our main motivation for using a PQC is to explore how the latent conductivity is embedded and processed in a quantum register, and characterise the nonlinear mapping encoded through Chebyshev embedding within the PQC. Figure~\ref{fig:intro} summarises this construction alongside the high-fidelity finite element pipeline it emulates. The resulting model is embedded within a stochastic variational inference scheme~\cite{Hoffman2013SVI, povala2022variational} that operates entirely in latent space, so that posterior inference never invokes a full finite element solve and never manipulates fields at their full resolution. We demonstrate the framework on one- and two-dimensional elliptic problems. As a forward surrogate it reproduces the finite element response across these problems, and within a Bayesian inversion scheme it recovers single and dual localised inclusions, structured inclusions, and unstructured fields drawn from a Gaussian process, under both informative and non-informative priors. We then evaluate QLSM under realistic hardware noise, using a trapped-ion noise-aware simulator (IonQ) and a physical superconducting device (Rigetti Cepheus-1-108Q), and we introduce a lightweight learned denoising stage that compensates for hardware noise in the spirit of learning-based error mitigation~\cite{Temme2017ErrorMitigation, Strikis2021LearningEM, Cai2023QEMreview}.

This paper establishes QLSM as a functional component of a complete Bayesian inversion pipeline, from the latent compression of the conductivity and response fields, through a quantum-assisted forward map realised by a parameterised quantum circuit, to posterior inference performed wholly in latent space. The aim is to establish the NISQ-suitable pipeline itself, and in particular how the high-dimensional parameter and response fields are compressed to fit a small register, where the quantum component can be inserted, and how its readout is corrected for hardware noise, so that the scaffolding required by any future quantum-assisted inversion is in place and its constraints are identified empirically rather than assumed. The PQC-based design is adopted in preference to solving the discretised elliptic system directly on the device, which allows us to sidestep the need for state preparation and full state readout in each iteration. Confining the parameterised circuit to a small latent-to-latent map instead targets the part of the problem for which a near-term circuit is best suited. Within this scope the study is framed to isolate and characterise the behaviour of the quantum component, and two questions are examined in detail. The first concerns the PQC as an approximator of the nonlinear latent forward map, and in particular the way in which circuit depth and design govern the balance between expressivity and trainability. The second concerns the resilience of the pipeline under realistic hardware noise, and the accuracy that can be recovered once a lightweight classical correction is applied. Across one- and two-dimensional elliptic problems the model attains accuracy comparable to a classical neural-network surrogate of equivalent role, and it remains robust when executed under realistic NISQ noise. 
These results provide a concrete and reproducible pathway for the deployment of quantum-assisted surrogates in PDE-based inverse problems, and they identify the principal levers, namely circuit expressivity and hardware noise, that govern the extension of the approach to larger and more strongly nonlinear regimes.

The remainder of the paper is organised as follows. Section 2 describes the forward problem and the latent surrogate construction. Section 3 details the quantum latent forward map. Section 4 presents the variational inference framework. Section 5 reports numerical experiments in one and two dimensions. Section 6 reports the deployment on NISQ hardware. Section 7 concludes and outlines directions for future work.

%% file: forward-map.tex
\section{Forward problem and quantum-assisted latent surrogate model (QLSM)}
\label{sec:forward-map}

\subsection{Model problem}
\label{sec:poisson-problem}

\begin{figure}[]
	\centering
	\includegraphics[width=0.9\textwidth]{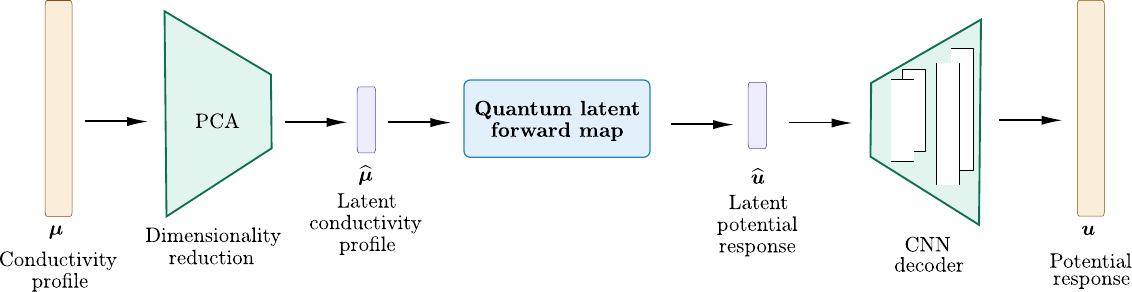}
	\caption{Forward map architecture of our proposed quantum-assisted latent surrogate model (QLSM).}
	\label{fig:hybrid-architecture}
\end{figure}

As a model problem, we consider the one-dimensional Poisson--Dirichlet equation on the bounded domain $\Omega = (0,1) \subset \mathbb{R}$,
\begin{equation}
	\label{eq:bvp-1d}
	-\frac{\D}{\D x} \left( \mu(x) \, \frac{\D u(x)}{\D x} \right) = s(x)
	\quad \text{in } \Omega \, ,
	\qquad
	u(0) = u(1) = 0 \, ,
\end{equation}
with a unit source $s(x) = 1$ and a positive conductivity field $\mu(x)$ controlled by a set of anchor parameters $\vec \mu^{(a)} \in \mathbb R^{n_a}$ introduced at a set of chosen control points $\{x_j^{(a)}\}_{j=1}^{n_a}$ as described later in Section~$\ref{sec:svi}$. Discretising the weak form of the boundary value problem in Eq.~\eqref{eq:bvp-1d} yields the discrete system
\begin{equation}
	\vec{A}(\vec{\mu}) \, \vec{u} = \vec{s} \, ,
\end{equation}
where $\vec{\mu} \in \mathbb{R}^{n_e}$ is the discretised conductivity field, $\vec{A}(\vec{\mu}) \in \mathbb{R}^{n_{\text{dof}} \times n_{\text{dof}}}$ is the parameter-dependent stiffness matrix, and $\vec{s} \in \mathbb{R}^{n_{\text{dof}}}$ is the discrete source vector. Here $n_e$ indicates the number of elements and $n_{\text{dof}}$ denotes the number of degrees of freedom. 
Although higher-order or other smooth discretisations could equally be employed~\cite{febrianto2021mollified, febrianto2024three}, it is not the key interest of this work and is therefore not pursued further. The solution of this system defines the classical finite element method (FEM) forward map
\begin{equation}
	\vec{u} = \vec{A}(\vec{\mu})^{-1} \vec{s} = \mathcal{F}(\vec{\mu}) \, ,
	\label{eq:fem-forward-map}
\end{equation}
which maps a given conductivity $\vec \mu$ to its corresponding potential response $\vec u$. Note that as discussed in~\cite{Girolami2021, povala2022variational}, this map is nonlinear in $\vec \mu$ even though the PDE operator is linear. Each posterior evaluation in the inverse problem requires one such solve, and it is this map that the surrogate is designed to approximate. 

The proposed surrogate model provides an alternative route to this FEM solve through a sequence of reduced-space operations. We refer to this surrogate as the quantum-assisted latent surrogate model (QLSM). The conductivity field is first compressed to a latent representation $\hat{\vec{\mu}}$ by principal component analysis (PCA), a quantum latent forward map then propagates $\hat{\vec{\mu}}$ to a latent potential response $\hat{\vec{u}}$, and a decoder finally reconstructs the full response field $\vec{u}$. Both FEM and surrogate map share the same input and output spaces, and they differ only in the manner in which the parameter-to-response map is realised. The remainder of this section constructs the two reduced spaces and the map between them. The latent representation of the potential response $\hat{\vec u}$ is obtained from the autoencoder of Section~\ref{sec:autoencoder}, the latent representation of the conductivity from the principal component analysis of Section~\ref{sec:pca}, and the quantum-assisted forward map that connects them is described in Section~\ref{sec:latent-surrogate}. Figure~\ref{fig:hybrid-architecture} provides the schematic of the proposed quantum-assisted latent surrogate model (QLSM).

\subsection{Latent representation of potential response}
\label{sec:autoencoder}

\begin{figure}[]
	\centering
	\includegraphics[width=0.73\textwidth]{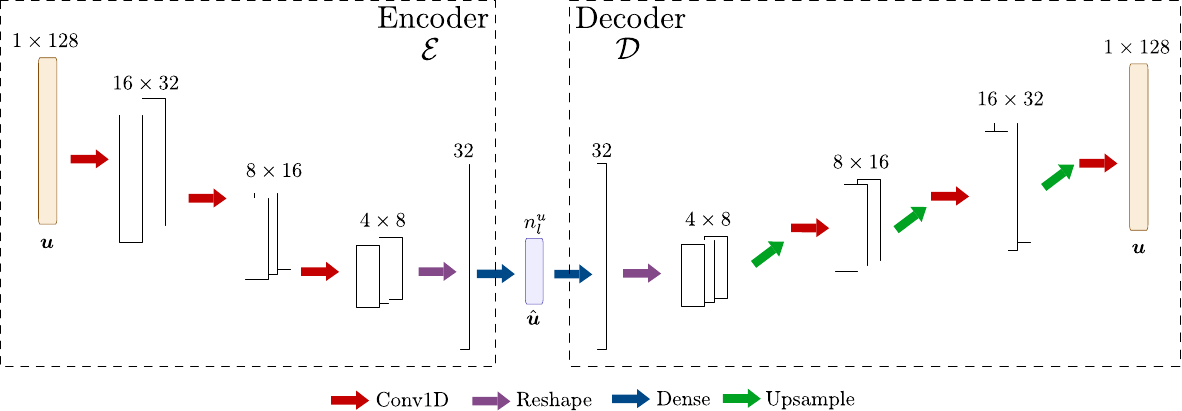}
	\caption{Autoencoder architecture with encoder $\mathcal{E}$ and decoder $\mathcal{D}$, in which $\vec{u} \in \mathbb{R}^{n_{\text{dof}}}$ is the response field in physical space and $\hat{\vec{u}} \in \mathbb{R}^{n_l^{u}}$ is the response field in the latent space.}
	\label{fig:autoencoder-architecture}
\end{figure}

To obtain a compact latent representation of the response field, we train a convolutional encoder--decoder architecture. The encoder $\mathcal{E}(\cdot)$ maps the high-dimensional input to a low-dimensional latent representation
\begin{equation}
	\hat{\vec{u}} = \mathcal{E}(\vec{u}) \, ,
\end{equation}
where $\hat{\vec{u}} \in \mathbb{R}^{n_l^{u}}$ with the potential latent dimension $n_l^{u} \ll n_{\text{dof}}$. The decoder $\mathcal{D}(\cdot)$ reconstructs the field from the latent space
\begin{equation}
	\tilde{\vec{u}} = \mathcal{D}(\hat{\vec{u}}) \, ,
\end{equation}
yielding an approximation $\tilde{\vec{u}} \approx \vec{u}$.

The encoder is constructed as a sequence of convolutional downsampling operators denoted as $\{\mathcal{C}_\ell\}_{\ell=1}^{L}$ followed by a linear layer,
\begin{equation}
	\hat{\vec{u}}
	=
	\vec{W}_e \,
	\text{vec} \!\left(
	\mathcal{C}_L \circ \cdots \circ \mathcal{C}_1 (\vec{u})
	\right)
	+ \vec{b}_e \, ,
\end{equation}
where $\text{vec}(\cdot)$ denotes vectorisation and $(\vec{W}_e,\vec{b}_e)$ are the weight and bias parameters of the final linear projection. Each convolutional block consists of a convolutional operator followed by a pointwise nonlinear activation, chosen here as the hyperbolic tangent ($\tanh$) function. The decoder mirrors this process by lifting the latent representation back to the original spatial domain. Starting from the latent vector, a linear layer first maps $\hat{\vec{u}}$ to a structured feature tensor, which is then processed through a sequence of upsampling and convolutional operators. Denoting by $\{\mathcal{U}_\ell\}_{\ell=1}^{L}$ the upsampling operators and by $\{\tilde{\mathcal{C}}_\ell\}_{\ell=1}^{L}$ the corresponding convolutional layers, the decoder reads
\begin{equation}
	\tilde{\vec{u}}
	=
	\tilde{\mathcal{C}}_1 \circ \mathcal{U}_1 \circ \cdots \circ
	\tilde{\mathcal{C}}_L \circ \mathcal{U}_L
	\left(
	\text{reshape}\!\left(
	\tanh(\vec{W}_d \, \hat{\vec{u}} + \vec{b}_d)
	\right)
	\right) \, ,
\end{equation}
where $(\vec{W}_d,\vec{b}_d)$ are the parameters of the decoder linear layer and $\text{reshape}(\cdot)$ maps the latent vector to the initial feature tensor. The encoder--decoder pair is trained together by minimising the reconstruction error among $n_s$ training data
\begin{equation}
	\mathcal{L}_{\text{AE}}
	=
	\frac{1}{n_s}
	\sum_{i=1}^{n_s}
	\left\|
	\vec{u}_{i} - \mathcal{D}(\mathcal{E}(\vec{u}_{i}))
	\right\|_2^2 \, ,
\end{equation}
which ensures that the latent representation retains the dominant spatial features of the original field. This compressed representation serves as the interface between the classical preprocessing stage and the quantum latent forward map.

Figure~\ref{fig:autoencoder-architecture} depicts the proposed convolutional autoencoder where the one-dimensional response field is described using $n_{\text{dof}} = 128$. The encoder
applies three \texttt{Conv1D} layers that progressively coarsen the spatial
resolution while adjusting the channel count, $(1,128) \rightarrow (16,32) \rightarrow (8,16) \rightarrow (4,8)$, where each tuple denotes (channels, length). The resulting feature map is flattened and
projected by a single linear layer onto the latent space of dimension $n_l$
(namely $n_l^\mu$ for the conductivity field and $n_l^u$ for the potential
field). The decoder mirrors this structure: a linear layer lifts the latent
vector back to a $(4,8)$ feature map, followed by three
upsampling--convolution blocks that invert the encoder,
$(4,8) \rightarrow (8,16) \rightarrow (16,32) \rightarrow (1,128)$. All
convolutional layers use a kernel of width three.

\subsection{Latent representation of material conductivity}
\label{sec:pca}

\begin{figure}[]
	\centering
    \includegraphics[width=0.3\textwidth]{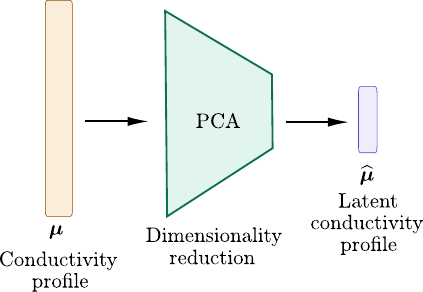}
	\caption{Principal component analysis mapping the conductivity field $\vec{\mu}$ in the physical space to its reduced representation $\hat{\vec{\mu}}$ in the latent space.}
	\label{fig:pca-architecture}
\end{figure}

Whereas the response field is compressed through the nonlinear autoencoder of Section~\ref{sec:autoencoder}, the conductivity field is reduced by a linear
projection. Prior to constructing the quantum surrogate, the high-dimensional conductivity field $\vec{\mu} \in \mathbb{R}^{n_e}$ is projected onto a
low-dimensional latent space using principal component analysis (PCA), see Figure~\ref{fig:pca-architecture}, which captures the dominant modes of variation in the dataset while reducing the dimensionality seen by the surrogate. Beyond projection, PCA is used to inform the inherent size of the latent dimension. Its linearity also allows us to delegate more of the nonlinearity to the quantum element, that is, the PQC, which performs the nonlinear mapping between latent spaces.

For a set of conductivity samples $\{\vec{\mu}_{i}\}_{i=1}^{n_s}$ with empirical mean $\overline{\vec{\mu}} = \tfrac{1}{n_s}\sum_{i=1}^{n_s} \vec{\mu}_{i}$, the empirical covariance matrix is
\begin{equation}
	\vec{C}_\mu = \frac{1}{n_s} \sum_{i=1}^{n_s}
	\left( \vec{\mu}_{i} - \overline{\vec{\mu}} \right)
	\left( \vec{\mu}_{i} - \overline{\vec{\mu}} \right)^\trans \, .
\end{equation}
The eigen decomposition of the covariance matrix yields
\begin{equation}
	\vec{C}_\mu \, \vec{v}_j = \lambda_j \vec{v}_j \, ,
\end{equation}
where $\lambda_j$ and $\vec{v}_j$ denote the eigenvalues and corresponding
orthonormal eigenvectors. The eigenvectors associated with the largest
eigenvalues define the principal directions of variation. Retaining the first $n_l^{\mu}$ dominant modes (i.e., principal components), the reduced representation of the conductivity field is obtained as
\begin{equation}
	\hat{\vec{\mu}} =
	\vec{V}^\trans \left( \vec{\mu} - \overline{\vec{\mu}} \right) \, ,
\end{equation}
where $\vec{V} = \left(\vec{v}_1, \dots, \vec{v}_{n_l^{\mu}} \right) \in \mathbb{R}^{n_e \times n_l^{\mu}}$ is the matrix of leading eigenvectors. The resulting vector $\hat{\vec{\mu}} \in \mathbb{R}^{n_l^{\mu}}$ constitutes the low-dimensional latent representation used as input to the quantum-assisted latent model.

\subsection{Quantum-assisted latent surrogate model}
\label{sec:latent-surrogate}

The two dimensionality reduction stages introduced above provide the latent spaces on which the proposed QLSM surrogate operates. The conductivity field is represented in the parametric latent space through the principal component projection $\hat{\vec{\mu}} \in \mathbb{R}^{n_l^{\mu}}$, while the response field is represented in the latent space of the autoencoder through $\hat{\vec{u}} = \mathcal{E}(\vec{u}) \in \mathbb{R}^{n_l^{u}}$. The forward map is then constructed as a mapping between these two latent representations, using a parameterised quantum circuit (PQC), and is denoted as $\mathcal{Q}(\cdot)$, i.e., 
\begin{equation}
	\hat{\vec{u}} = \mathcal{Q}(\hat{\vec{\mu}}) \, .
\end{equation}
The full response field is recovered by decoding the predicted latent response, so that the complete QLSM forward map reads
\begin{equation}
	\widetilde{\mathcal{F}}(\vec{\mu})
	=
	\mathcal{D}\!\left(
	\mathcal{Q}\!\left(
	\vec{V}^\trans (\vec{\mu} - \overline{\vec{\mu}})
	\right)
	\right)
	\approx
	\mathcal{F}(\vec{\mu}) \, ,
	\label{eq:hybrid-forward-map}
\end{equation}
which approximates the finite element map in  Eq.~\eqref{eq:fem-forward-map} while operating wholly in the reduced spaces. The latent dimensions $n_l^{\mu}$ and $n_l^{u}$ are selected from the decay of the explained variance of the principal component analysis and the reconstruction error of the autoencoder, and both are far smaller than the full field dimension $n_\text{dof}$. It is this compression that renders the forward map compatible with a quantum circuit acting on a small qubit register, since the circuit need only represent the relationship between two low-dimensional latent vectors rather than between the full fields. The internal structure of the quantum latent forward map $\mathcal{Q}$, comprising the scaling of the latent inputs, the parameterised quantum circuit, and the transfer of the measured quantum outputs into the latent response, is described in Section~\ref{sec:qlf}.

%% file: quantum-latent.tex
\section{Quantum latent forward map}
\label{sec:qlf}

\subsection{Overview}
\label{sec:algorithm}

\begin{figure}[]
	\centering
    \includegraphics[width=0.85\textwidth]{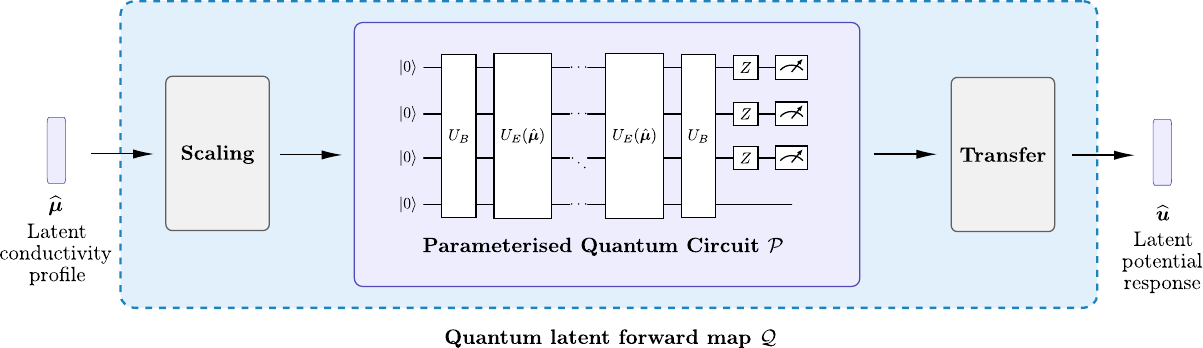}
	\caption{Quantum-assisted forward map in the latent space. First the latent conductivity is scaled and then encoded into the quantum circuit. The measured quantum output is subsequently transferred into the latent response space.}
	\label{fig:quantum-latent-map}
\end{figure}

As discussed in Section~\ref{sec:forward-map}, one of the key components of QLSM is the quantum latent forward map $\mathcal{Q}$ which provides an approximation of the forward mapping between the latent conductivity representation and the corresponding latent potential response. We note here that both latent spaces have different dimensions, $n_l^\mu$ and $n_l^u$, for the conductivity and potential, respectively, and that the mapping between these spaces is nonlinear. This is because the PCA is linear, while the decoder, although nonlinear, has no information about the conductivity input. As illustrated in Figure~\ref{fig:quantum-latent-map}, the map operates on a compressed input $\hat{\vec{\mu}}$, obtained from the prior dimensionality reduction stage, and produces a latent output $\hat{\vec{u}}$ that is subsequently decoded to reconstruct the full response field. The overall mapping is composed of three sequential components, namely a scaling transformation $\mathcal{S}$, a parameterised quantum circuit $\mathcal{P}$, and a transfer layer $\mathcal{T}$. 

First, the input latent vector $\hat{\vec{\mu}}$ is normalised through the scaling operation so that its components lie within a suitable range for quantum angle encoding. The scaled representation is then embedded into the parameterised quantum circuit, which realises a nonlinear mapping embedded in a $2^{n_l^\mu}$ Hilbert space. This circuit captures the essential structure of the forward operator while operating on a significantly reduced number of degrees of freedom. The output of the quantum circuit is obtained through expectation value measurements, yielding a vector of quantum features in $n_l^u$-dimension. Since these measurements do not directly correspond to the latent representation required by the decoder, the transfer stage is introduced as a classical mapping that projects the quantum outputs onto the latent space of the response field. The resulting vector $\hat{\vec{u}}$ is then passed to the decoder to reconstruct the spatial response. In compact form, the quantum latent  forward map reads
\begin{equation}
	\hat{\vec{u}}
	=
	\mathcal{Q}(\hat{\vec{\mu}})
	=
	\mathcal{T}\!\left(
	\mathcal{P}\!\left(
	\mathcal{S}(\hat{\vec{\mu}})
	\right)
	\right) \, ,
\end{equation}
where $\mathcal{S}(\cdot)$ denotes the scaling operation, $\mathcal{P}(\cdot)$ represents the parameterised quantum circuit, and $\mathcal{T}(\cdot)$ is the transfer layer. The detailed formulation of the scaling and transfer modules is provided in Section~\ref{sec:scaling-transfer}. The construction of the parameterised quantum circuit (PQC) used in this work is detailed next in Section~\ref{sec:pqc}. 

\subsection{Parameterised quantum circuit (PQC)}
\label{sec:pqc}

The parameterised quantum circuit (PQC), denoted as operator $\mathcal{P}$, acts on an $n_q$-qubit register. In this work, the number of qubits matches the latent conductivity dimension, $n_q = n_l^{\mu}$ since we do not require ancillary qubits as temporary registers. The PQC considered in this work consists of three stages, namely an embedding $U_B(\vec \beta_I)$ that changes the basis from the reference state to a data-dependent state, a variational layer $U_V$ that encodes the scaled latent conductivity and modifies the corresponding quantum state, and a final basis change $U_B(\vec \beta_O)$ followed by measurement. 

Starting from the reference state $\ket{0}^{\otimes n_q}$, the embedding $U_B (\vec \beta_I)$ performs the map $U_B (\vec \beta_I) \, : \, \ket{0}^{\otimes n_q} \mapsto \ket{\psi_{I} }$, 
%
%
where the unitary $U_B$ performs tensor product of single-qubit rotations 
\begin{equation}
	U_B (\vec \beta_I) =
	\bigotimes_{j=1}^{n_q} R_y(\beta_{I, j})  \, ,
\end{equation}
where $\vec \beta_{I} = \{\beta_j^I \}_{j=1}^{n_q}$ are trainable scaling parameters.
Although similar effect can be obtained by, for example, tensor product of Hadamard gates, we emphasise here that we intend to learn a basis that aligns best with the data rather than prescribe the weight of superposition~\cite{Kreplin2024}. 

\begin{figure}[]
	\centering
	\scalebox{0.9}{
		\input{kyriienko-ec}
	}
	\caption{One layer of $U_E$ used in this study illustrated for a four-qubit system ($n_q = 4$). The layer consists of parameterised single-qubit gates~\cite{Kyriienko2021DiffEq, Kandala2017HEA} with an $R_z$, $R_x$, $R_z$ structure, together with CNOT gates in a ring topology for entanglement.}
	\label{fig:pqc}
\end{figure}
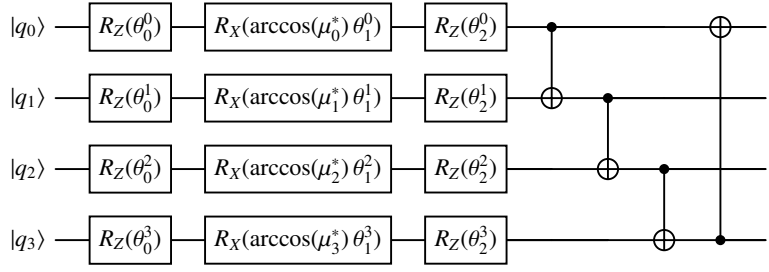

The data-dependent evolution is carried out by the variational stage, which maps
the embedded state through $U_V (\vec{\mu}^*, \vec{\theta}) \, : \, \ket{\psi_{I}} \mapsto \ket{\psi_V}$, 
%
%
where $\ket{\psi_V} = U_V (\vec{\mu}^*, \vec{\theta}) \ket{\psi_{I}}$ and the
variational unitary is the ordered composition of $n_v$ layers,
\begin{equation}
	U_V (\vec{\mu}^*, \vec{\theta})
	=
	U_E(\vec{\mu}^*, \vec{\theta}_{n_v}) \cdots U_E(\vec{\mu}^*, \vec{\theta}_1) \, .
\end{equation}
Here $\vec{\mu}^* \in \mathbb{R}^{n_q}$ is the scaled latent input, which will be described in Section~\ref{sec:scaling-transfer}. Each layer applies a trainable single-qubit rotation to every qubit, followed by a fixed entangling operation,
\begin{equation}
	U_E(\vec{\mu}^*, \vec{\theta}_\ell)
	=
	U_{\text{ent}}
	\left(
	\bigotimes_{j=1}^{n_q}
	R_z \left(\theta_{\ell, j}^2 \right) \,
	R_x\! \left(\theta_{\ell, j}^1 \arccos(\mu^*_j) \right) \,
	R_z \left( \theta_{\ell, j}^0 \right)
	\right) \, ,
\end{equation}
as shown in Figure~\ref{fig:pqc}. $\vec{\theta}_\ell$ collects the trainable parameters of layer $\ell$. The
per-qubit product $R_z \, R_x \, R_z$ is an arbitrary single-qubit rotation that mixes
the amplitudes of that qubit, and the central $R_x$ gate re-uploads the latent
conductivity through the angle $\arccos(\mu^*_j)$. This angle encoding is the
origin of the nonlinearity in $\vec{\mu}^*$ by leveraging Chebyshev polynomial embedding~\cite{Kreplin2024, Kyriienko2021DiffEq}. 
Re-uploading the scaled latent conductivity in every layer raises the polynomial order that the circuit can
represent~\cite{PerezSalinas2020Quantum, SchuldSwekeMeyer2021}. Additionally, mixed component nonlinearities, such as $\mu_j^* \mu_k^*$, will arise due to the tensor product construction. 

Without the entanglement unitary $U_{\text{ent}}$, the construction remains separable. The basis change $U_B^I$ and the single-qubit rotations within each variational layer all act on one qubit at
a time.
Even though the quantum state is in superposition containing higher-order and nonlinear mixed components, its measurement expectation on qubit $j$ depends only on the
rotations and the encoded value of that same qubit. In other words, each output feature would be a function of the single latent component $\mu^*_j$ in isolation. A
separable circuit therefore reduces to a collection of independent univariate
maps and cannot represent any coupling between distinct latent components. By contrast, the operator we intend to approximate couples the features. That is, the solution of
the elliptic problem at any point depends on the entire conductivity field
through the dense inverse $\vec{A}(\vec{\mu})^{-1}$, so each latent response
component depends jointly on all latent conductivity components. Reproducing this
requires correlations between qubits, which the entangling operator
$U_{\text{ent}}$ introduces. 

In this work we use for $U_{\text{ent}}$ the controlled-NOT (CX) gates in a ring
topology, following~\cite{Kyriienko2021DiffEq}. Each qubit acts as the control for its successor, with qubit $j$ coupled to qubit $j+1$ for $j = 1,\dots,n_q-1$, and the register is closed by coupling
qubit $n_q$ back to qubit $1$,
\begin{equation}
	U_{\text{ent}}
	=
	\text{CX}_{n_q,1}
	\prod_{j=1}^{n_q-1}
	\text{CX}_{j, j+1} \, ,
\end{equation}
where $\text{CX}_{a,b}$ denotes a controlled-NOT with control qubit $a$ and target
qubit $b$. A single layer entangles neighbouring qubits, and closing the ring
removes the open boundary of a linear chain so that correlations can propagate in
both directions around the register rather than terminating at the two end
qubits. Repeating the entangling layer across the $n_v$ variational layers spreads
these correlations over the whole register, so that after a few layers each
output feature becomes a polynomial that mixes several latent components and
carries cross-terms such as $\mu^*_j \mu^*_k$. This joint dependence is precisely
what allows the circuit to approximate the coupled, nonlinear
conductivity-to-response map, while the ring uses only $n_q$ CX gates
per layer, which keeps the entangling cost linear in the number of qubits and
well suited to near-term hardware.

The final stage is a trainable basis change $U_B (\vec \beta_O)$ that rotates the evolved
state into the measurement basis, $U_B (\vec \beta_O) \, : \, \ket{\psi_V} \mapsto \ket{\psi_O}$, 
%
%
where the unitary is again a tensor product of single-qubit rotations,
\begin{equation}
	U_B (\vec \beta_O) = \bigotimes_{j=1}^{n_q} R_y(\beta_{O, j}) \, ,
\end{equation}
with trainable parameters $\vec{\beta}_O = \{\beta_{O, j}\}_{j=1}^{n_q}$. The three
stages compose into the circuit unitary $U_B (\vec \beta_O) \, U_V(\vec{\mu}^*, \vec{\theta}) \,
U_B (\vec \beta_I)$ acting on the initial state $\ket{0}^{\otimes n_q}$, so that
\begin{equation}
	\ket{\psi_{O}}
	=
	U_B (\vec \beta_O) \, U_V(\vec{\mu}^*, \vec{\theta}) \,
	U_B (\vec \beta_I) \, \ket{0}^{\otimes n_q} \, .
\end{equation}
The latent output is read out as the expectation values of the single-qubit
Pauli-$Z$ operators,
\begin{equation}
	q_j
	=
	\bra{\psi_{O}} Z_j \ket{\psi_{O}}
	=
	p(\ket{0}_j) - p(\ket{1}_j) \, ,
	\qquad j = 1,\dots,n_l^u \, ,
	\label{eq:pauli-z}
\end{equation}
where $p(\ket{0}_j)$ and $p(\ket{1}_j)$ are the probabilities of measuring qubit
$j$ in the states $\ket{0}$ and $\ket{1}$. 
We measure only the first $n_l^u$ qubits due to the lower latent space dimension of the response. Measuring a lower number of qubits, together with shallow, compact PQC may help avoid the problem of barren plateaus during learning~\cite{Cerezo2021cost, McClean2018BarrenPlateaus, Larocca2025barren}. The resulting vector $\vec{q} \in \mathbb{R}^{n_l^u}$ is the latent output of the circuit, and defines the action of the PQC map $\vec{q} = \mathcal{P}(\vec{\mu}^*)$
used in the decomposition $\mathcal{Q} = \mathcal{T} \circ \mathcal{P} \circ
\mathcal{S}$ of Section~\ref{sec:algorithm}.

\subsection{Scaling and transfer}
\label{sec:scaling-transfer}

Recall the reduced representation of the input conductivity field $\hat{\vec{\mu}}$ $\in \mathbb{R}^{n_l^{\mu}}$. Before entering the parameterised quantum circuit, the latent vectors are first normalised through an affine rescaling,
\begin{equation}
	\vec z
	=
	\frac{\hat{\vec{\mu}}  - \overline{\vec{\mu}}^l}{\vec{\sigma}_{\mu}^l} \, ,
	\label{eq:pca_rescaling}
\end{equation}
where $\overline{\vec{\mu}}^l \in \mathbb{R}^{n_l^{\mu}}$ and $\vec{\sigma}_{\mu}^l \in \mathbb{R}^{n_l^{\mu}}$ denote the empirical mean and standard deviation of the latent input population from the training set, and the division is understood element-wise. This transformation centres the latent variables around zero and rescales them to have variance of order unity, thereby improving numerical stability in both the classical optimisation and the quantum encoding.

The rescaled latent vector is then passed through a nonlinear squashing map, $\vec{\mu}^* = \tanh(\vec z)$, so that each component is bounded in the interval $(-1,1)$. This step regularises the magnitude of the latent coordinates prior to quantum embedding, ensuring that the subsequent angle encoding remains within a controlled range and reducing sensitivity to outlying PCA coefficients. The bounded vector $\vec{\mu}^*$ is the encoded input embedded into the quantum circuit of Section~\ref{sec:pqc} through angle encoding.

To interface the measured quantum output with the decoder, the expectation values $\vec{q} \in \mathbb{R}^{n_l^u}$ obtained from Eq.~\eqref{eq:pauli-z} are passed through a transfer block consisting of a single fully connected linear layer,
\begin{equation}
	\hat{\vec{u}}
	=
	\vec{W}_{\text{tr}} \, \vec{q} + \vec{b}_{\text{tr}} \, ,
	\label{eq:transfer_layer}
\end{equation}
where $\vec{W}_{\text{tr}} \in \mathbb{R}^{n_l^{u} \times n_l^u}$ and $\vec{b}_{\text{tr}} \in \mathbb{R}^{n_l^{u}}$ are trainable parameters. This transfer layer maps the measured quantum outputs into the latent response representation required by the decoder, compensating for the scale mismatch between the quantum expectation values and the target latent space.

%% file: kyriienko-ec.tex
\begin{quantikz}[row sep=0.35cm, column sep=0.5cm]
\lstick{$\ket{q_0}$}
  & \gate{R_Z(\theta_0^0)}
  & \gate{R_X(\arccos(\mu^*_0)\,\theta_1^0)}
  & \gate{R_Z(\theta_2^0)}
  & \ctrl{1}
  & \qw
  & \qw
  & \targ{}
  & \qw\\
\lstick{$\ket{q_1}$}
  & \gate{R_Z(\theta_0^1)}
  & \gate{R_X(\arccos(\mu^*_1)\,\theta_1^1)}
  & \gate{R_Z(\theta_2^1)}
  & \targ{}
  & \ctrl{1}
  & \qw
  & \qw
  & \qw \\
\lstick{$\ket{q_2}$}
  & \gate{R_Z(\theta_0^2)}
  & \gate{R_X(\arccos(\mu^*_2)\,\theta_1^2)}
  & \gate{R_Z(\theta_2^2)}
  & \qw
  & \targ{}
  & \ctrl{1}
  & \qw 
  & \qw \\
\lstick{$\ket{q_3}$}
  & \gate{R_Z(\theta_0^3)}
  & \gate{R_X(\arccos(\mu^*_3)\,\theta_1^3)}
  & \gate{R_Z(\theta_2^3)}
  & \qw
  & \qw
  & \targ{}
  & \ctrl{-3}
  & \qw
\end{quantikz}

%% file: stochastic-vi.tex
\section{Stochastic variational inference}
\label{sec:svi}

\subsection{Statistical model}
\label{sec:stat-model}

\begin{figure}[]
	\centering
	\includegraphics[width=0.6\textwidth]{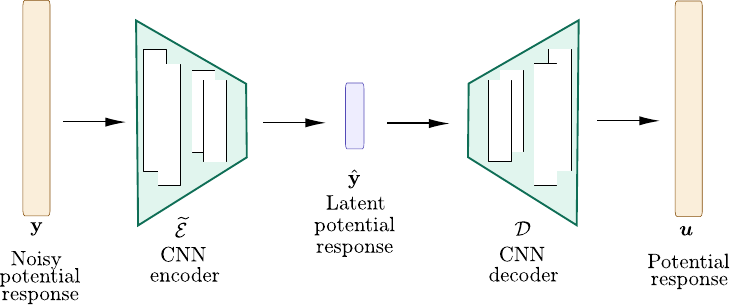}
	\caption{Modified architecture with observation-aware encoder $\tilde{\mathcal{E}}$ and decoder $\mathcal{D}$, in which $\vec{y} \in \mathbb{R}^{n_{\text{dof}}}$ is the noisy response field in physical space $\Omega$ and $\hat{\vec{y}} \in \mathbb{R}^{n_l^{u}}$ is the denoised response field in the latent space $\Omega^l$.}
	\label{fig:autoencoder-noisy-architecture}
\end{figure}

We aim in this work to obtain an estimate of anchor conductivity parameters $\vec \mu^{(a)}$ from measurements of the potential response by means of Bayesian inference. Since exact Bayesian inference is typically intractable and sampling methods can be challenging for high-dimensional inverse problems, in this work we use a variational inference (VI) framework to approximate the posterior distribution. We assume a statistical model connecting conductivity and the data at a set of measurement locations
\begin{equation}
	\vec{y} = \vec{u}(\vec \mu^{(a)}) + \vec{\varepsilon} \, ,
	\qquad
	\vec{\varepsilon} \sim \mathcal{N}(\vec{0}, \sigma_e^2 \vec{I})  \, ,
\end{equation}
where $\vec y$ is the observed data, $\sigma_e$ is the noise parameter, and $\vec u$ is the forward model result that is ultimately governed by the anchor conductivity parameters $\vec \mu^{(a)}$ defined at a set of control points $\{x^{(a)}_j\}_{j=1}^{n_a}$ as introduced in Section~\ref{sec:poisson-problem}. Specifically, the inhomogeneous conductivity field is obtained through interpolation involving the parameters 
\begin{equation}
	\label{eq:gp-basis}
	\mu(x) = \sum_j \phi_j(x) \, \mu_j^{(a)} = \vec \phi (x)^\trans \,  \vec \mu^{(a)}  \, ,
\end{equation}
where in this work $\phi_j$ is the interpolating basis function obtained by conditioning a Gaussian process at the anchor points; see, for example,~\cite{Girolami2021, Rasmussen2005} for details on constructing GP basis functions. We take the sampled conductivity field at the finite element centroids to yield the input vector $\vec \mu \in \mathbb R^{n_e}$. Following this understanding that the potential depends ultimately on the anchor parameters, in this section we explicitly write $\vec u(\vec \mu^{(a)})$, which also simplifies the notation. 

The forward model maps the conductivity field to the corresponding potential response. In the present work this map is provided by the quantum-assisted forward map $\mathcal{Q}$ acting in latent space, rather than by a full finite element solve, as illustrated in Figure~\ref{fig:hybrid-architecture}. Given a set of parameters $\vec \mu^{(a)}$, the discretised conductivity field $\vec \mu$ is reduced to its latent representation $\hat{\vec{\mu}}$ by the principal component projection of Section~\ref{sec:pca}, and the latent potential response follows as $\hat{\vec{u}} = \mathcal{Q}(\hat{\vec{\mu}}(\vec{\mu}^{(a)}))$. 

The learned autoencoder is not robust with respect to the measurement noise contained in the data. To improve its resilience, we obtain an observation-aware encoder $\tilde{\mathcal{E}}$ by retraining the encoder on the noise-augmented potential responses $\vec y$, as illustrated in Figure~\ref{fig:autoencoder-noisy-architecture}. The observation-aware encoder $\tilde{\mathcal{E}}$ is a denoising autoencoder~\cite{vincent2008extracting,vincent2010stacked} trained on noisy input and clean target. It denoises each observation toward the latent code of the underlying clean response. The decoder $\mathcal{D}$ then reconstructs the corresponding clean response field,
\begin{equation}
	\tilde{\vec{u}} = \mathcal{D}(\hat{\vec{y}}) \, ,
\end{equation}
so that the pair learns a denoising mapping from noisy measurements to the latent representation of the underlying response. This requires retraining the autoencoder on pairs of noisy observations and clean responses,
\begin{equation}
	\mathcal{L}_{\text{AE}}^{(y \rightarrow u)}
	=
	\frac{1}{n_s}
	\sum_{i=1}^{n_s}
	\big\|
	\vec{u}_{i} - \mathcal{D}\big(\tilde{\mathcal{E}}(\vec{y}_{i})\big)
	\big\|_2^2 \, ,
\end{equation}
so that the latent space remains consistent with the forward surrogate while being robust to observational noise. The key motivation for considering such a noise-resilient encoder is that we intend to perform Bayesian inversion in the latent space of the response rather than in the physical space. 

\subsection{Latent space inference}
\label{sec:latent-infer}


For each measurement $\vec y$, the latent representation is obtained as $\hat{\vec{y}} = \tilde{\mathcal{E}}(\vec{y})$. We also denote the forward map connecting $\vec \mu^{(a)}$ and $\hat{\vec u}$ is written as $\hat{\vec u} (\vec {\hat \mu^{(a)}}) = \mathcal Q \big( \vec V^{\trans} \big(\vec \mu(\vec{\mu}^{(a)}) - \overline{\vec \mu}\big) \big)$. 
%
%
The likelihood of the latent data given the anchor parameters $\vec \mu^{(a)}$ then reads
\begin{equation}
	\label{eq:likelihood-multireading}
	p(\hat{\vec{y}} \,|\, \vec{\mu}^{(a)})
	=
	\mathcal{N}\big(\hat{\vec{y}}; \, \hat{\vec u}(\vec \mu^{(a)}), \, \sigma_e^2 \vec{I}\big) \, .
\end{equation}
For settings in which $n_o$ repeated observations are available, the likelihood factorises as
\begin{equation}
	p(\hat{\vec{Y}} \,|\, \vec{\mu}^{(a)})
	=
	\prod_{k=1}^{n_o}
	\mathcal{N}\big(\hat{\vec{y}}^k; \, \hat{\vec u} (\vec \mu^{(a)}), \, \sigma_e^2 \vec{I}\big) \, .
\end{equation}
Although the nonlinear encoding $\tilde{\mathcal E}$ does not in general preserve a Gaussian isotropic noise, when the magnitude of $\sigma_e$ is small, the propagated noise remains approximately Gaussian~\cite{bishop2006}.

The inference target is the vector of anchor coefficients $\vec{\mu}^{(a)} \in \mathbb{R}^{n_a}$ that parameterises the conductivity field, as introduced in Section~\ref{sec:poisson-problem}. In variational inference the true posterior $p(\vec{\mu}^{(a)} \,|\, \hat{\vec{y}})$ is approximated by a variational distribution $q_{\vec{\xi}}(\vec{\mu}^{(a)})$, parameterised by $\vec{\xi}$, which is obtained by minimising the Kullback--Leibler (KL) divergence,
\begin{equation}
	\vec{\xi}^\ast
	=
	\arg\min_{\vec{\xi}} \,
	\text{KL}\big(q_{\vec{\xi}}(\vec{\mu}^{(a)}) \,\|\, p(\vec{\mu}^{(a)} \,|\, \hat{\vec{y}})\big) \, ,
\end{equation}
which is equivalent to maximising the evidence lower bound (ELBO),
\begin{equation} \label{eq:elbo}
	\mathcal{L}(\vec{\xi})
	=
	\mathbb{E}_{q_{\vec{\xi}}(\vec{\mu}^{(a)})} \left[ \log p(\hat{\vec{y}} \,|\, \vec{\mu}^{(a)}) \right]
	- \text{KL} \left(q_{\vec{\xi}}(\vec{\mu}^{(a)}) \,\|\, p(\vec{\mu}^{(a)})\right) \, .
\end{equation}
%
We place a mean-field Gaussian prior on the anchor coefficients,
\begin{equation}
	p(\vec{\mu}^{(a)}) = \mathcal{N} \left(\overline{\vec{\mu}}^{(a)}, \, \sigma_\mu^2 \vec{I} \right) \, ,
\end{equation}
where $\overline{\vec{\mu}}^{(a)}$ is the prior mean and $\sigma_\mu$ controls the prior uncertainty about the background conductivity level. 
The guide adopts a mean-field Gaussian form,
\begin{equation}
	q_{\vec{\xi}}(\vec{\mu}^{(a)}) =
	\mathcal{N} \left(\vec{m}_q, \, \text{diag}(\vec{\sigma}_q^2) \right) \, ,
\end{equation}
with variational parameters $\vec{\xi} = (\vec{m}_q, \vec{\sigma}_q)$. This mean-field form assumes the anchor coefficients are independent and so does not represent their posterior correlations. Other parameterisations recover this correlation structure~\cite{povala2022variational}, but are not considered here.
Because the latent surrogate is fitted on the training samples, its accuracy is assured only on the support of that population. During inference the guide may propose anchor coefficients outside this support, particularly in the early iterations. To limit this, we apply two safeguards. The mean of each variational factor is constrained to an interval $[\mu_{\min}, \mu_{\max}]$ inside the range used to generate the training set, and any guide draw that yields a non-positive conductivity field is rejected and redrawn. Together with the prior, these keep the sampled fields physically admissible and close to the training distribution in the experiments reported here. Verifying that the posterior mass remains within the surrogate domain of validity is advisable when the approach is applied to broader priors.

\subsection{Inference algorithm}
\label{sec:algo-infer}

We implement the inference using the stochastic variational inference (SVI) framework of \texttt{Pyro}  \citep{bingham2019}. The procedure uses stochastic gradient ascent to maximise the ELBO, sampling from the guide and evaluating the model and guide log-probabilities through automatic differentiation. \texttt{Pyro} requires the specification of a model, which defines the joint distribution $p(\vec{\mu}^{(a)}, \vec{y})$, and a guide, which defines the variational distribution $q_{\vec{\xi}}(\vec{\mu}^{(a)})$. During training, samples of $\vec{\mu}^{(a)}$ are drawn from the guide, the corresponding latent responses are evaluated through the surrogate to form the likelihood, and the ELBO is estimated by Monte Carlo sampling and maximised with the Adam optimiser~\cite{kingma2017adam}. The complete procedure is summarised in Algorithm~\ref{alg:svi-latent}.

\begin{algorithm}[]
	\caption{Stochastic variational inference in latent space}
	\label{alg:svi-latent}
	\begin{algorithmic}[1]
		\Require Observations $\{\vec{y}^k\}_{k=1}^{n_o}$, PCA basis $\vec{V}$ and mean $\overline{\vec{\mu}}$, trained forward map $\mathcal{Q}$, observation-aware encoder $\tilde{\mathcal{E}}$, prior $p(\vec{\mu}^{(a)}) = \mathcal{N}(\overline{\vec{\mu}}^{(a)}, \sigma_\mu^2 \vec{I})$, noise level $\sigma_e$, learning rate $\eta$, Monte Carlo sample size $n_{SVI}$, number of iterations $T$, training region $\mu_{\min}$, $\mu_{\max}$
		\Ensure Variational parameters $\vec{\xi} = (\vec{m}_q, \vec{\sigma}_q)$
		\State Encode each observation to the latent space, $\hat{\vec{y}}^k \gets \tilde{\mathcal{E}}(\vec{y}^k)$ for $k = 1,\dots,n_o$
		\State Initialise the variational parameters $\vec{m}_q$ and $\vec{\sigma}_q$, with each component of $\vec{m}_q$ constrained to $[\mu_{\min}, \mu_{\max}]$
		\For{$t = 1$ to $T$}
		\For{$s = 1$ to $n_{SVI}$}
		\State Draw $\vec{\epsilon}^s \sim \mathcal{N}(\vec{0}, \vec{I})$ and set $\vec{\mu}^{(a),s} \gets \vec{m}_q + \vec{\sigma}_q \odot \vec{\epsilon}^s$ 
        \Comment{reparameterisation}
		\State Interpolate the conductivity field $\vec{\mu}^s$ from $\vec{\mu}^{(a),s}$ using Eq.~\eqref{eq:gp-basis}
		\State If $\mu^s(x) \leq 0$ for any spatial coordinate $x$, reject $\vec{\mu}^{(a),s}$ and return to step~5; otherwise accept.
		\State Reduce to the latent conductivity, $\hat{\vec{\mu}}^s \gets \vec{V}^\trans(\vec{\mu}^s - \overline{\vec{\mu}})$
		\State Predict the latent response, $\hat{\vec{u}}^s \gets \mathcal{Q}(\hat{\vec{\mu}}^s)$
		\EndFor
		\State Estimate the ELBO,
		\Statex \hspace{\algorithmicindent}$\displaystyle \tilde{\mathcal{L}}(\vec{\xi}) \gets \frac{1}{n_{SVI}}\sum_{s=1}^{n_{SVI}}\sum_{k=1}^{n_o} \log \mathcal{N}\!\big(\hat{\vec{y}}^k; \, \hat{\vec{u}}^s, \, \sigma_e^2 \vec{I}\big) - \text{KL}\!\big(q_{\vec{\xi}}(\vec{\mu}^{(a)}) \,\|\, p(\vec{\mu}^{(a)})\big)$
		\State Update the variational parameters, $\vec{\xi} \gets \vec{\xi} + \eta \, \nabla_{\vec{\xi}} \tilde{\mathcal{L}}(\vec{\xi})$ \Comment{Adam~\cite{kingma2017adam}}
		\EndFor
		\State \Return $\vec{\xi}$
	\end{algorithmic}
\end{algorithm}

This approach enables scalable inference over high-dimensional fields with limited data while maintaining a principled treatment of uncertainty. The learned variational distribution $q_{\vec{\xi}}(\vec{\mu}^{(a)})$ can be used to draw posterior samples and to construct credible intervals for downstream predictions, and the observation-aware latent mapping improves the robustness and consistency between the surrogate and the likelihood. 

The gradient of the ELBO estimate with respect to the variational parameters $\vec{\xi}$ is formed differently depending on where the forward map is evaluated. On the noiseless quantum simulator the surrogate $\mathcal{Q}$ is a differentiable operation, so the reparameterisation used in Algorithm~\ref{alg:svi-latent} lets $\nabla_{\vec{\xi}} \tilde{\mathcal{L}}$ propagate through the whole pathway from the guide sample to the likelihood by automatic differentiation, which is the default backpropagation route in \texttt{Pyro} through \texttt{Trace\_ELBO}. 
On NISQ hardware the device returns expectation values as a black box and cannot be differentiated on-device, so this pathwise gradient is unavailable. We therefore mark the guide sites as non-reparameterisable, for which \texttt{Pyro} falls back to the score-function (REINFORCE) estimator~\cite{Schulman2015},
\begin{equation}
	\label{eq:score-function-grad}
	\nabla_{\vec{\xi}} \, \mathbb{E}_{q_{\vec{\xi}}} \!\left[ \log p(\hat{\vec{y}} \,|\, \vec{\mu}^{(a)}) \right]
	=
	\mathbb{E}_{q_{\vec{\xi}}} \!\left[ \left( \log p(\hat{\vec{y}} \,|\, \vec{\mu}^{(a)}) - b \right) \nabla_{\vec{\xi}} \log q_{\vec{\xi}}(\vec{\mu}^{(a)}) \right] \, ,
\end{equation}
with the Gaussian KL term differentiated in closed form. The baseline $b$, a decaying average of past log-likelihood values provided natively by \texttt{Pyro}, reduces the variance of the estimator without biasing it~\cite{mnih2014neuralVI}. Since the hardware likelihood enters only as a detached constant, the estimator reuses the forward evaluations already drawn for the ELBO, avoiding the repeated device calls per latent coordinate that the parameter-shift rule would require.

%% file: examples.tex
\section{Examples}
\label{sec:examples}

\subsection{One-dimensional forward problem}
\label{sec:1d-forward}

In the first example, we assess the quality of the proposed quantum latent surrogate model against finite element (FE) analysis results. We adopt the one-dimensional Poisson--Dirichlet problem of Section~\ref{sec:poisson-problem}. 
The spatially varying conductivity is parameterised as in Eq.~\eqref{eq:gp-basis}
with the interpolating basis functions obtained by conditioning a Gaussian process at $11$ anchor points with uniform spacing $h = 0.1$. The kernel is a stationary squared exponential with lengthscale $\ell \in \{0.1, 0.2, 0.3\}$ and scaling parameter $\sigma_f = 0.1$. The anchor coefficients $\mu_j^{(a)}$ are sampled independently from a uniform distribution between $0.4$ and $2.0$, yielding a training dataset of $n_s = 10^4$ field samples. The corresponding potential response $u(x)$ is computed using linear finite elements with $n_{\mathrm{dof}} = 128$ degrees of freedom, giving the input--output pairs $\{\vec{\mu}^{(i)},\, \vec{u}^{(i)}\}_{i=1}^{n_s}$. 

Prior to training, the encoder--decoder architecture requires the selection of appropriate latent dimensions for the input conductivity and output potential fields. From the $n_s = 10^4$ training samples, the explained variance saturates rapidly for both fields, retaining $n_l^{\mu} = 9$ components captures the essential variability of the conductivity-coefficient ensemble, while $n_l^u = 6$
components suffice for the potential field. This analysis confirms that a
low-dimensional latent representation is suitable for this problem and informs
the choice of input and output dimensions of the intrinsic latent space. 
On the other hand, the architecture of the convolutional autoencoder used for the 
response vector $\vec u$ is shown previously in Section~\ref{sec:autoencoder},  Figure~\ref{fig:autoencoder-architecture}. The autoencoder is trained using Adam optimiser~\cite{kingma2017adam} with a learning rate of $0.001$ for 200 epochs. 

\begin{figure}[]
	\centering
	\subfloat[Validation loss history\label{fig:validation-loss-1d}]{
		\includegraphics[width=0.48\textwidth]{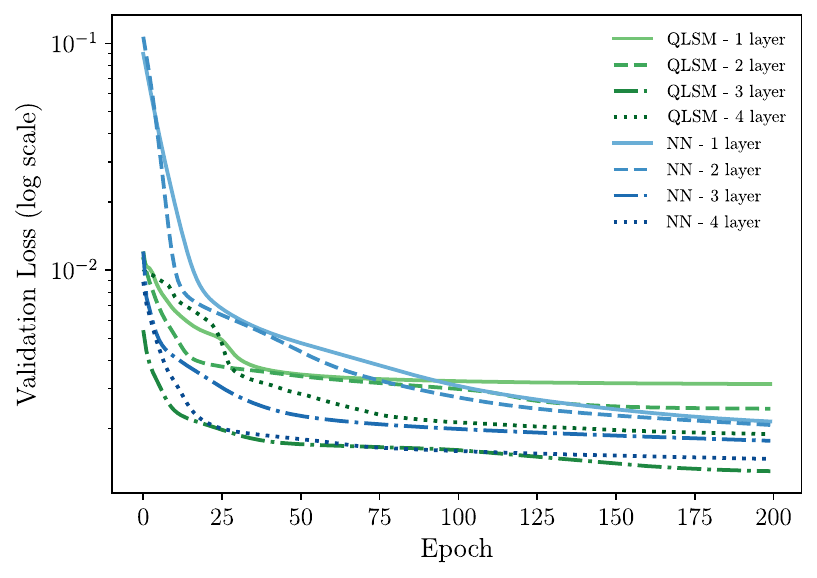}
	}
	\hfill
	\subfloat[Trainable parameter count\label{fig:param-1d-sigma}]{%
		\includegraphics[width=0.48\textwidth]{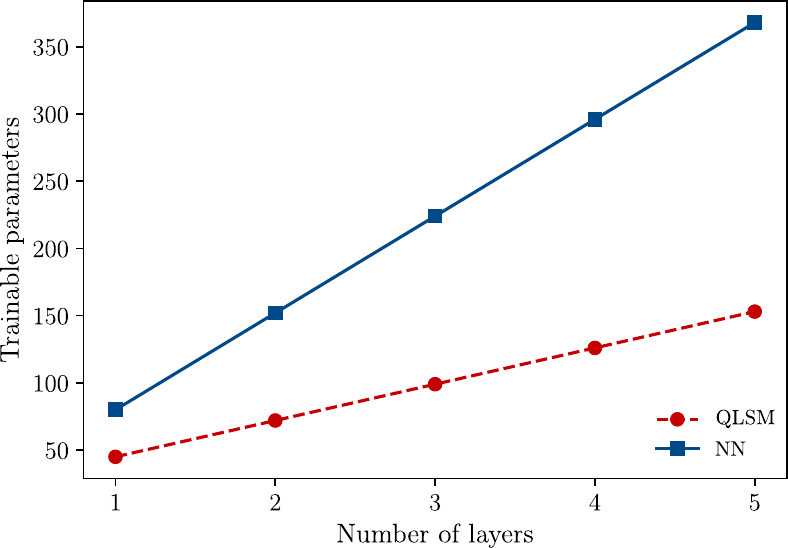}%
	}
	\caption{Validation losses and parameter count for the one-dimensional
		Poisson forward prediction problem using the QLSM and classical
		neural-network forward map across model depths.}
	\label{fig:1d-training-validation-loss}
\end{figure}

With the latent dimensions configured, we next study the effect of QLSM architecture on the approximation error. For this purpose, we consider QLSM models with depths ranging
from one to four PQC layers and compared against a classical fully connected
neural-network latent map of equivalent depth. Both model classes share an
identical encoder--decoder module, and the distinction arises solely in the
latent-space mapping, where the QLSM employs parameterised quantum circuits (PQCs),
while the classical network uses fully connected layers with tanh activation
functions. All models are constructed with identical latent input and output
dimensions $n_l^{\mu} = 9$ and $n_l^{u} = 6$, with the classical NN layers using
a fixed hidden dimension of $8$ neurons. All models are trained under identical
conditions using the Adam optimiser~\cite{kingma2017adam} with a learning rate of
$0.0001$ for $200$ epochs, using the mean squared error (MSE) as the objective
function. The training data pairs are transformed into their latent
representation, where PCA is used to obtain $\hat{\vec \mu}^{(i)}$ and the encoder is
used to produce $\hat{\vec u}^{(i)}$. The projected dataset then comprises
$\{\hat{\vec \mu}^{(i)},\, \hat{\vec u}^{(i)}\}_{i=1}^{n_s}$. We reserve $10\%$ of the
data pairs for validation and $10\%$ for testing. The validation loss histories
across the various network depths are presented in
Figure~\ref{fig:validation-loss-1d}, and the parameter counts for each
configuration are shown in Figure~\ref{fig:param-1d-sigma}. All models exhibit a
sharp initial decrease in MSE during the early epochs, followed by gradual
convergence towards a steady plateau. For the classical NN models, increasing the
depth improves performance up to the four-layer configuration, which achieves the
lowest validation loss among the NN models. For the QLSM, accuracy improves with depth up to three layers, beyond which the four-layer circuit degrades. Based on
these observations, the three-layer QLSM and the four-layer neural network are
selected as the optimal representatives of their respective model classes and
retained for all subsequent analyses.

\begin{figure}[]
	\centering
	\subfloat[Conductivity test samples $\mu(x)$\label{fig:mu-1d}]{%
	\includegraphics[width=0.48\textwidth]{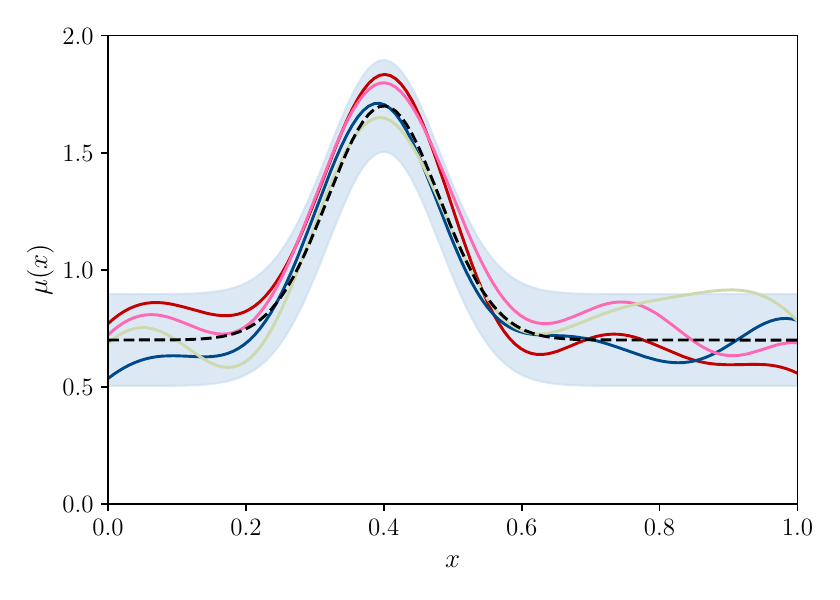}%
	}
	\hfill
	\subfloat[Mean solution comparison\label{fig:u-comparison-1d}]{%
		\includegraphics[width=0.48\textwidth]{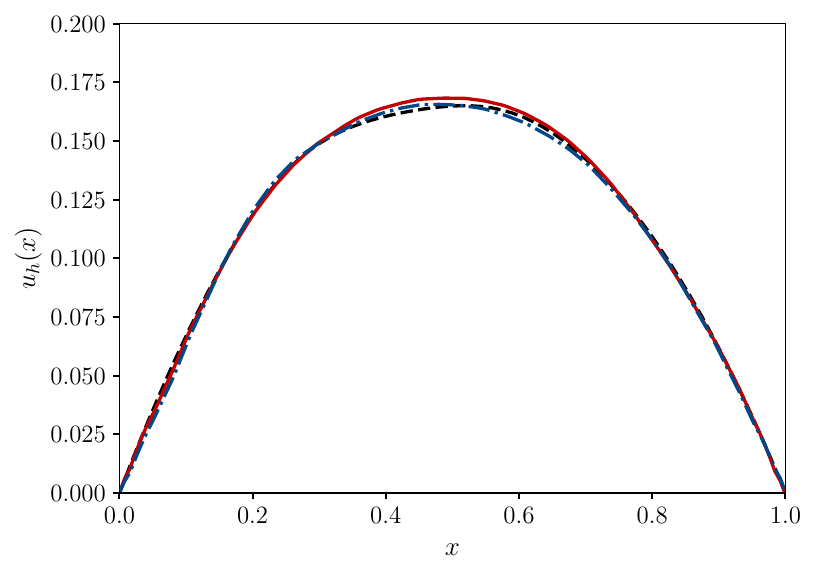}%
	}
	\\
	\subfloat[QLSM forward solutions $u_q(x)$\label{fig:u-qalm-1d}]{%
		\includegraphics[width=0.48\textwidth]{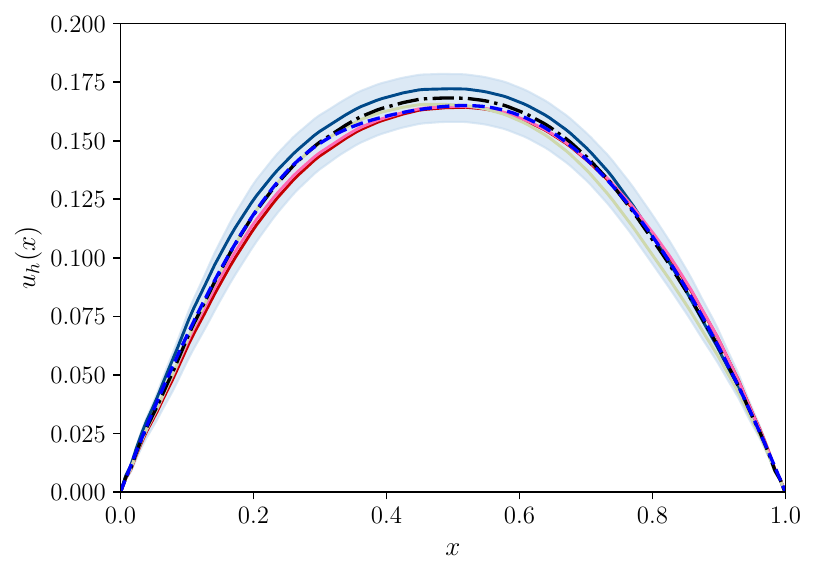}%
	}
	\hfill
	\subfloat[Reference forward solutions obtained from FE and NN \label{fig:u-nn-1d}]{%
		\includegraphics[width=0.48\textwidth]{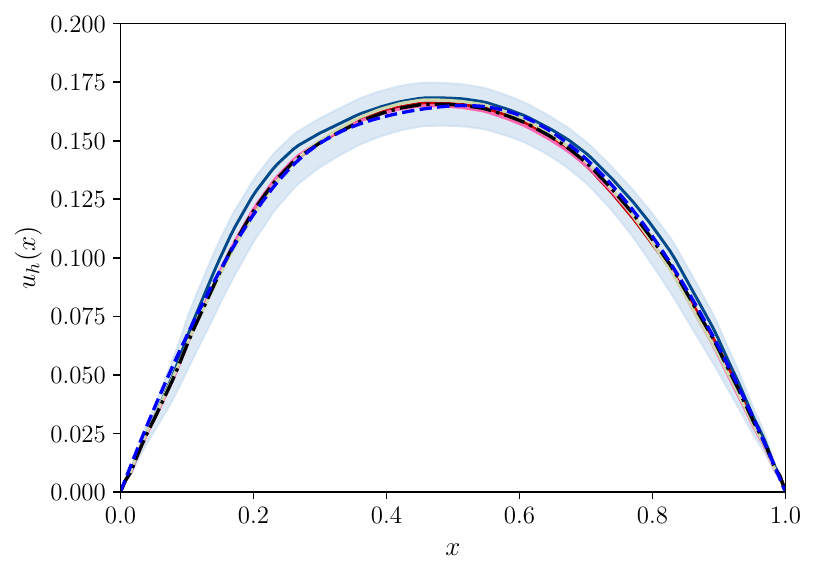}%
	}
	\caption{One-dimensional model forward problem using the Poisson equation
		with spatially varying material conductivity $\mu(x)$.
		(a) Five sample draws of the conductivity field from the test distribution.
		(b) Comparison of the mean forward solutions from the QLSM (red solid),
		classical neural network (blue dash-dot), and reference finite element
		solution (black dashed). (c) Push-forward distribution of the potential
		response using the proposed QLSM. (d) Corresponding push-forward
		distribution using the classical FE reference and NN surrogate.}
	\label{fig:1d_mu_and_u}
\end{figure}

The optimal configurations identified in the hyperparameter study are then
deployed as forward surrogate maps and evaluated on the test ensemble. Test
samples of the conductivity field are drawn from a GP with a prescribed mean
function
\begin{equation}
	\label{eq:gp-mean-basis-shifted}
	\mu(x)^{\text{test}} = \overline{\mu}(x) + \vec \phi(x)^\trans \,
	\vec{\mu}^{(a),\,\text{test}} \, ,
\end{equation}
with the prescribed mean
\begin{equation}
	\label{eq:test-mean-1d}
	\overline{\mu}(x) = 0.7 + \exp\left(-\frac{(x - 0.4)^2}{0.01314}\right) \, ,
\end{equation}
where the kernel is a squared exponential with lengthscale $\ell = 0.1$ and
scaling factor $\sigma_f = 0.1$. The coefficients $\vec{\mu}^{(a),\,\text{test}}$ are
sampled from $\mathcal{N}(\vec{0},\, 0.01\,\vec{I})$ at the training anchor points
$\{x^{(a)}\}$. Figure~\ref{fig:mu-1d} shows samples of the test conductivity field, 
with the shaded region indicating the $95\%$ confidence interval over
$n_s^{\text{test}} = 1000$ samples. Figure~\ref{fig:u-comparison-1d} presents a
direct comparison of the mean forward solutions from the QLSM and the classical NN
against the reference finite element solution. The QLSM is seen to provide a
reasonable approximation to the reference mean field obtained through finite
elements. The push-forward distributions of the potential response shown in
Figure~\ref{fig:u-qalm-1d} confirm that the proposed quantum-assisted surrogate
produces distributions with comparable uncertainty bands to those of the finite
element reference.

We further quantify the accuracy of the QLSM through the relative $L^2$-norm error
with respect to the reference finite element solution over
$n_s^{\text{test}} = 1000$ samples,
\begin{equation}
	\epsilon_{L^2} = \frac{1}{n_s^{\text{test}}}\sum_{i=1}^{n_s^{\text{test}}}
	\frac{\| u_{\mathrm{FE}}^{(i)} - u_q^{(i)} \|}
	{\| u_{\mathrm{FE}}^{(i)} \|} \, .
\end{equation}
The potential $u_q$ predicted by the three-layer QLSM achieves a mean relative $L^2$ error of $\epsilon_{L^2} = 0.0153$ with respect to the FE reference, where the error is averaged over all test samples. This is comparable to the optimal NN configuration, which gives $\epsilon_{L^2} = 0.0156$. The corresponding trainable-parameter counts across all depths are summarised in Figure~\ref{fig:param-1d-sigma}. At every depth the QLSM uses fewer trainable parameters than the classical network, with the optimal three-layer QLSM using $99$ parameters against $224$ for the most accurate NN configuration. We report this as an observation rather than as a claim of efficiency. 
Trainable-parameter count is in any case not a like-for-like measure of cost, since each evaluation of the quantum map is estimated from repeated measurement shots, here $n^{\text{shots}} = 1000$ per circuit evaluation, whereas a network forward pass is deterministic. We therefore read this experiment as confirming the viability of the quantum-assisted route at accuracy comparable to the classical surrogate. 

\begin{figure}[]
	\centering
	\includegraphics[width=0.48\textwidth]{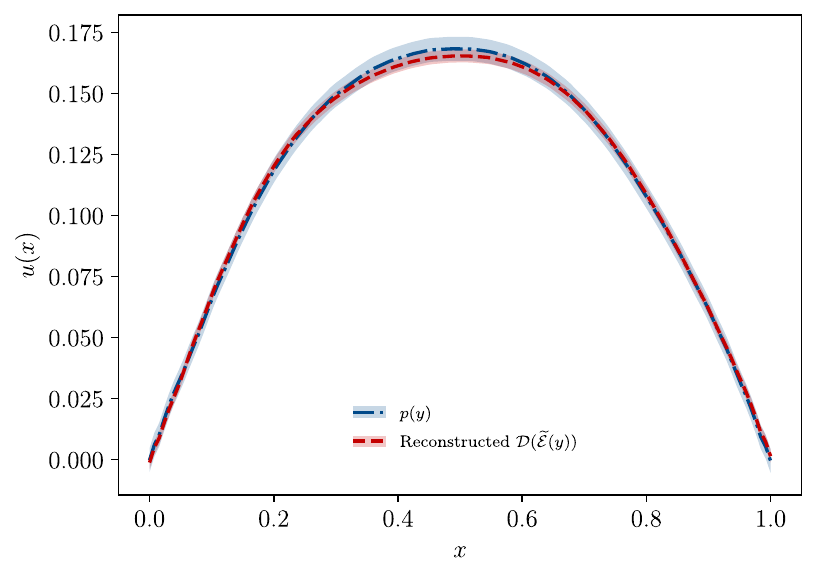}
	\caption{Comparison between the observation profile $p(\vec Y)$ and the reconstructed samples through the observation-aware encoder $\mathcal{D}(\tilde{\mathcal{E}}(\vec Y))$ for the test dataset.}
	\label{fig:1d-noisy-enc-dec}
\end{figure}

Before moving to the inversion, we assess the performance of the observation-aware encoder described in Section~\ref{sec:stat-model} that will be used to map the observation into the latent space for inference. Specifically, the observation-aware encoder shares the same architecture with the one described in Section~\ref{sec:autoencoder} and trained using the noise-augmented training dataset, with Gaussian white noise of $\sigma_e = 0.0025$. In Figure~\ref{fig:1d-noisy-enc-dec}, the distribution mean of the test observation is shown as a blue dashed line, with the respective confidence interval in blue shaded region. As a sanity check, we reconstruct the population of the test observation using the learned encoder, i.e., $\mathcal{D}(\tilde{\mathcal{E}}(\vec y))$. Figure~\ref{fig:1d-noisy-enc-dec} shows that the reconstructed profile matches the original observation distribution quite well, with the means having an $L^2$-norm relative discrepancy of $\Delta \epsilon_{L^2} = 0.0143$. 

\subsection{One-dimensional inverse problem}
\label{sec:1d-inverse}

We consider a one-dimensional inverse problem aimed at identifying the
spatially varying conductivity field from noisy observations of
the potential field. As the forward problem, we adopt the Poisson--Dirichlet
problem studied in Section~\ref{sec:1d-forward}. The objective is to infer the
conductivity coefficients $\vec{\mu}^{(a)}$ at the anchor points that describe the
conductivity field in Eq.~\eqref{eq:gp-basis}. It is worth emphasising that, for
efficiency, we work in the latent space of the potential field rather than in the
original observation space, as discussed in Section~\ref{sec:svi}. Specifically,
we first generate the observations according to the data-generating model
\begin{equation}
	\label{eq:stat-model-1d}
	\vec{y}^k = \vec{u} + \vec{\varepsilon}^k \, ,
	\qquad \vec{\varepsilon}^k \sim \mathcal{N}(\vec{0}, \sigma_e^2 \vec{I}) \, ,
\end{equation}
where $k = \{1, \ldots, n_o\}$. Note that, to generate the observations, the true
potential $\vec u$ is computed using the QLSM forward propagation of the true coefficients $\vec{\mu}^{(a),\,\text{true}}$, and the measurements
are taken at the discretisation nodes. We use a noise parameter
$\sigma_e = 0.0025$, which is fixed and not inferred. The latent-space observations for the proposed Bayesian inversion are obtained using the pre-trained noise-resilient encoder $\tilde{\mathcal E}$ discussed in Section~\ref{sec:svi}, and are collected in the measurement matrix $\hat{\vec{Y}} = \left(\hat{\vec{y}}^1, \ldots, \hat{\vec{y}}^{n_o}\right)
\in \mathbb{R}^{n_l^u \times n_o}$, where $n_l^u = 6$ denotes the number of latent
dimensions of the potential and $n_o \in \{1, 5, 10, 25\}$ denotes the number of
independent repeated measurements. Although it is not considered here, sparser or
denser observations could be accommodated by introducing a projection operator
connecting the data and model spaces; see, e.g.,~\cite{Girolami2021, Febrianto2022}.

Next, the likelihood of all observation data for a given set of parameters
$\vec{\mu}^{(a)}$ follows from the statistical model in Eq.~\eqref{eq:stat-model-1d} according to Eq.~\eqref{eq:likelihood-multireading}.
%
In this section, we use a prior distribution for the material coefficients
\begin{equation}
	p(\vec{\mu}^{(a)}) = \mathcal{N} \, \left(
	1.2 \,\vec{1}_{11}, \, 0.2^2 \,\vec{I}_{11} \right) \, ,
\end{equation}
with a constant prior mean and moderate uncertainty about the background
conductivity level, while remaining agnostic about the spatial structure. The
prior mean of $1.2$ is deliberately offset from the true background value of
$0.7$ used in the test cases below, so that the inference is examined under a
mildly misspecified and structurally non-informative prior rather than one centred
on the truth. The target posterior is solved using the stochastic variational inference framework described in Section~\ref{sec:svi}, where we use $40,000$ SVI steps with Adam optimiser with learning rate of 0.001 and $n_{SVI} = 2$ Monte Carlo samples per gradient step, for each value of $n_o$.

\subsubsection{Single inclusion}
\label{sec:1d-poisson-inv-singleint}
\begin{figure}[]
	\centering
	    \subfloat[Variational inference \label{fig:pred-sing-1d}]{%
		\includegraphics[width=0.48\textwidth]{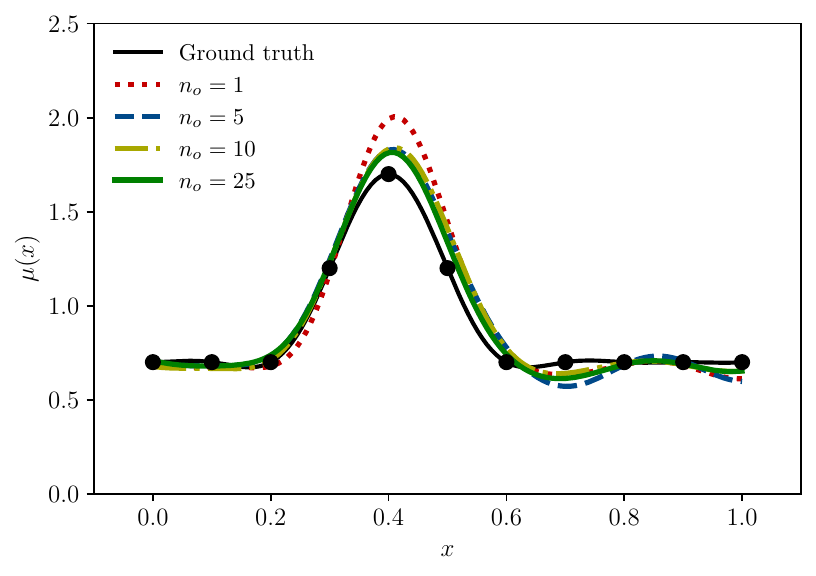}%
	}
	\subfloat[Comparison with MCMC \label{fig:pred-sing-mcmc-1d}]{%
		\includegraphics[width=0.48\textwidth]{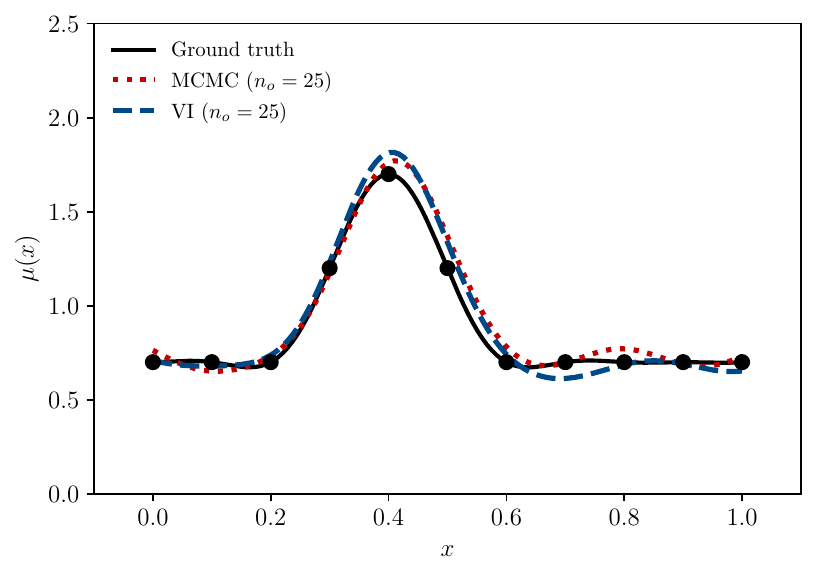}%
	}
	\caption{Posterior mean conductivity profile $\mu(x)$ recovered by the variational 
		inference framework for increasing numbers of repeated observations 
		$n_o \in \{1, 5, 10, 25\}$. The solid black line indicates the ground-truth 
		conductivity profile and the black dots mark the anchor locations 
		$\{x_j^{(a)}\}$ and their corresponding true coefficient values.}
	\label{fig:1d-prediction-singleint}
\end{figure}

\begin{figure}[]
	\centering
	\subfloat[$\mu^{(a)}_0$]{%
		\includegraphics[width=0.24\textwidth]{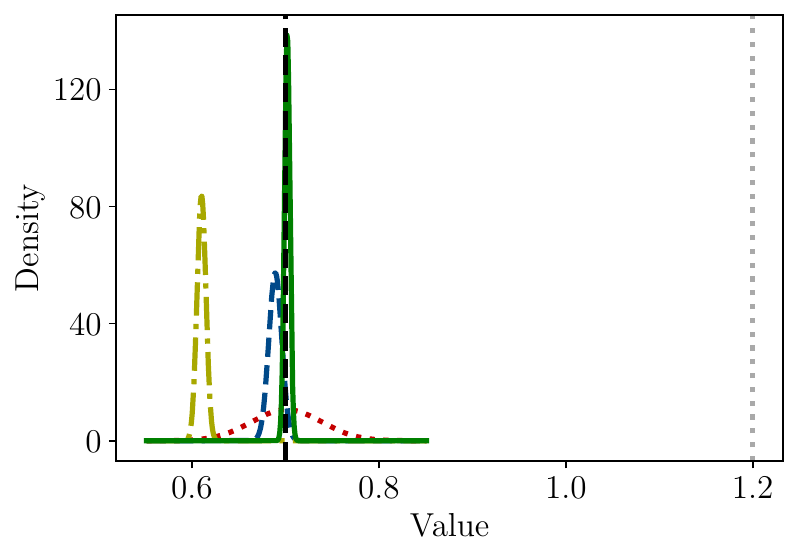}%
	}
	\subfloat[$\mu^{(a)}_1$]{%
		\includegraphics[width=0.24\textwidth]{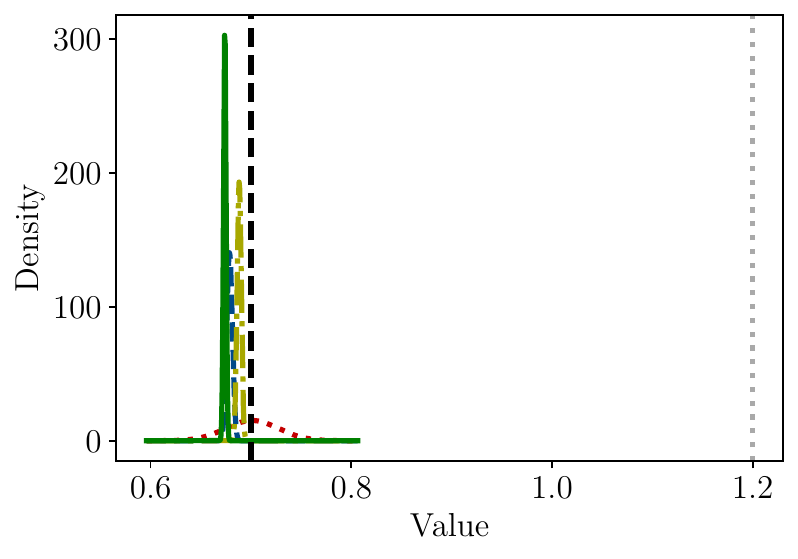}%
	}
	\subfloat[$\mu^{(a)}_2$]{%
		\includegraphics[width=0.24\textwidth]{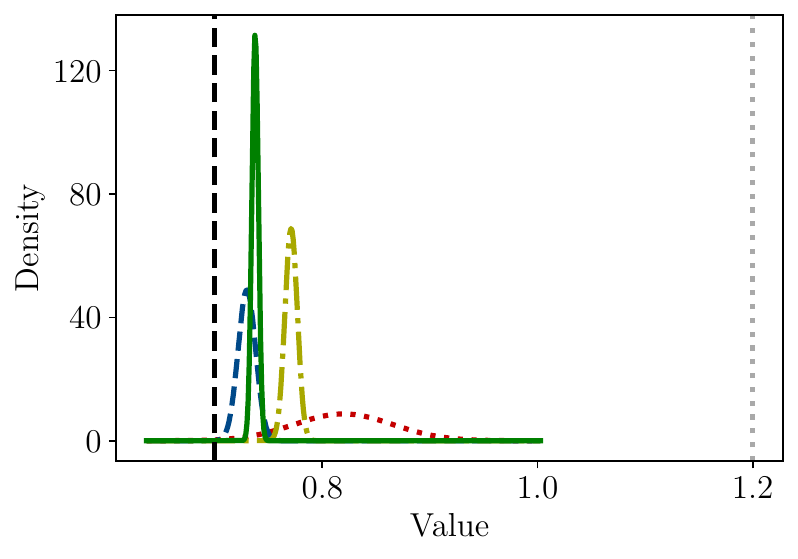}%
	}
	\subfloat[$\mu^{(a)}_3$]{%
		\includegraphics[width=0.24\textwidth]{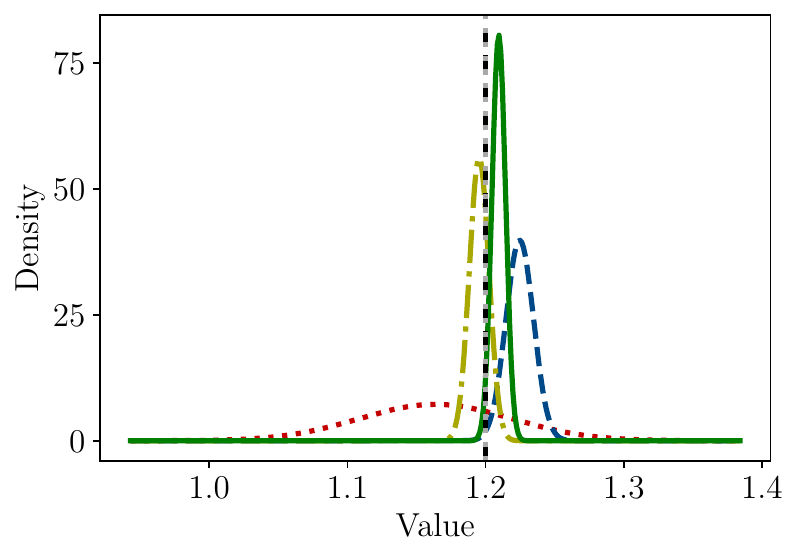}%
	}
	
	\vspace{0.5em}
	
	\subfloat[$\mu^{(a)}_4$]{%
		\includegraphics[width=0.24\textwidth]{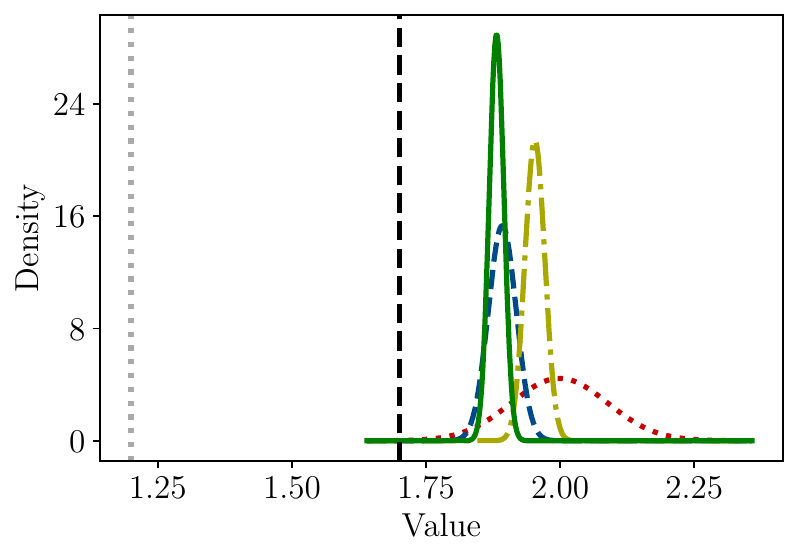}%
	}
	\subfloat[$\mu^{(a)}_5$]{%
		\includegraphics[width=0.24\textwidth]{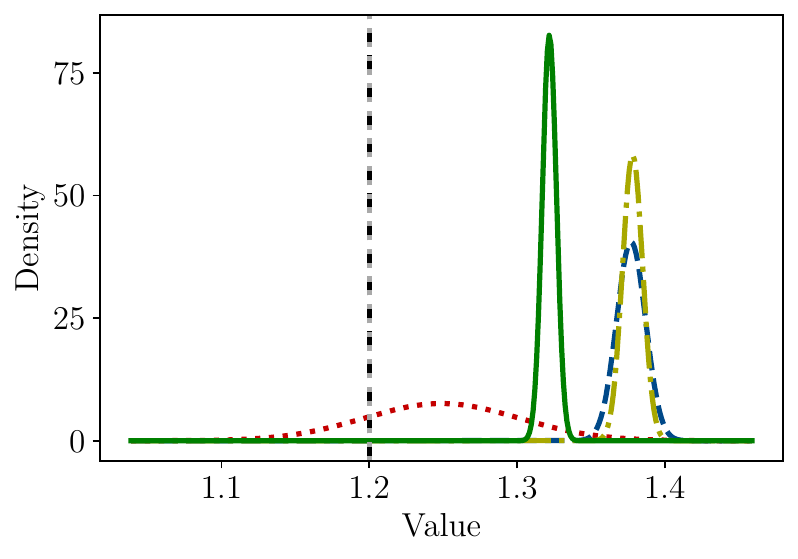}%
	}
	\subfloat[$\mu^{(a)}_6$]{%
		\includegraphics[width=0.24\textwidth]{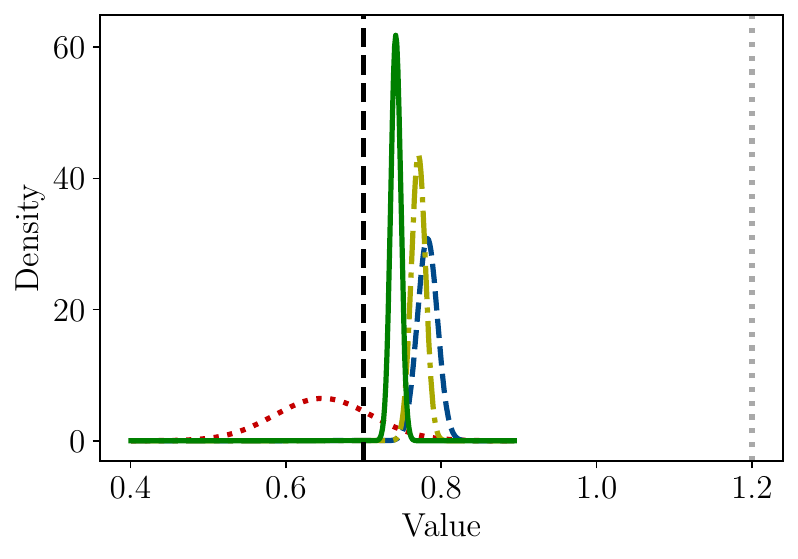}%
	}
	\subfloat[$\mu^{(a)}_7$]{%
		\includegraphics[width=0.24\textwidth]{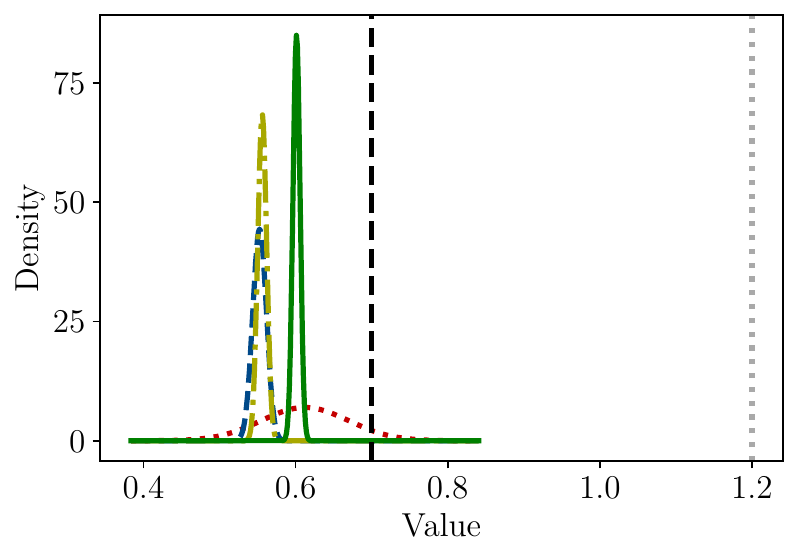}%
	}
	
	\vspace{0.5em}
	
	\subfloat[$\mu^{(a)}_8$]{%
		\includegraphics[width=0.24\textwidth]{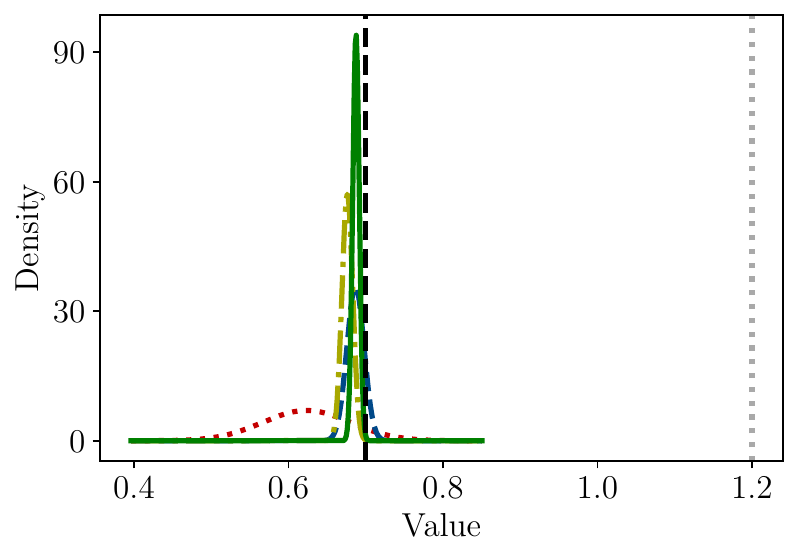}%
	}
	\subfloat[$\mu^{(a)}_9$]{%
		\includegraphics[width=0.24\textwidth]{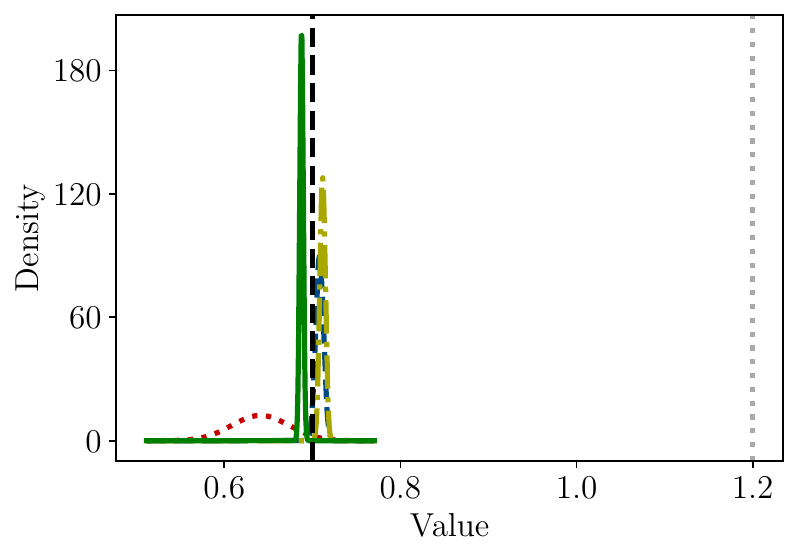}%
	}
	\subfloat[$\mu^{(a)}_{10}$]{%
		\includegraphics[width=0.24\textwidth]{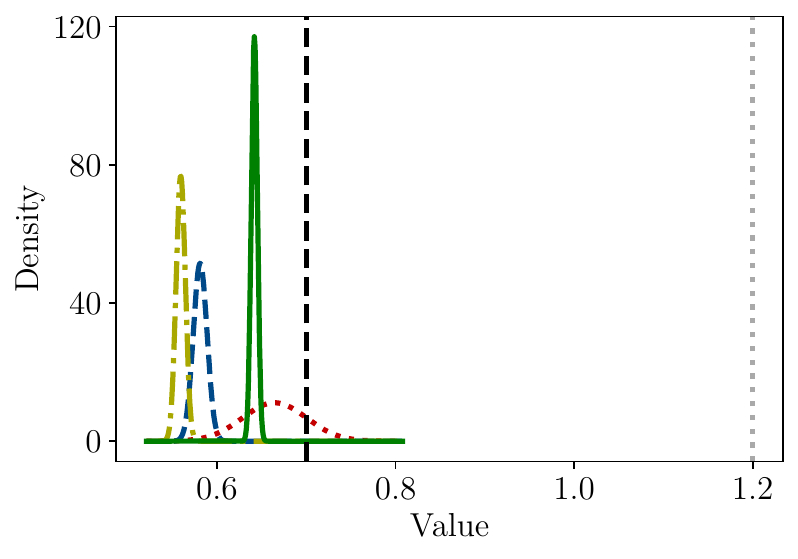}%
	}
	\subfloat{%
		\includegraphics[width=0.24\textwidth]{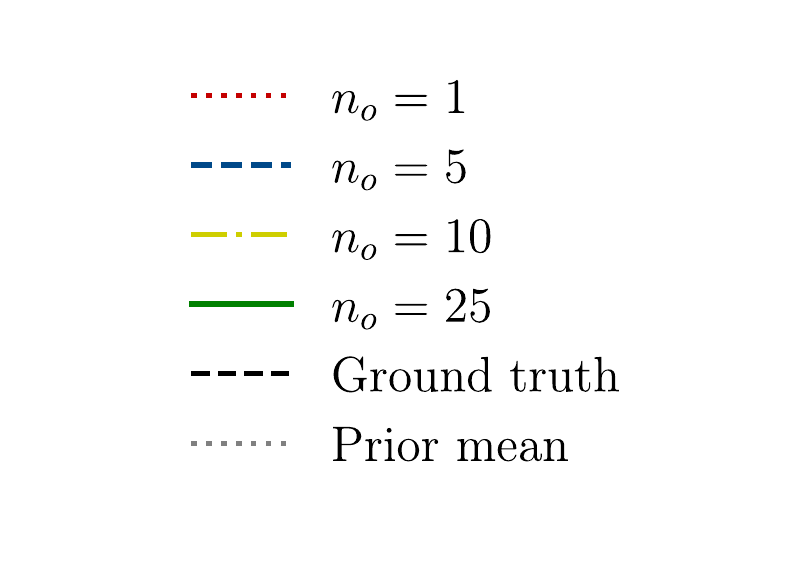}%
	}
	\caption{Posterior distributions of the anchor parameters $\mu_j^{(a)}$ for the single-inclusion inverse problem for different numbers of observations $n_o$.}
	\label{fig:1d-posterior_singleint}
\end{figure}

We consider the inversion of a single localised inclusion, characterised by a
region of elevated conductivity centred on the anchor point $x^{(a)} = 0.4$.
Specifically, the ground-truth coefficient vector is
\begin{equation}
	\vec{\mu}^{(a), \, \text{true}} = \left( 
	0.7,\, 0.7,\, 0.7,\, 1.2,\, 1.7,\, 1.2,\, 
	0.7,\, 0.7,\, 0.7,\, 0.7,\, 0.7 
	\right)^\trans \, ,
\end{equation}
where the background conductivity is assigned a uniform value of $0.7$ and the
inclusion anchors take the elevated values around $x^{(a)} = 0.4$. We use this
parameter set to obtain the latent observation $\hat{\vec{Y}}$ as described in
Section~\ref{sec:1d-inverse}. Figure~\ref{fig:pred-sing-1d} shows the
posterior mean conductivity field recovered for increasing numbers of repeated
measurements $n_o \in \{1, 5, 10, 25\}$. The values at the anchor points are
obtained by taking the mean of the posterior distribution
$p(\vec{\mu}^{(a)} \mid \hat{\vec{Y}})$ and the field over space is reconstructed
using Eq.~\eqref{eq:gp-basis}. For $n_o = 1$, the posterior mean deviates noticeably
from the true profile, but as $n_o$ increases, the predicted curves move closer
to the ground truth. This trend is corroborated quantitatively by the relative
$L^2$ error between the posterior mean and the ground-truth profile, which
decreases monotonically with $n_o$, from $\epsilon_{L^2} = 0.1310$ at
$n_o = 1$, to $\epsilon_{L^2} =0.0948$ at $n_o = 5$, $\epsilon_{L^2} = 0.0900$ at $n_o = 10$, and $\epsilon_{L^2} = 0.0697$ at
$n_o = 25$. It is noticeable that artefacts appear, especially near the
boundary. This behaviour arises from overfitting of the Gaussian process
interpolator, which permits more degrees of freedom than the data can constrain.
Figure~\ref{fig:1d-posterior_singleint} shows the marginal posterior
distributions of all eleven anchor coefficients for each value of $n_o$, along
with the ground-truth values $\vec{\mu}^{(a),\text{true}}$ indicated by
dashed black vertical lines and the prior mean by the dotted grey line. It is
evident that more observations lead to better inference of the mean values at the
anchor points. It is also noticeable that the inferred standard deviation decreases as the number of observations increases. 

Figure~\ref{fig:pred-sing-mcmc-1d} compares the inference using SVI and sampling approach using MCMC for $n_o = 25$ case. We use NUTS sampler within the MCMC module in \texttt{Pyro} for $50,000$ steps, with the first $10,000$ samples discarded to account for the warm-up stage. The resulting posterior mean of the MCMC has a relative $L^2$ error of $\epsilon_{L^2} = 0.0715$, which is slightly higher than the $\epsilon_{L^2} = 0.0697$ obtained with SVI in this specific case.

\subsubsection{Dual inclusion}
\label{sec:1d-poisson-inv-twoint}

\begin{figure}[]
	\centering
	
    \subfloat[Variational inference]{%
		\includegraphics[width=0.48\textwidth]{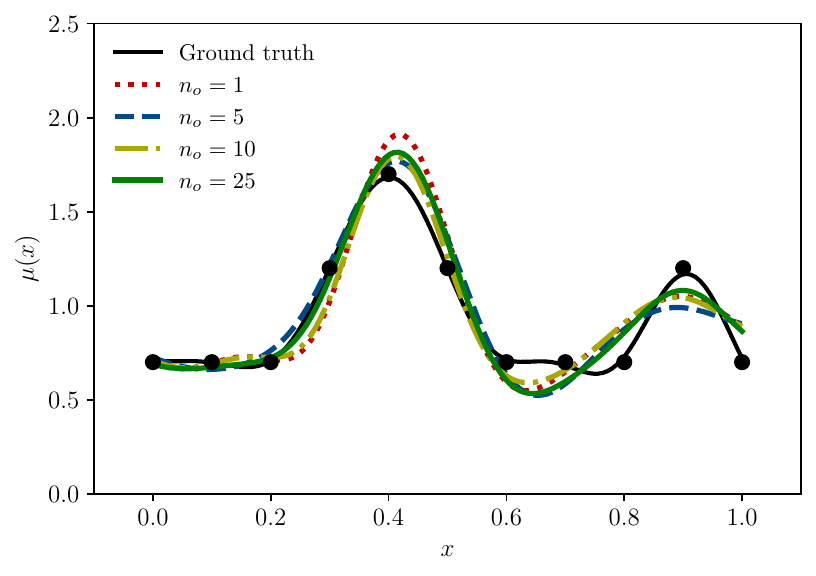}%
	}
	\subfloat[MCMC comparison]{%
		\includegraphics[width=0.48\textwidth]{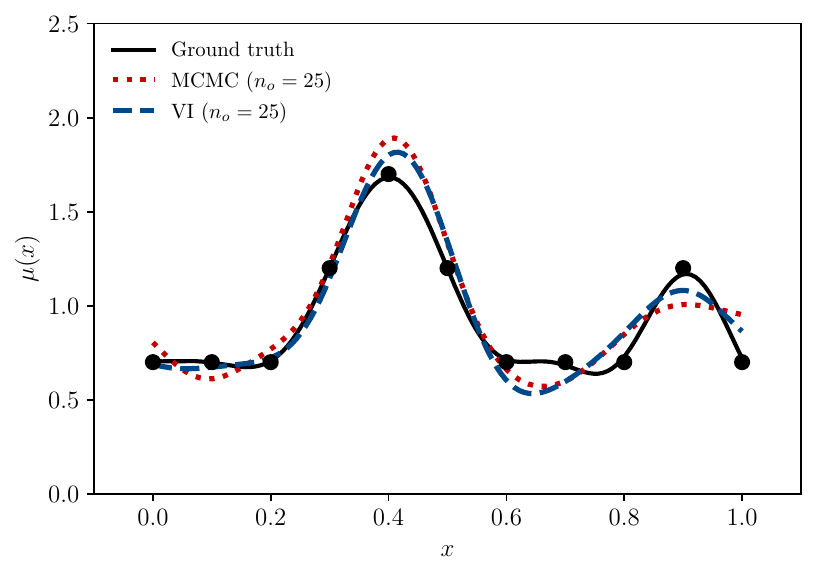}%
	}
	\caption{Posterior mean conductivity $\mu(x)$ recovered by the variational 
		inference framework for increasing numbers of repeated observations 
		$n_o \in \{1, 5, 10, 25\}$. The solid black line indicates the ground-truth 
		conductivity profile and the black dots mark the anchor locations 
		$\{x_j^{(a)}\}$ and their corresponding true coefficient values. The primary 
		inclusion is centred at $x_4^{(a)}$ and the secondary inclusion is located at $x_9^{(a)}$.}
	\label{fig:1d-prediction-twoint}
\end{figure}

\begin{figure}[]
	\centering
	\subfloat[$\mu^{(a)}_0$]{%
		\includegraphics[width=0.24\textwidth]{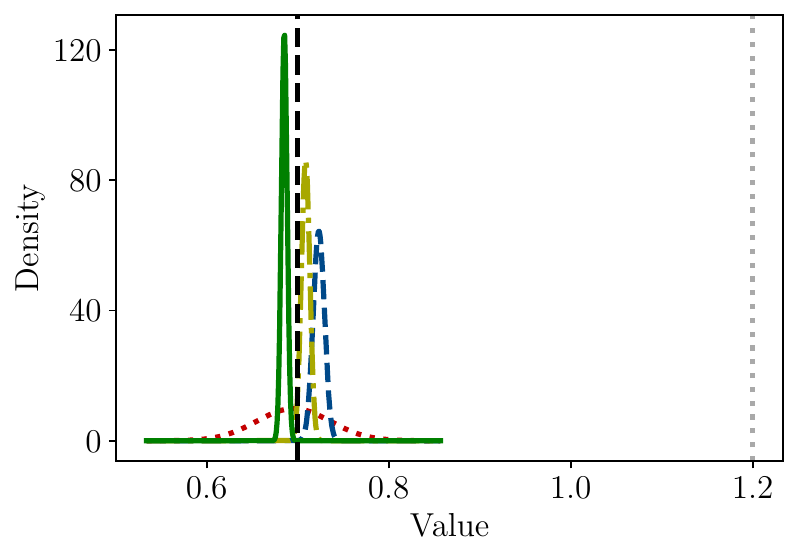}%
	}
	\subfloat[$\mu^{(a)}_1$]{%
		\includegraphics[width=0.24\textwidth]{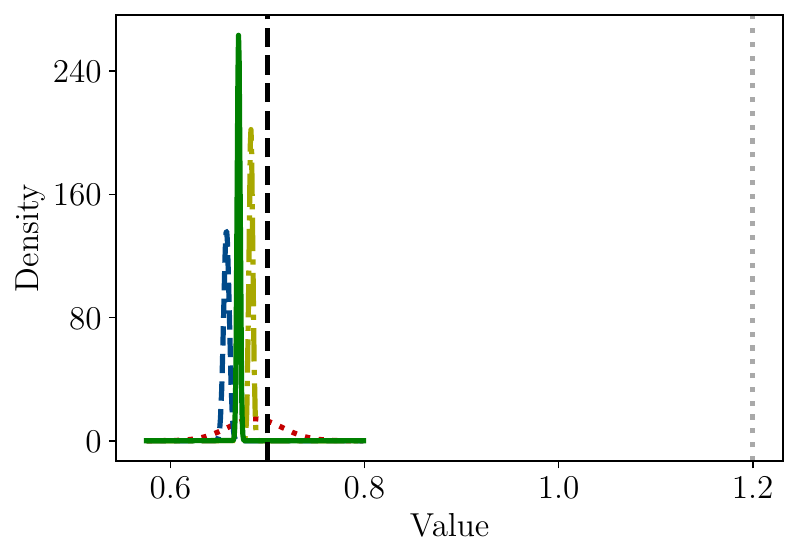}%
	}
	\subfloat[$\mu^{(a)}_2$]{%
		\includegraphics[width=0.24\textwidth]{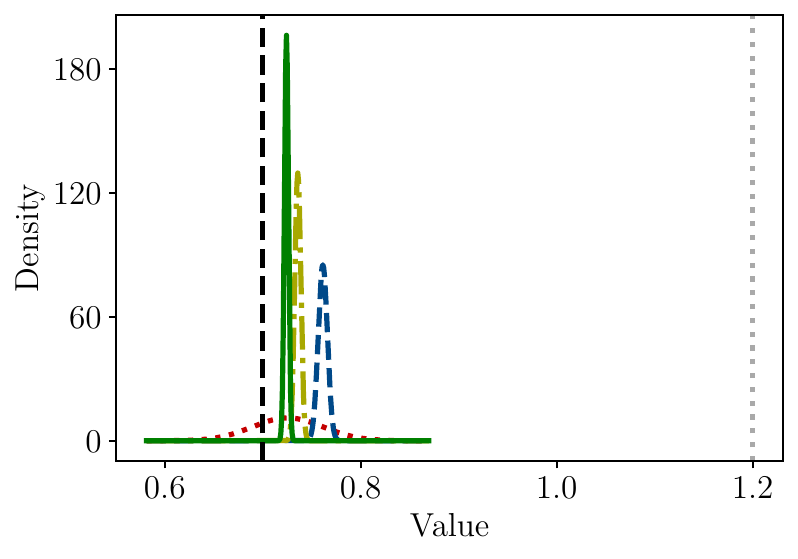}%
	}
	\subfloat[$\mu^{(a)}_3$]{%
		\includegraphics[width=0.24\textwidth]{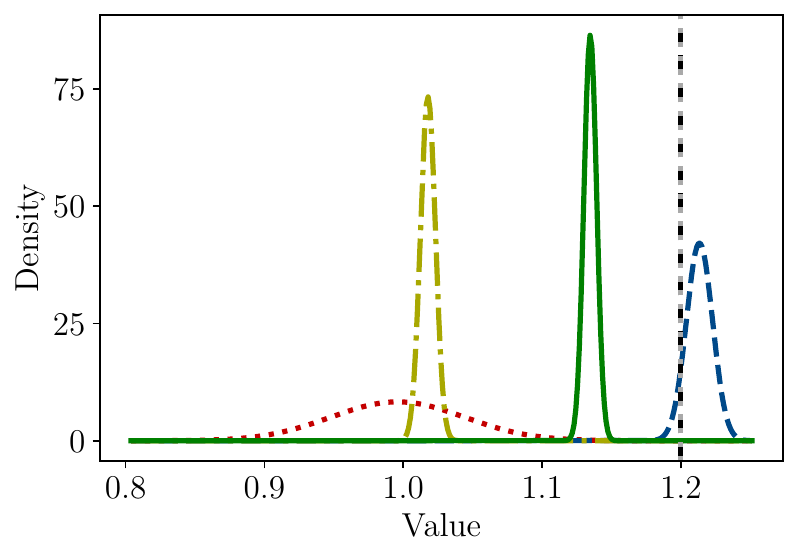}%
	}
	\vspace{0.5em}
	
	\subfloat[$\mu^{(a)}_4$]{%
		\includegraphics[width=0.24\textwidth]{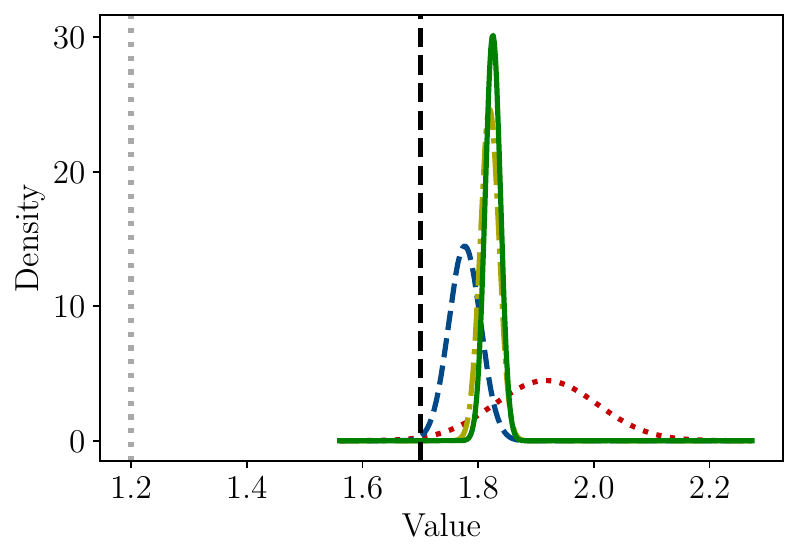}%
	}
	\subfloat[$\mu^{(a)}_5$]{%
		\includegraphics[width=0.24\textwidth]{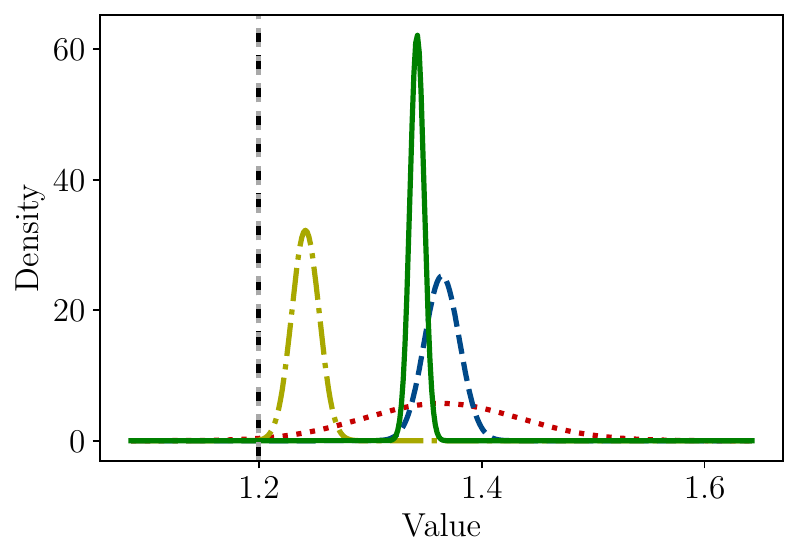}%
	}
	\subfloat[$\mu^{(a)}_6$]{%
		\includegraphics[width=0.24\textwidth]{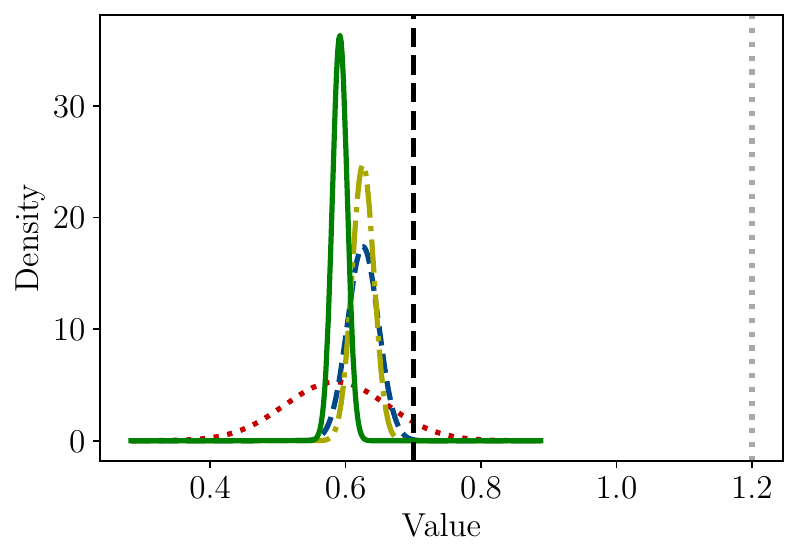}%
	}
	\subfloat[$\mu^{(a)}_7$]{%
		\includegraphics[width=0.24\textwidth]{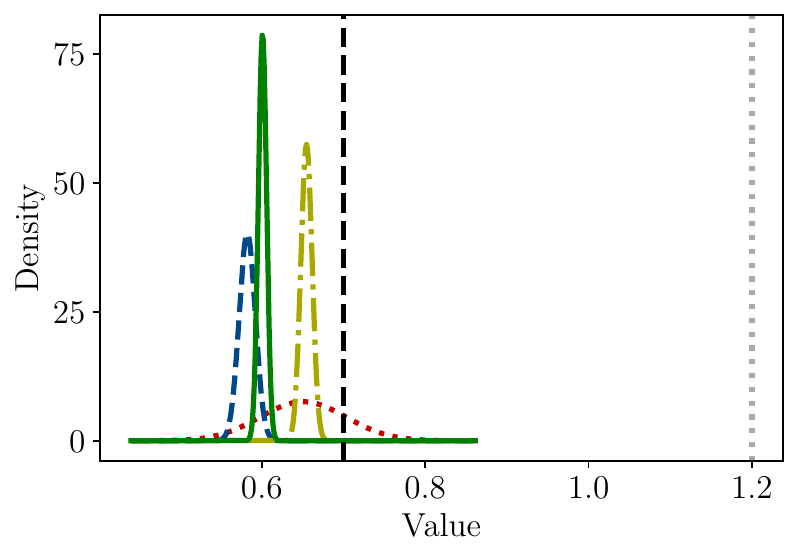}%
	}
	\vspace{0.5em}
	
	\subfloat[$\mu^{(a)}_8$]{%
		\includegraphics[width=0.24\textwidth]{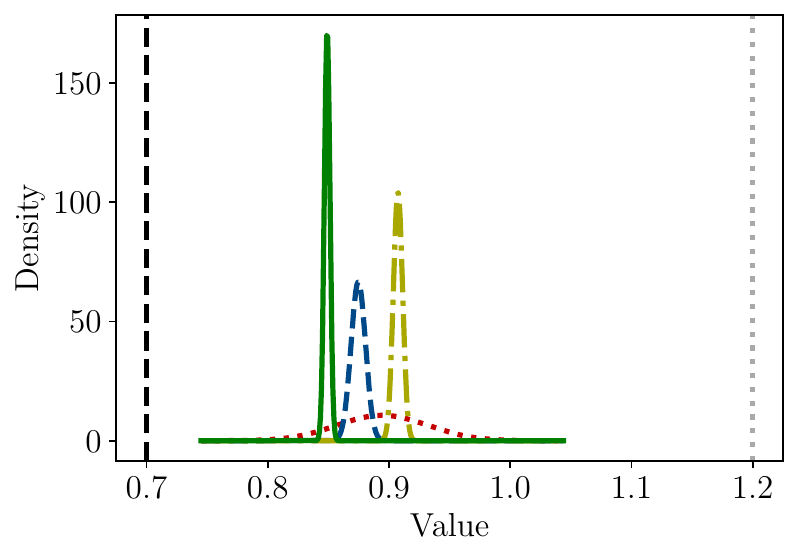}%
	}
	\subfloat[$\mu^{(a)}_9$]{%
		\includegraphics[width=0.24\textwidth]{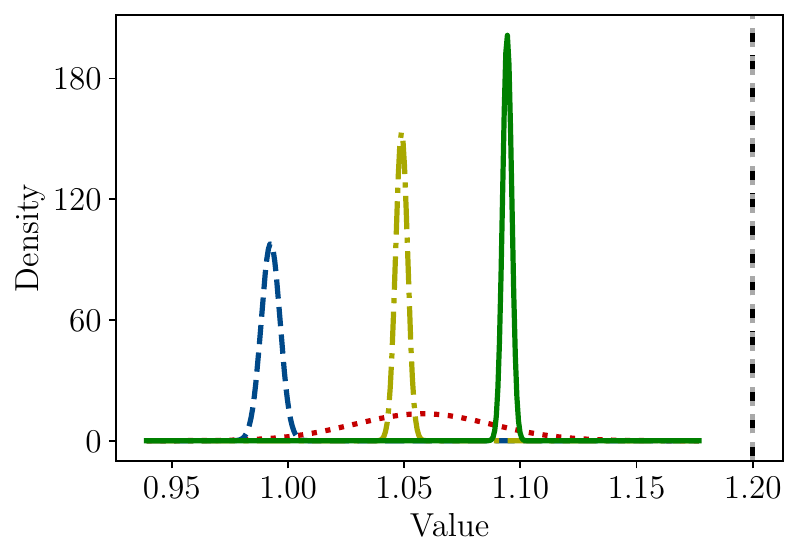}%
	}
	\subfloat[$\mu^{(a)}_{10}$]{%
		\includegraphics[width=0.24\textwidth]{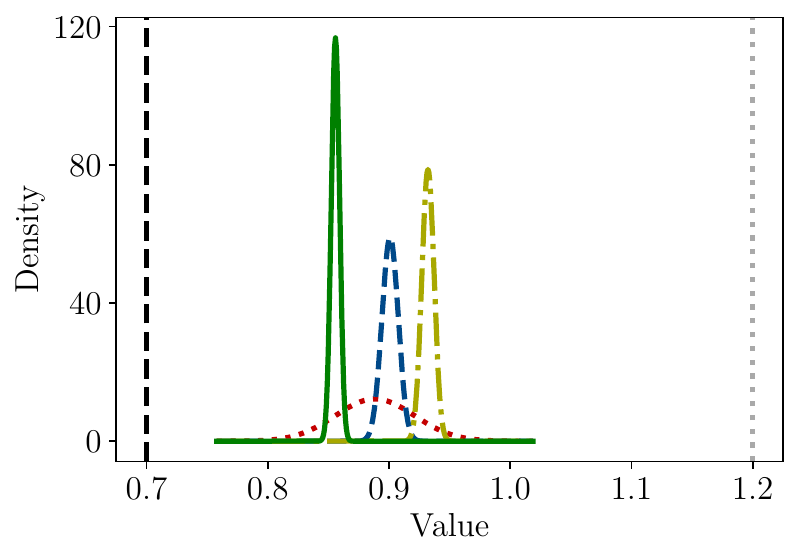}%
	}
	\subfloat{%
		\includegraphics[width=0.24\textwidth]{legends-1d.pdf}%
	}
	\caption{Posterior distributions of the anchor parameters $\mu_j^{(a)}$ for the dual-inclusion inverse problem for different numbers of observations $n_o$.}
	\label{fig:1d-posterior_twoint}
\end{figure}

We next consider a more challenging inverse problem involving two spatially
distinct inclusions of differing amplitude and spatial extent. The
primary inclusion is centred on the anchor point $x_4^{(a)}$
and a secondary inclusion of milder amplitude
is introduced at $x_9^{(a)}$, while the remaining anchor coefficients are held at the
background level, corresponding to homogeneous regions outside the inclusion
zones. The ground-truth coefficient vector is
\begin{equation}
	\vec{\mu}^{(a), \, \text{true}} = \left(
	0.7,\, 0.7,\, 0.7,\, 1.2,\, 1.7,\, 1.2,\,
	0.7,\, 0.7,\, 0.7,\, 1.2,\, 0.7
	\right)^\trans \, .
\end{equation}
Figure~\ref{fig:1d-prediction-twoint} shows the posterior mean conductivity field
$\mu(x)$ recovered for $n_o \in \{1, 5, 10, 25\}$. For all $n_o$, the main inclusion
position $x_4^{(a)} = 0.4$ is correctly recovered. However, the smaller inclusion is not well estimated. The relative $L^2$ error between the posterior mean and the ground-truth (evaluated at the discretisation nodes), decreases slightly as the number of observations increases, i.e., 
$\epsilon_{L^2} = 0.1229$ at $n_o = 1$, to $\epsilon_{L^2} = 0.0981$ at $n_o = 5$, $\epsilon_{L^2} = 0.0921$
at $n_o = 10$, and $\epsilon_{L^2} = 0.0889$ at $n_o = 25$. 
Figure~\ref{fig:1d-posterior_twoint} shows the marginal posterior distributions
of all eleven anchor coefficients $\vec{\mu}^{(a)}$ for each value
of $n_o$. At most anchor points, the inferred coefficients match the ground
truth, with a larger number of observations yielding closer estimates. As
deduced from Figure~\ref{fig:1d-prediction-twoint}, at $x_8^{(a)}$ and $x_9^{(a)}$ 
the inferred value still lies away from the true value. These artefacts may appear due to the use of a non-informative prior and the fixed GP lengthscale. Note that the GP lengthscale learning is not studied here and alternative parameterisation for the material variability is one of our primary interests for future work. As in the single-inclusion case, we also compare the SVI inference with MCMC-NUTS sampler using the same computational setup as described in Section~\ref{sec:1d-poisson-inv-singleint}.  The variational posterior mean yields a relative $L^2$ error of $\epsilon_{L^2} = 0.0889$, while the NUTS posterior mean yields $\epsilon_{L^2} = 0.1056$ which affirms our confidence that the proposed QLSM surrogate can be used for inference using either variational inference as well as sampling-based approach. 

\subsection{Two-dimensional forward problem}
\label{sec:2d-forward}

We next consider the two-dimensional Poisson--Dirichlet problem over the domain
$\Omega = (0,1)^2 \subset \mathbb{R}^2$. Following the one-dimensional construction, the
spatially varying conductivity field $\mu(\vec{x})$ is parameterised
as in Eq.~\eqref{eq:gp-basis}, now over a set of uniformly spaced $11 \times 11$ anchor
points $\{\vec{x}_j^{(a)}\}$ with grid spacing $h = 0.1$ in each coordinate
direction. The kernel is a stationary squared exponential with lengthscale
$\ell = 0.5$ and scaling parameter $\sigma_f = 0.3$. To generate the training
dataset, the anchor coefficients $\mu_j^{(a)}$ are sampled independently and
uniformly on the interval $(0.05, \, 1.05)$, and the corresponding full-field
conductivity $\mu(\vec{x})$ is constructed by GP interpolation. The conductivity field and the associated solution field are discretised using a $128 \times 128$ finite element mesh. We collate $n_s = 10^4$ input--output data pairs into the dataset for training and validation $\{\vec{\mu}^{(i)},\, \vec{u}^{(i)}\}_{i=1}^{n_s}$.

Operating the quantum surrogate model directly on the $128^2$-dimensional field
representations is computationally intractable for quantum circuit architectures. To
reduce the dimensionality of the problem, we obtain latent representations of the
conductivity field $\hat{\vec{\mu}}$ and the corresponding potential response
$\hat{\vec{u}}$. As in the one-dimensional case, we employ a convolutional
autoencoder, learning encoder and decoder networks that map between the full-field and
latent representations of the response field, while we use PCA to determine the latent representation of the material conductivity. The choice of retained latent dimensions is
informed by the explained variance ratio of the respective PCA decompositions.
For the conductivity field, the explained variance saturates rapidly with the number of components, and $n_l^{\mu} = 16$ components suffice to capture the dominant spatial variability of the ensemble (95\% explained variance). 

\begin{figure}[]
	\centering
	\includegraphics[width=0.7\textwidth]{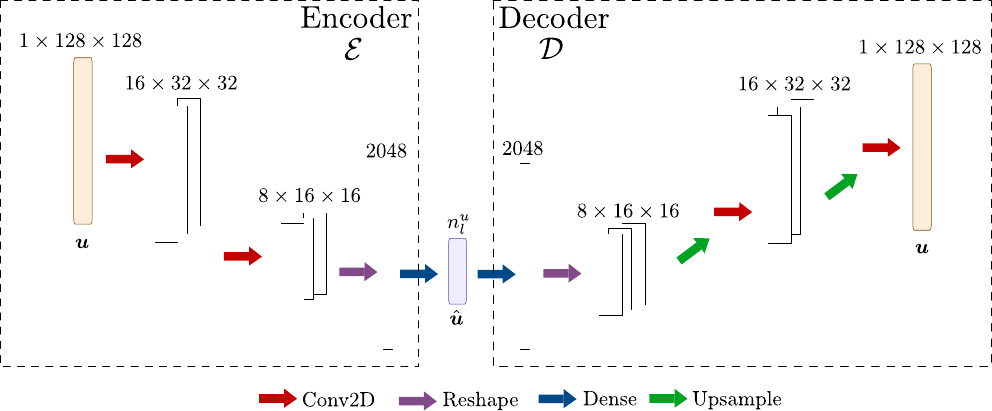}
	\caption{Autoencoder architecture with encoder $\mathcal{E}$ and decoder $\mathcal{D}$, in which $\vec{u} \in \mathbb{R}^{n_{\text{dof}}}$ is the response field in physical space and $\hat{\vec{u}} \in \mathbb{R}^{n_l^{u}}$ is the response field in the latent space.}
	\label{fig:autoencoder-architecture-2d}
\end{figure}

For the potential field, the solution manifold is of lower intrinsic dimensionality, owing to the smoothing action of the elliptic operator, and $n_l^u = 13$ components suffice to represent the ensemble (99\% explained variance). To obtain the mapping between physical and latent space, we use an autoencoder structure following the one-dimensional construction discussed in Section~\ref{sec:1d-forward}, adapted to the two-dimensional fields, see Figure~\ref{fig:autoencoder-architecture-2d}. 
The encoder applies two \texttt{Conv2D} layers that
progressively coarsen the spatial resolution while adjusting the channel count,
$(1,128,128) \rightarrow (16,32,32) \rightarrow (8,16,16)$, where each tuple denotes
(channels, height, width). The resulting feature map is flattened and projected by a
single linear layer onto the latent space of dimension $n_l$ (namely $n_l^{\mu} = 16$
for the conductivity field and $n_l^u = 13$ for the potential field). The decoder
mirrors this structure: a linear layer lifts the latent vector back to an $(8,16,16)$
feature map, followed by two upsampling--convolution blocks that invert the encoder,
$(8,16,16) \rightarrow (16,32,32) \rightarrow (1,128,128)$. All convolutional layers use
a $3 \times 3$ kernel. 

\begin{figure}[]
	\centering
		\includegraphics[width=0.48\textwidth]{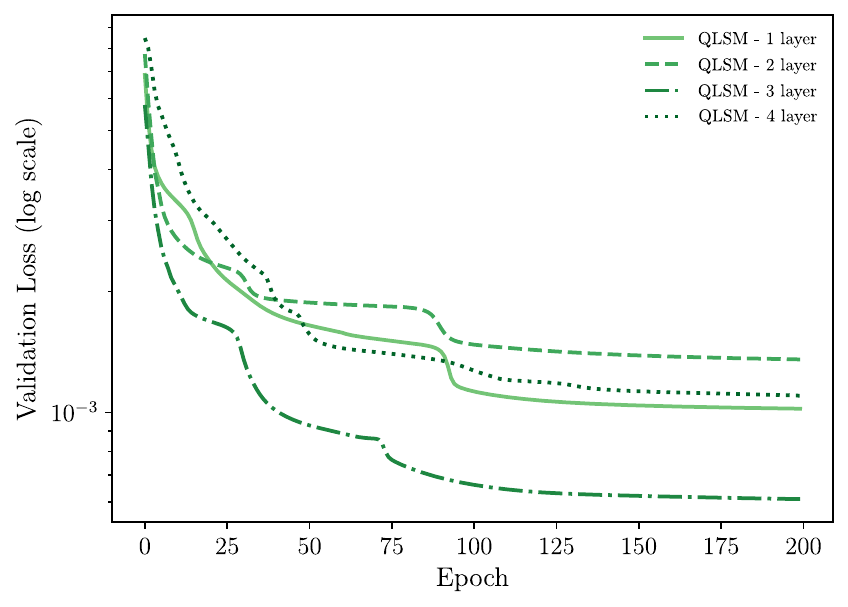}
	\caption{Validation loss for the two-dimensional Poisson
		forward prediction problem using the QLSM across model depths of one to
		four layers.}
	\label{fig:2d-validation-loss}
\end{figure}
\begin{figure}[]
	\centering
	{%
		\includegraphics[width=1.0\textwidth]{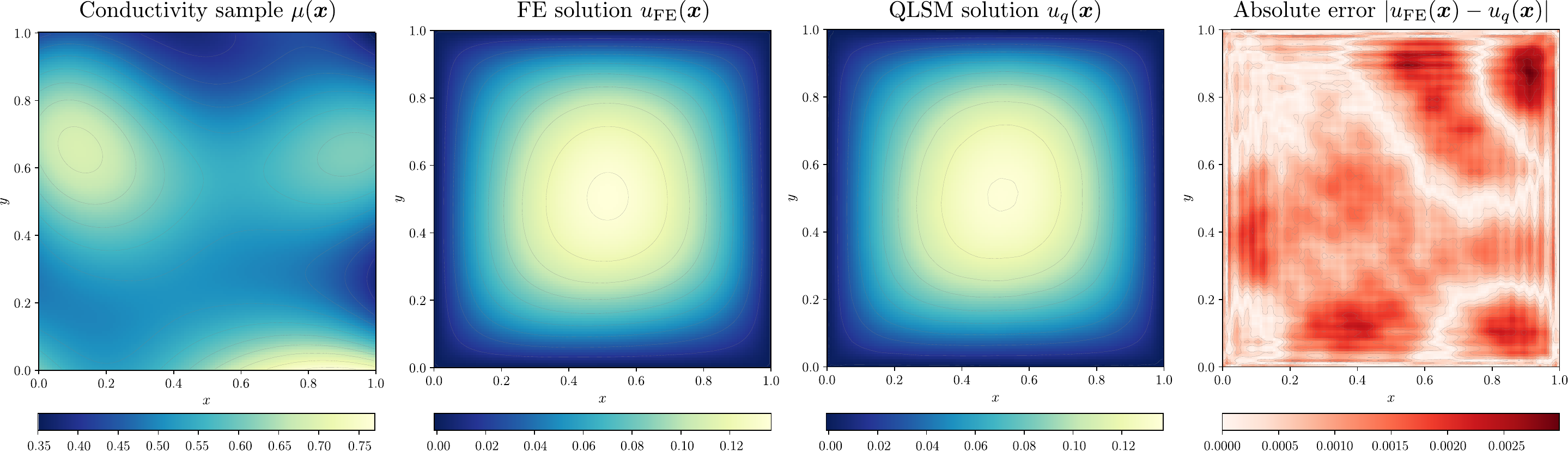}%
	}
	\caption{Two-dimensional Poisson problem
		$-\nabla \cdot (\mu(\vec{x})\,\nabla u(\vec{x})) = 1$ on
		$\Omega = (0,1)^2$, for a randomly drawn test sample from the Gaussian process parameterisation with covariance parameters $\sigma_f = 0.3$
		and $\ell = 0.5$. (a) Conductivity field $\mu(\vec{x})$ constructed
		via GP interpolation from the $11 \times 11$ anchor grid to the
		$128 \times 128$ finite element mesh. (b) Reference finite element solution
		$u_{\text{FE}}(\vec{x})$ obtained on the $128 \times 128$ mesh. (c) QLSM-predicted
		solution $u_q(\vec{x})$ obtained via the forward pipeline
		and reconstructed to the full field.
		(d) Pointwise absolute error $|u_{\text{FE}}(\vec{x}) - u_q(\vec{x})|$ between the QLSM
		prediction and the finite element reference, with a relative $L^2$ error of
		$\epsilon_{L^2} = 0.0137$ for this sample.}
	\label{fig:2d_mu_and_u_sample}
\end{figure}

We first study the optimum architecture parameter, specifically the number of PQC layers, for approximating the mapping between latent representations $\hat{\vec \mu}$ and $\hat{\vec u}$. The QLSM architecture and training protocol follow those described in Section~\ref{sec:1d-forward}, with PQC depth ranging from one to four
layers. The validation loss histories across all depths are shown in
Figure~\ref{fig:2d-validation-loss}. The three-layer configuration achieves the lowest
validation loss with stable convergence, while the four-layer model shows minor
degradation attributable to increased optimisation complexity and model overfitting. The three-layer QLSM is therefore selected as the optimal architecture and retained for all subsequent
two-dimensional experiments. The predictive fidelity of the optimal three-layer QLSM is first assessed on a single randomly drawn test sample, shown in Figure~\ref{fig:2d_mu_and_u_sample}.
The conductivity field in panel~(a) exhibits spatial variability characteristic
of the GP ensemble, with the inclusion feature modulated by the random anchor
coefficients. The QLSM-predicted solution in panel~(c) closely matches the finite
element reference in panel~(b), capturing the spatial structure of the response
field across the full domain. The pointwise absolute error in panel~(d) remains
small throughout, with a relative $L^2$ error of $\epsilon_{L^2} = 0.0137$ for this sample, confirming that the full two-dimensional pipeline generalises to unseen inputs.

\begin{figure}[]
	\centering
	{%
		\includegraphics[width=1.0\textwidth]{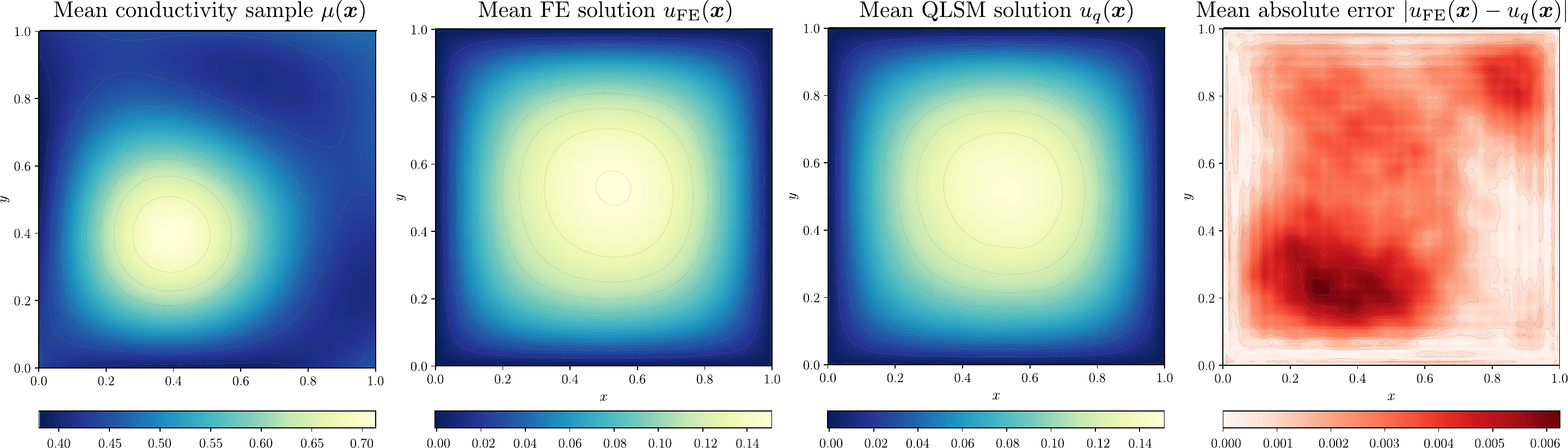}%
	}
	\caption{Two-dimensional Poisson problem
		$-\nabla \cdot (\mu(\vec{x})\,\nabla u(\vec{x})) = 1$ on
		$\Omega = (0,1)^2$, for a set of inclusion test sampled from the Gaussian process parameterisation with covariance parameters $\sigma_f = 0.3$
		and $\ell = 0.5$. (a) Mean conductivity field $\mu(\vec{x})$ constructed
		via GP interpolation from the $11 \times 11$ anchor grid to the
		$128 \times 128$ finite element mesh. (b) Reference mean finite element solution
		$u_{\text{FE}}(\vec{x})$ obtained on the $128 \times 128$ mesh. (c) Mean QLSM
		solution $u_q(\vec{x})$ obtained via the forward pipeline
		and reconstructed to the full field.
		(d) Mean pointwise absolute error $|u_{\text{FE}}(\vec{x}) - u_q(\vec{x})|$ between the QLSM
		prediction and the finite element reference, with a mean relative $L^2$ error of
		$\epsilon_{L^2} = 0.0298$.}
	\label{fig:2d_mu_and_u_mean}
\end{figure}
Next we assess the statistical reliability of the trained QLSM over a broader input
ensemble. The surrogate is further evaluated over $n_s^{\text{test}} = 1000$ test
samples. Similarly to the one-dimensional case, we use a shifted function
\begin{equation}
	\label{eq:gp-mean-shifted-2d}
	\mu(\vec x)^{\text{test}} = \overline{\mu}(\vec x) + \vec \phi(\vec x)^{\trans} \,
	\vec \mu^{(a),\,\text{test}} \, ,
\end{equation}
with GP basis functions $\vec \phi(\vec x)$ and mean conductivity field according to a circular inclusion
\begin{equation}
	\overline{\mu}(\vec{x})
	= \mu_{\text{out}}
	+ (\mu_{\text{in}} - \mu_{\text{out}})
	\left(
	1 - \frac{1}{2}
	\left(
	1 + \tanh\!\left(\frac{d(\vec{x})}{\varepsilon}\right)
	\right)
	\right)  \, .
	\label{eq:mu_levelset}
\end{equation}
Here $d(\vec{x})$ denotes the signed distance function, where a negative value $d(\vec{x}) < 0$ indicates the interior of the inclusion and a positive value $d(\vec{x}) > 0$ indicates the exterior. The shape parameter $\varepsilon = 0.1$ controls the width of the interfacial transition
layer, and $\mu_{\text{in}} = 0.8$, $\mu_{\text{out}} = 0.4$ are the interior and
exterior conductivity values, respectively. For a circular inclusion with
centroid $\vec{x}_c = (0.4, 0.4)$ and radius $r = 0.1$, the signed distance function is
$d(\vec{x}) = \|\vec{x} - \vec{x}_c\|_2 - r$. See Figure~\ref{fig:2d_mu_and_u_mean} for the contour depiction of the mean function $\overline{\mu}$. Each test sample
is obtained by uniformly sampling the perturbation coefficients according to $\mu_j^{(a),\,\text{test}} \sim \mathcal{U}(-0.1, 0.1)$, which are then propagated through the GP basis functions in Eq.~\eqref{eq:gp-mean-shifted-2d} and evaluated at the $128 \times 128$ mesh nodes. 

Figure~\ref{fig:2d_mu_and_u_mean}  presents the resulting ensemble statistics
over $n_s^{\text{test}} = 1000$ test samples. The mean QLSM-predicted solution agrees 
well with the finite element reference. Note here that the solution profile is similar to the one obtained from the random conductivity sample, emphasising the ill-conditioning of the inverse problem. The mean
absolute error field remains small across the domain, with a mean
relative $L^2$ error of $\epsilon_{L^2} = 0.0298$, confirming that the two-dimensional QLSM forward pipeline maintains consistent predictive accuracy across the test ensemble.

\begin{figure}[]
	\centering
    {%
		\includegraphics[width=0.745\textwidth]{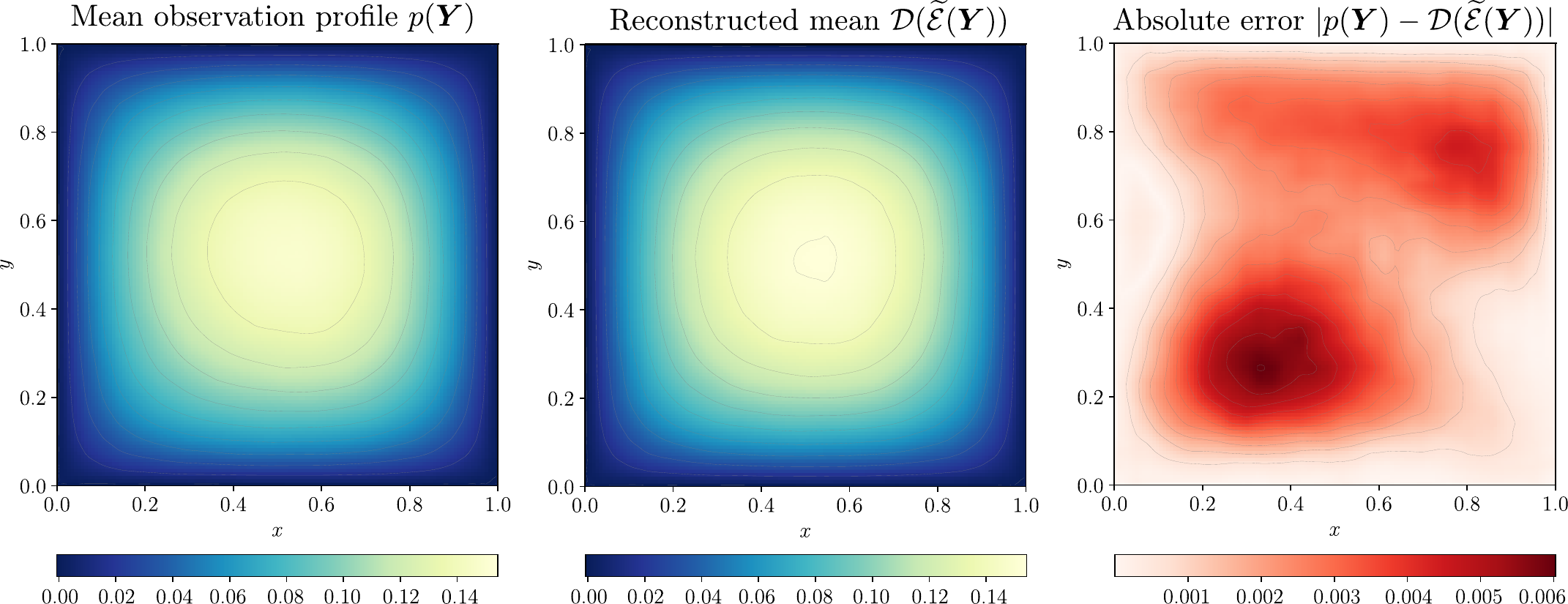}%
	}
	\caption{Comparison between the mean observation profile and the reconstructed mean through the observation-aware encoder $\mathcal{D}(\tilde{\mathcal{E}}(\vec Y))$ for the test dataset.}
	\label{fig:2d_noisy_forward}
\end{figure}

Prior to the inversion, we assess the performance of the observation-aware encoder similarly to our one-dimensional example in Section~\ref{sec:1d-forward}. Specifically, the observation-aware encoder shares the same architecture with the one described in Section~\ref{sec:autoencoder} and trained using the noise-augmented training dataset, with Gaussian white noise of $\sigma_e = 0.0025$. In Figure~\ref{fig:2d_noisy_forward}, the distribution mean of the test observation is shown in the left panel and the mean of the reconstructed observation population, $\mathcal{D}(\tilde{\mathcal{E}}(\vec Y))$, is shown in the middle panel. As evident in Figure~\ref{fig:2d_noisy_forward}, the reconstructed profile matches the original observation distribution quite well, with the means have $L^2$-norm relative discrepancy of $\Delta \epsilon_{L^2} =0.008$. As the error field in the right panel shows, this residual is not distributed uniformly but largely coincides with regions of steep potential gradient. This points to a field-dependent effect, though a contribution from the autoencoder cannot be fully ruled out. A systematic study across autoencoder architectures and latent dimensions would be needed to isolate the contributions, but lies beyond the scope of the present paper.

\subsection{Two-dimensional inverse problem}
\label{sec:2d-inverse}

In this example, we consider a two-dimensional inverse problem aimed at identifying
a spatially varying conductivity field from noisy observations. 
The objective is to infer the anchor coefficients $\vec{\mu}^{(a)}$ that define the conductivity field from indirect noisy measurements of the corresponding potential field. Here we use the parameterisation introduced in Section~\ref{sec:2d-forward} with anchor coefficients
$\vec{\mu}^{(a)} \in \mathbb{R}^{121}$ on the uniform $11 \times 11$ grid. The ground-truth conductivity fields used to generate synthetic observations are
defined via the smooth level-set formulation in Eq.~\eqref{eq:mu_levelset}, with the specific
inclusion geometries, conductivity values, and smoothing parameters specified in each of the following cases. As noted in Section~\ref{sec:2d-forward}, this construction is used solely for
data generation and the inference model has no access to the inclusion geometry and
instead recovers the field through the GP parameterisation, introducing a deliberate
model-form mismatch that reflects the realistic scenario in which the analyst has no
prior knowledge of the structural form of the unknown field.

The observation model relates the forward solution to a measurement matrix
$\vec{Y} \in \mathbb{R}^{n_y \times n_o}$, where $n_y$ is the number of
observation locations and $n_o \in \{1, 5, 10\}$ is the number
of independent repeated measurements at each location. Each observation is
generated according to
\begin{equation}
	\vec{y}^k = \vec{u} + \vec{\varepsilon}^k \, ,
	\qquad \vec{\varepsilon}^k \sim \mathcal{N}(\vec{0}, \sigma_e^2 \vec{I}) \, ,
\end{equation}
where $\vec{u}$ is the QLSM forward solution evaluated at the observation locations, $\sigma_e = 0.0025$ is the fixed observational noise, and the index $k = \{1, \ldots, n_o\}$. Prior to inversion, each observation replicate $\vec{y}^k$ is encoded to the latent potential space via the
noise-resilient encoder $\tilde{\mathcal{E}}$, giving $\hat{\vec{y}}^k =
\tilde{\mathcal{E}}(\vec{y}^k)$. The final observation matrix is denoted as $\hat{\vec{Y}} = \left(\hat{\vec{y}}^1, \ldots, \hat{\vec{y}}^{n_o}\right)
\in \mathbb{R}^{n_l^u \times n_o}$, where $n_l^u = 13$ according to Section~\ref{sec:2d-forward}. The resulting latent likelihood and the full variational inference
procedure are described in Section~\ref{sec:svi}. Throughout this section, we use $40,000$ SVI steps with Adam optimiser with learning rate of 0.001 and $n_{SVI} = 2$ Monte Carlo samples per gradient step, for each value of $n_o$.

\subsubsection{Single inclusion}
\label{sec:2d-inverse-singleint}

We consider the inversion of a single localised inclusion of elevated
conductivity embedded in a homogeneous background. The ground-truth field is
defined via the level-set formulation in Eq.~\eqref{eq:mu_levelset} with 
interior and exterior values $\mu_{\text{in}} = 0.8$ and $\mu_{\text{out}}
= 0.4$ and a circular inclusion of centroid $\vec{x}_c = (0.4, 0.4)$ and radius $r = 0.1$, with smoothing parameter $\varepsilon = 0.1$. The
inclusion radius $r = 0.1$ is substantially smaller than the GP kernel lengthscale
$\ell = 0.5$, so the posterior is therefore expected to recover a spatially
smoothed approximation rather than the exact geometry. The prior and guide are spatially uniform, encoding a non-informative initial estimate with no structural bias towards the inclusion geometry. 
The prior and guide distributions have uniform mean and uncorrelated covariance $p(\vec{\mu}^{(a)}) = \mathcal{N}\!\left( 0.6 \,\vec{1}_{121}, \, 0.2^2\,\vec{I}_{121} \right)$.

\begin{figure}[]
	\centering
	\includegraphics[width=\linewidth]{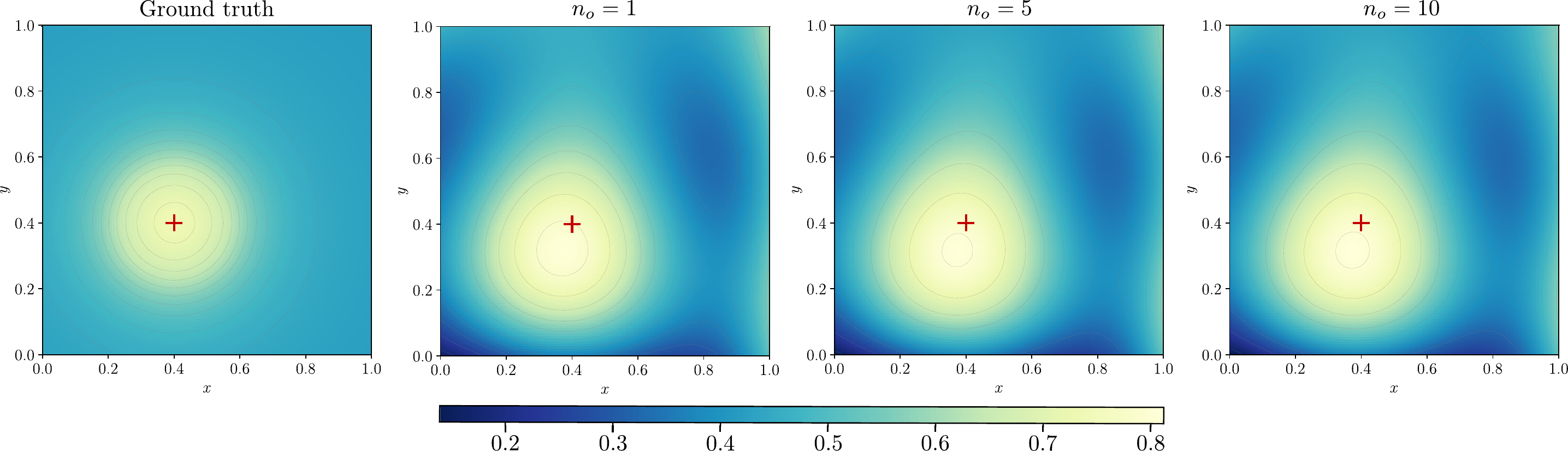}
	\caption{Posterior mean conductivity for the 2D single-inclusion case; the
		prior and guide are flat at $\mu = 0.6$.}
	\label{fig:2d-inferred-fields-QALM-singleintrusion}
\end{figure}

The inversion is performed for $n_o \in \{1, 5, 10\}$ repeated observations for a systematic assessment of how the volume of observational data affects the
reconstruction. For each $n_o$, the posterior mean field is evaluated against the ground truth, assessing recovery of the inclusion
centroid, the conductivity contrast, and the spatial extent of the reconstructed
region. Figure~\ref{fig:2d-inferred-fields-QALM-singleintrusion} shows the posterior mean
conductivity field recovered for all considered $n_o$, with the true inclusion
boundary overlaid in each panel. For $n_o = 1$, the posterior mean correctly
identifies the inclusion location but yields a spatially diffuse reconstruction
consistent with the under-resolved regime, with a relative $L^2$ error of
$\epsilon_{L^2} = 0.0679$. As $n_o$ increases to $5$ and $10$, the
reconstruction improves slightly and becomes progressively
more compact and better localised about the true centroid $\vec{x}_c = (0.4, 0.4)$,
the recovered conductivity contrast approaches the ground truth, and the background
field converges towards $\mu_{\mathrm{out}} = 0.4$, with the relative $L^2$ error
reducing to $\epsilon_{L^2} = 0.0671$ and $\epsilon_{L^2} = 0.0670$, respectively. The
marginal reduction in error across $n_o$ reflects the resolution limit imposed by the
GP kernel lengthscale, beyond which additional observations cannot further improve
the reconstruction, which indicates that the dominant residual error is the kernel smoothing rather than a limitation of the inference procedure.

\subsubsection{Dual inclusions}
\label{sec:2d-inverse-doubleint}

\begin{figure}[]
	\centering
	\includegraphics[width=\linewidth]{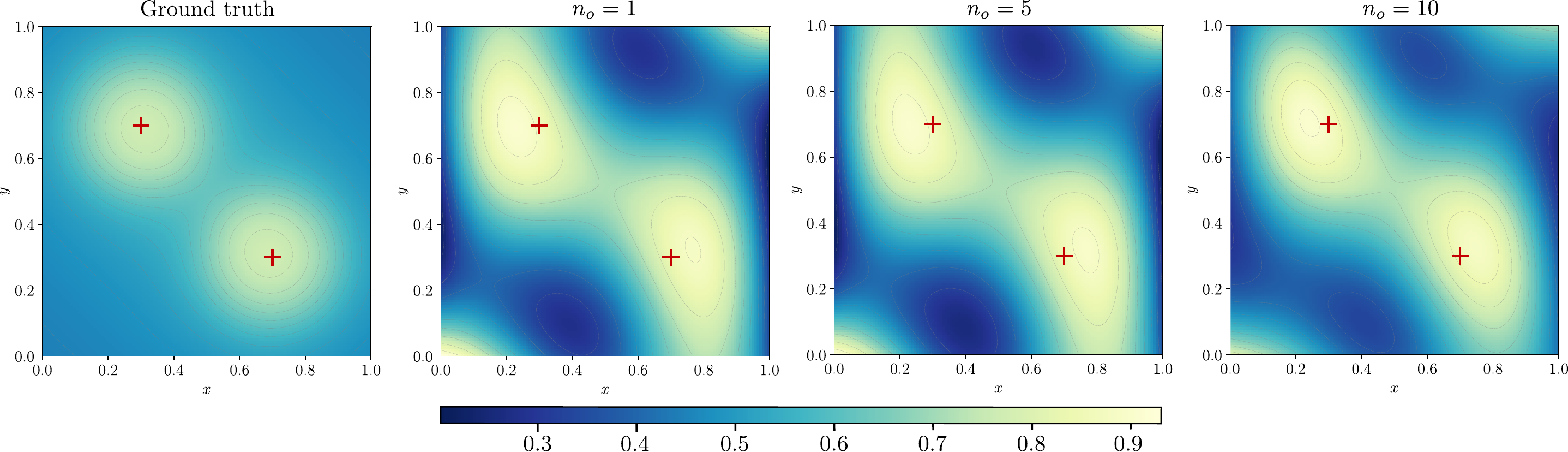}
	\caption{Posterior mean conductivity for the 2D two-inclusion case (cross
		configuration). The prior and guide mean are flat throughout the domain at $0.6$.}
	\label{fig:2d-inferred-fields-QALM-doubleintrusion-cross}
\end{figure}

We next consider a more challenging problem with two spatially separated inclusions
in a homogeneous background. The ground-truth field uses the level-set
formulation in Eq.~\eqref{eq:mu_levelset} with two inclusions carrying the same interior values
$\mu_{\text{in}} = 0.8$, and the surrounding conductivity of $\mu_{\text{out}} = 0.4$. We first consider the two circular inclusions of radius $r = 0.1$ and centroids
$\vec{x}_{c_1} = (0.3, 0.7)$ and $\vec{x}_{c_2} = (0.7, 0.3)$. As in the
single-inclusion case, $\ell = 0.5$ is used so each inclusion is expected to be
recovered as a spatially smoothed approximation. We use the prior and guide identical to the single-inclusion case $p(\vec{\mu}^{(a)}) = \mathcal{N}\!\left( 0.6 \,\vec{1}_{121}, \, 0.2^2\,\vec{I}_{121} \right)$, where no insight of the inclusion location is given.  Figure~\ref{fig:2d-inferred-fields-QALM-doubleintrusion-cross} shows the posterior
mean conductivity field for $n_o \in \{1, 5, 10\}$, with the true inclusion
centroids shown as red crosses. All $n_o$ cases yield correctly identified inclusion centroids and general shape. For $n_o = 1$, the reconstructed region exhibits a relative $L^2$ error
of $\epsilon_{L^2} = 0.0740$ with respect to the ground truth. As $n_o$ increases to $5$ and $10$, the relative $L^2$ error reduces to $\epsilon_{L^2} = 0.0717$ and $\epsilon_{L^2} = 0.0712$, respectively. Residual boundary discrepancies are
again attributable to kernel smoothing rather than the inference procedure, but it is evident that more observations yield weaker boundary artefacts. 

\begin{figure}[]
	\centering
	\includegraphics[width=\linewidth]{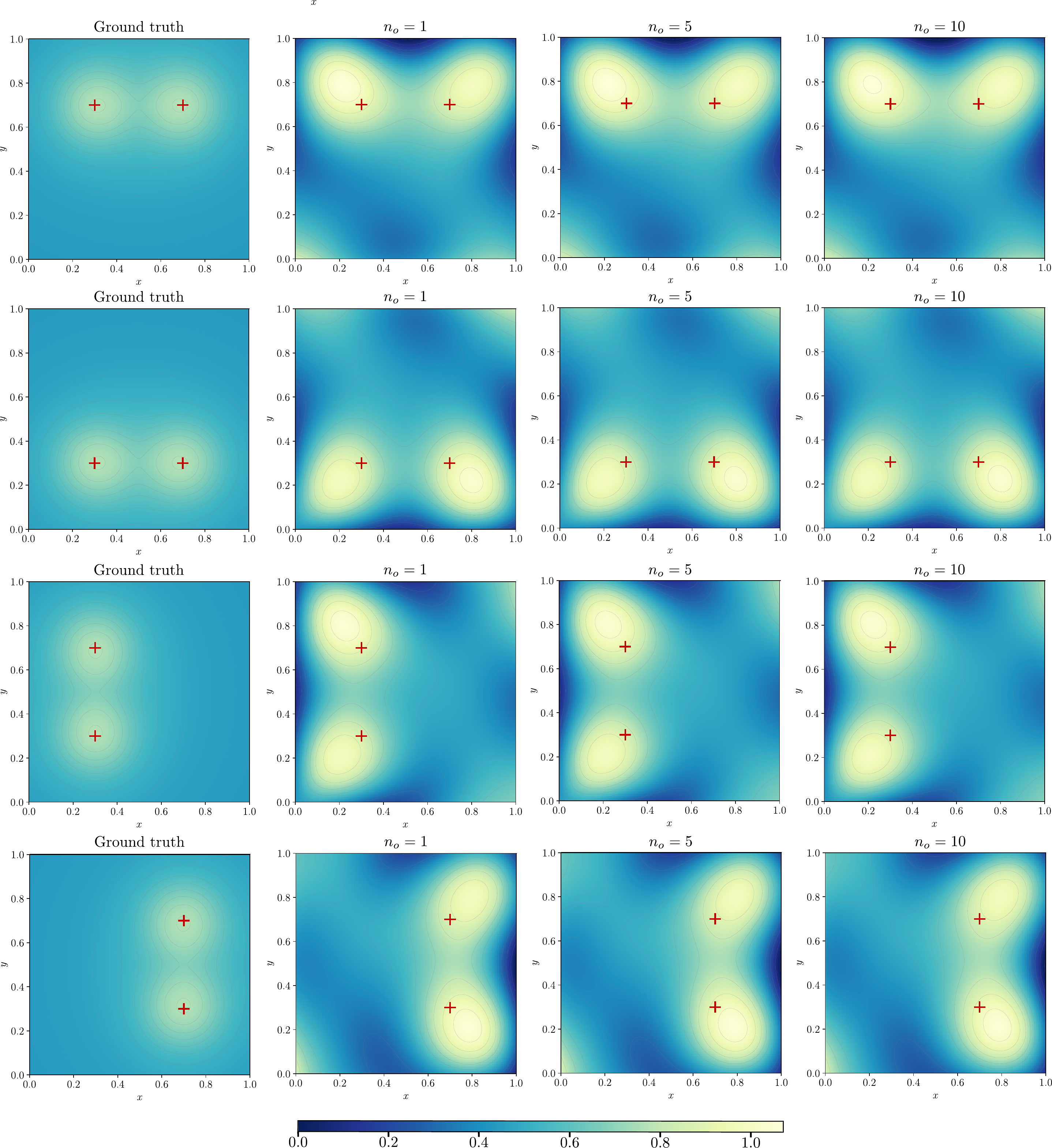}
	\caption{Posterior mean conductivity for the 2D two-inclusion case under
		the four semi-informed prior configurations of Eq.~\eqref{eq:prior-configs},
		with the suspected centroids shifted by $\pm 0.1$ from the ground truth.}
	\label{fig:2d-inferred-fields-QALM-doubleintrusion-side}
\end{figure}

We further investigate a semi-informed scenario in which partial prior knowledge of
the approximate inclusion locations is incorporated. This reflects a practically
relevant setting in which auxiliary information provides the correct number of inclusions, and their approximate locations, which may arise from alternative measurement modes, 
while the exact geometry, contrast, and extent of the inclusions remain unknown. The
GP parameterisation is retained over the same $121$ anchor points, but the prior mean and guide 
are modified to encode this partial information, that is, 
\begin{equation}
	\label{eq:prior-semi}
	p(\vec{\mu}^{(a)}) = \mathcal{N} \, \left( \overline{\vec{\mu}}^{(a)}, \, 0.2^2 \, \vec{I}_{121} \right) \, ,
	\qquad \text{where} \qquad 
	\overline{\mu}^{(a)}_j =
	\begin{cases}
		0.7, & \text{if } \| \vec{x}_j^{(a)} - \tilde{\vec{x}}_{c_k} \|_2 \leq r_k \, , \\
		0.3, & \text{otherwise} \, ,
	\end{cases}
\end{equation}
where $\tilde{\vec{x}}_{c_k}$ and $r_k$ are the prior centroid and radius of inclusion
$k$. To assess the influence of directional prior bias, each pre-identified centroid is displaced from its true counterpart by $0.1$ in each coordinate, away from the centre of the domain. Four configurations are considered to test the robustness of the framework with varying ground truth centroids
\begin{equation}
	\label{eq:prior-configs}
	\begin{aligned}
		\text{Top:}    &\quad \vec{x}_{c_1} = (0.3, 0.7) \, , \quad \vec{x}_{c_2} = (0.7, 0.7) \, , \\
		\text{Bottom:} &\quad \vec{x}_{c_1} = (0.3, 0.3)\, , \quad \vec{x}_{c_2} = (0.7, 0.3) \, , \\
		\text{Left:}   &\quad \vec{x}_{c_1} = (0.3, 0.7)\, , \quad \vec{x}_{c_2} = (0.3, 0.3) \, , \\
		\text{Right:}  &\quad \vec{x}_{c_1} = (0.7, 0.7)\, , \quad \vec{x}_{c_2} = (0.7, 0.3) \, ,
	\end{aligned}
\end{equation}
where in each case the prior centroids are displaced from the true inclusion locations
by a margin comparable to the inclusion radius and the GP anchor spacing. Figure~\ref{fig:2d-inferred-fields-QALM-doubleintrusion-side} shows the posterior mean
conductivity fields for all four configurations under $n_o \in \{1, 5, 10\}$, with the
true inclusion centroid depicted as red cross marks. Despite this initial misalignment, the proposed method is able to recover centroids close to the true locations,
showing that the likelihood contribution from the data is sufficient to override the
directional prior bias. This demonstrates the robustness of the proposed quantum-assisted latent space inference with respect to the position of the inclusions. 

\subsubsection{Inhomogeneous and unstructured conductivity field}
\label{sec:2d-complex}

\begin{figure}[]
	\centering
	\subfloat[Sample 1 of conductivity field]{%
		\includegraphics[width=\linewidth]{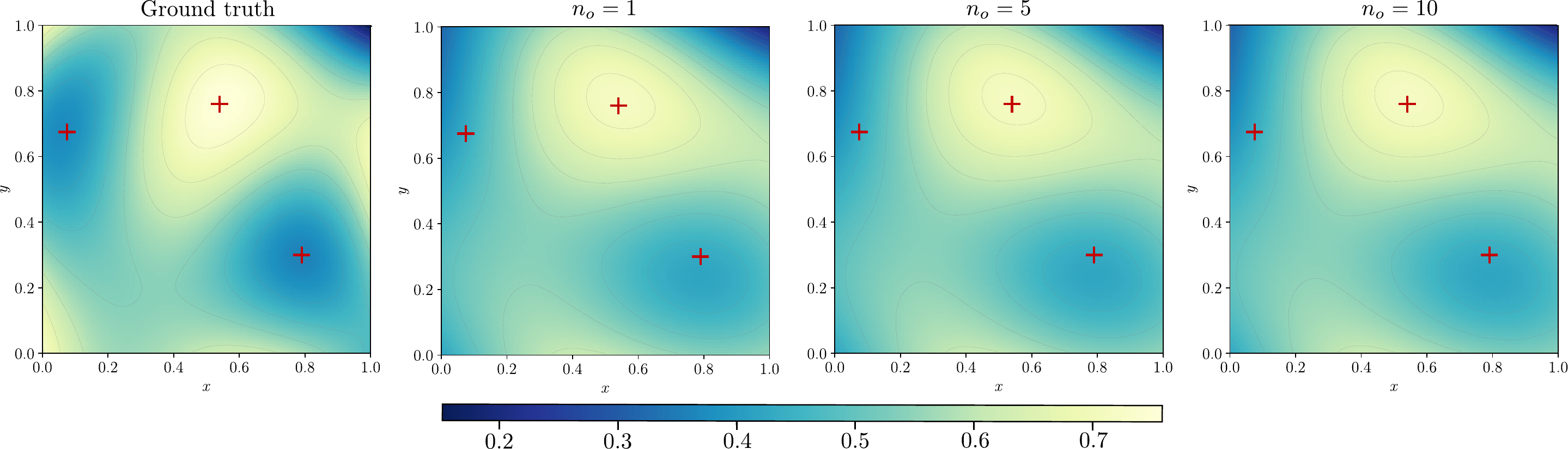}%
	} \\
	\subfloat[Sample 2 of conductivity field]{%
		\includegraphics[width=\linewidth]{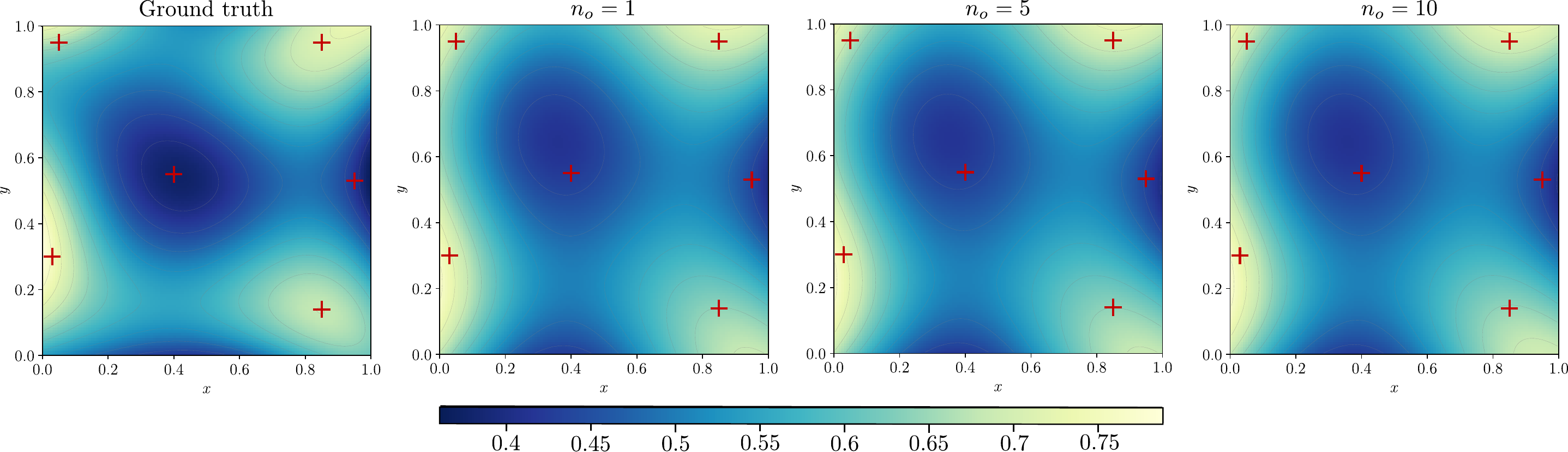}%
	}
	\caption{Posterior mean conductivity for the complex (GP-drawn) conductivity
		field exhibiting two peaks and flanking valleys.}
	\label{fig:2d-inferred-fields-QALM-doubleintrusion-complex-1}
\end{figure}

We consider a more general inverse problem in which the ground-truth field is no
longer a parametric level-set construction but a realisation drawn directly from the
Gaussian process parameterisation over the $11 \times 11$ anchor grid. This yields a
spatially complex field with no assumed geometric structure, exhibiting randomly structured high-conductivity peaks and low-conductivity valleys, as shown in the left-most ground-truth panel of
Figure~\ref{fig:2d-inferred-fields-QALM-doubleintrusion-complex-1}. This provides a more
demanding test relative to the structured inclusion cases, as the posterior must
recover an irregular, asymmetric distribution. The prior mean is again spatially uniform $p(\vec{\mu}^{(a)}) = \mathcal{N}\!\left( 0.6 \,\vec{1}_{121}, \, 0.2^2\,\vec{I}_{121} \right)$,
encoding no structural bias, so the reconstruction relies entirely on the observational data.

Figure~\ref{fig:2d-inferred-fields-QALM-doubleintrusion-complex-1} shows the posterior
mean conductivity field for $n_o \in \{1, 5, 10\}$, with the true peak and valley centroids
indicated by red crosses. For all $n_o$, the posterior mean captures the global spatial
structure of the ground truth, and correctly identifies the approximate locations of both the peaks and valleys, despite no prior knowledge of their positions being given. For the first sample, the relative $L^2$ errors are $\epsilon_{L^2} = 0.0760$, $\epsilon_{L^2} =0.0753$, and $\epsilon_{L^2} = 0.0751$ for $n_o = 1$, $5$, and $10$, respectively. For the second sample, the relative $L^2$ errors are $\epsilon_{L^2} = 0.0790$, $\epsilon_{L^2} = 0.0770$, and $\epsilon_{L^2} = 0.0730$ for $n_o = 1$, $5$, and $10$, a modest but consistent improvement with increasing data. The small error reduction reflects the increased complexity of the ground-truth field, whose irregular structure is inherently harder to recover from the $13$-dimensional latent observation space. Nevertheless, the framework identifies the dominant features of the conductivity distribution, demonstrating robustness to unstructured, geometrically complex ground-truth fields under a non-informative prior.

\subsection{Inversion under forward model discrepancy}
\label{sec:inverse-fe}

In Sections~\ref{sec:1d-inverse} and~\ref{sec:2d-inverse} the observations were generated by propagating the true coefficients $\vec{\mu}^{(a),\,\text{true}}$ through the QLSM, so the same forward map enters both the data-generating model and the likelihood. We now repeat the inversions with observations drawn from the finite element solver, so that the inversion is exposed to data the surrogate has not produced. The true potential $\vec u$ is computed with linear finite elements on the mesh of the corresponding problem, sampled at the discretisation nodes and corrupted with white noise at $\sigma_e = 0.0025$. The latent observations are obtained with the pre-trained encoder $\tilde{\mathcal E}$. We run $50{,}000$ SVI steps in 1D and $40{,}000$ in 2D, with the Adam optimiser at learning rate $0.001$ and $n_{SVI} = 2$ Monte Carlo samples per step; all other settings follow the sections above. The gap between these errors and those obtained with QLSM-generated observations measures the price of the surrogate discrepancy.

\subsubsection{One-dimensional inclusions}
\label{sec:1d-inverse-fe-singleint}

\begin{figure}[]
	\centering
	\subfloat[Single inclusion]{%
		\includegraphics[width=0.48\textwidth]{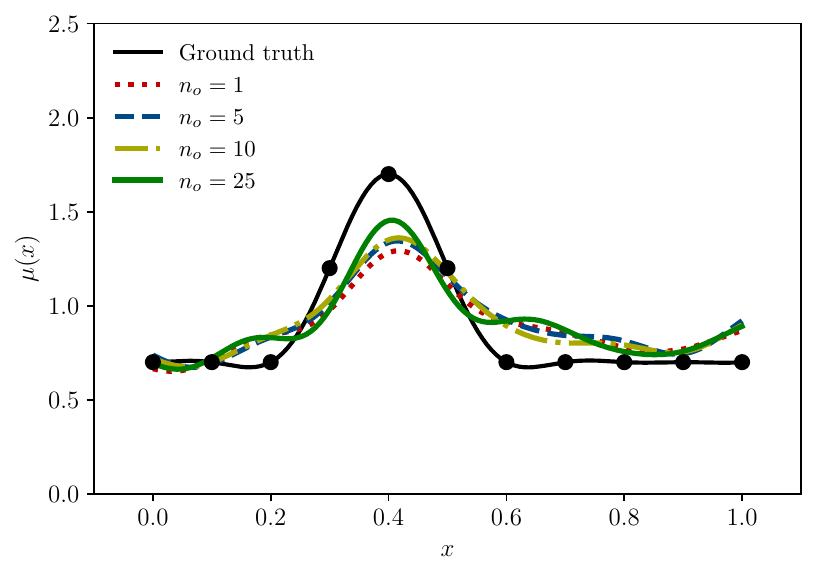}%
		\label{fig:1d-fe-single}%
	}
	\hfill
	\subfloat[Dual inclusion]{%
		\includegraphics[width=0.48\textwidth]{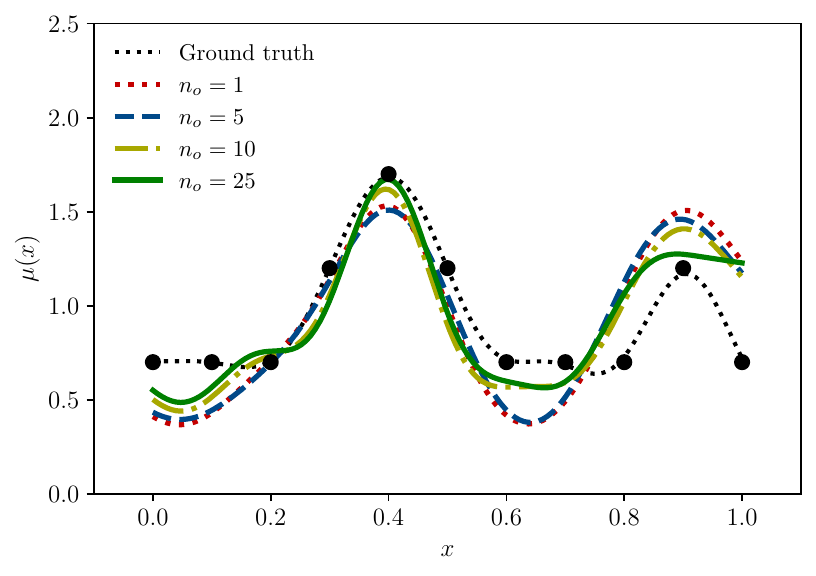}%
		\label{fig:1d-fe-dual}%
	}
	\caption{Posterior mean conductivity $\mu(x)$ recovered from finite element
			observations for the single-inclusion (a) and dual-inclusion (b) cases, at increasing
			numbers of repeated observations $n_o \in \{1, 5, 10, 25\}$. The solid black line is the
			ground truth and the black dots mark the anchor values.}
	\label{fig:1d-inverse-fe}
\end{figure}

We consider the single- and dual-inclusion ground truths of Sections~\ref{sec:1d-poisson-inv-singleint} and~\ref{sec:1d-poisson-inv-twoint}, with the same prior $p(\vec{\mu}^{(a)}) = \mathcal{N}(1.2\,\vec{1}_{11},\, 0.2^2\,\vec{I}_{11})$. Figure~\ref{fig:1d-inverse-fe} shows the recovered posterior means. In both cases the inclusions are localised correctly, but the peaks are underestimated and the reconstructions broader than the ground truth, and boundary artefacts appear. The relative $L^2$ error falls from $0.1982$ to $0.1632$ for the single inclusion and from $0.2576$ to $0.1795$ for the dual inclusion as $n_o$ increases from $1$ to $25$. These settle at roughly two to two and a half times the values obtained with QLSM-generated observations ($0.0697$ and $0.0889$ respectively). The error improves only marginally with $n_o$, since repeated measurements  average out the observation noise but not the systematic surrogate discrepancy, which dominates the residual. This is the behaviour a purely observational $\sigma_e$ cannot capture, and it motivates the discrepancy term to be modelled~\cite{kennedy2001bayesian} as written in the conclusion.

\subsubsection{Two-dimensional single inclusion}
\label{sec:2d-inverse-fe-singleint}

\begin{figure}[]
	\centering
	\includegraphics[width=\linewidth]{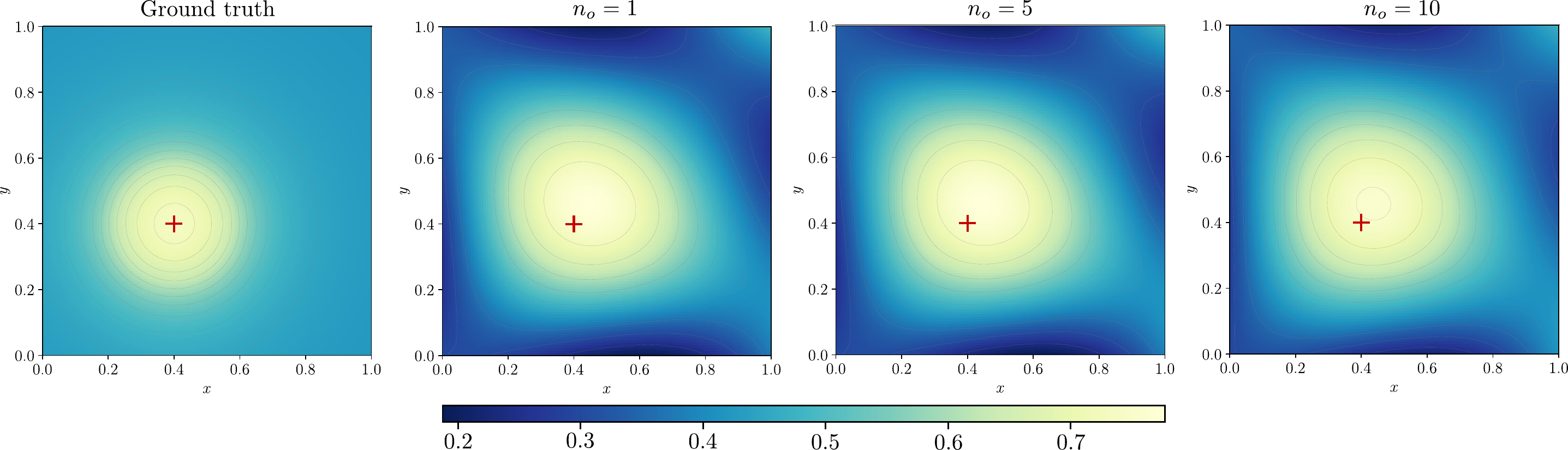}
	\caption{Posterior mean conductivity for the 2D single-inclusion case recovered from finite
		element observations; the prior and guide are flat at $\mu = 0.6$ and the red cross
		marks the true centroid $\vec{x}_c = (0.4, 0.4)$.}
	\label{fig:2d-inferred-fields-fe-singleint}
\end{figure}

We close with the 2D single inclusion of Section~\ref{sec:2d-inverse-singleint}, with the level-set ground truth ($\mu_{\text{in}} = 0.8$, $\mu_{\text{out}} = 0.4$, centroid $\vec{x}_c = (0.4, 0.4)$, radius $r = 0.1$) and the same prior $p(\vec{\mu}^{(a)}) = \mathcal{N}(0.6\,\vec{1}_{121},\, 0.2^2\,\vec{I}_{121})$. Figure~\ref{fig:2d-inferred-fields-fe-singleint} shows the posterior mean for $n_o \in \{1, 5, 10\}$. The high-conductivity region is placed close to the true centroid, displaced towards $(0.45, 0.45)$ and spread wider than the true inclusion, with low-conductivity corner artefacts that weaken as $n_o$ grows, as expected from the kernel lengthscale $\ell = 0.5$. The relative $L^2$ error decreases only slightly, from $0.1791$ to $0.1684$ as $n_o$ increases, settling at roughly two and a half times the $0.0670$ obtained with QLSM-generated observations.

\subsection{Computational cost and scalability}
\label{sec:cost}

We ran every component of the two-dimensional single inclusion study (Section~\ref{sec:2d-inverse-fe-singleint}) on an Apple Mac mini (M4 Pro, 24 GB unified memory, 2025) using CPU execution throughout, with \texttt{PyTorch} for the convolutional autoencoder, the \texttt{PennyLane} state-vector backend for the quantum layer, and \texttt{Pyro} for variational inference. The finite element solver generates $10^4$ samples in one minute. The autoencoder completes an epoch in $14.1$~s and the quantum layer in four minutes, the difference reflecting the classical simulation of the state vector that the quantum layer requires.

The forward map is dominated by the same simulation. One evaluation of $\mu \mapsto u$ spends $99\%$ of its time in the quantum circuit, leaving $0.8\%$ for the decoder, $0.2\%$ for the PCA projection, and under $0.1\%$ each for the normalisation and the linear output layer. One evaluation costs $55.1$~ms through the surrogate against $3$~ms through the finite element solver, about $18$ times slower at this problem size, a cost that belongs to the classical simulation of the circuit rather than to the method. The two inversions cannot be compared by their total cost. The QLSM updates the latent coordinates of $\mu$ and reconstructs no field, whereas the finite element inversion conditions on the full field and evaluates the residual at every degree of freedom. We therefore compare them on the forward map, where both evaluate the same $\mu \mapsto u$, as reported in Table~\ref{tab:cost2d}.

\begin{table}[]
	\centering
	\caption{Measured cost of the two-dimensional problem on the M4 Pro under CPU execution. The surrogate incurs the training cost once and the forward map per call.}
	\label{tab:cost2d}
	\begin{tabular}{llll}
		\toprule
		{Stage} & {Component} & {QLSM} & {FE} \\
		\midrule
		& Dataset generation          & 1 min        & {---} \\
		{Surrogate training} & Autoencoder training        & 14.1 s/epoch & {---} \\
		& QLSM training               & 4 min/epoch  & {---} \\
		\midrule
		Forward map
		& One evaluation, $\mu \mapsto u$ & 55.1 ms      & 3 ms \\
		\bottomrule
	\end{tabular}
\end{table}

Three quantities govern the extension to three dimensions. The first is the latent dimension, which principal component analysis fixes by retaining the modes that reach $95\%$ of the sampled conductivity variance, so the truncation follows from the character of the training set rather than from any choice we impose. The second is the convolutional autoencoder: a three-dimensional encoder compresses a volume rather than a slice, so the kernel becomes $3 \times 3 \times 3$ and each layer holds a larger activation for the same channel width, though this is a one-off offline cost. The third is dataset generation, which grows with the mesh as sparse direct solves scale towards $\mathcal{O}(n_{\mathrm{dof}}^{2})$ under nested dissection, though it remains a minor share of the budget. The latent dimension also sets the circuit width, since angle encoding assigns one qubit per retained mode. The shot budget scales with the number of measured observables $n_l^u$ and the target precision rather than with the grid, and robustness to NISQ noise falls as register width and circuit depth grow through barren-plateau effects.

\section{Implementation of QLSM on NISQ devices}
\label{sec:nisq}

We now demonstrate how we implement the proposed QLSM framework on noisy intermediate-scale quantum (NISQ) devices. Although the framework ultimately targets error-corrected machines, we describe here the modifications required for practical deployment on currently available NISQ devices. The principal challenge is the noise arising from gate imperfections and decoherence, which degrades the measured quantum expectation values and, in turn, the latent forward mapping. The ability of the framework to retain predictive fidelity under these conditions therefore constitutes a direct test of its practical viability as a quantum-assisted surrogate for inverse-problem applications. All hardware experiments are carried out on the one-dimensional problem, which permits a focused assessment of the model under device noise. In this setting, we consider a single centred inclusion analogous to that of Sections~\ref{sec:1d-forward} and~\ref{sec:1d-inverse}. The numerical experiments in this Section are carried out on the quantum computing platform $\mathtt{qBraid}$~(\url{https://www.qbraid.com}). 

\begin{figure}[]     
    \centering         
    \includegraphics[width=1.\textwidth]{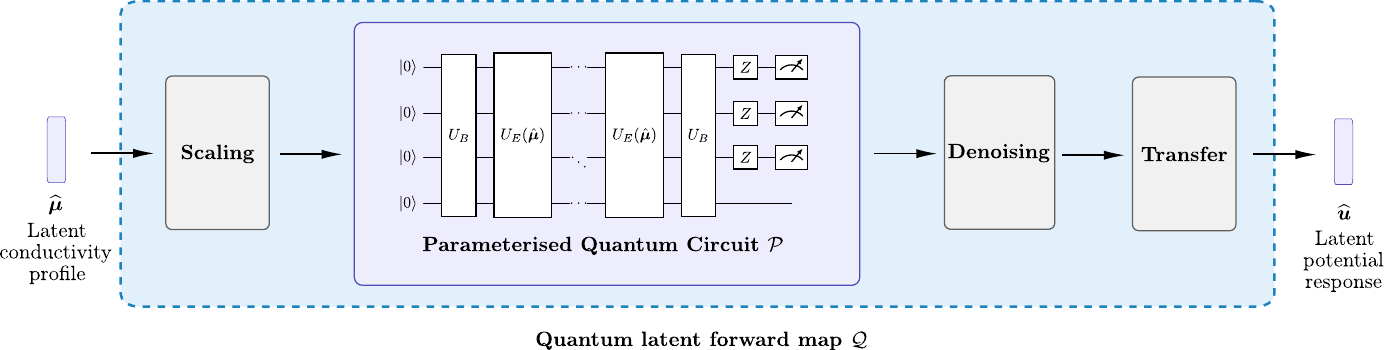}
    \caption{Quantum latent forward mapping in the latent space with NISQ          implementation. The latent material properties are first scaled before          encoding into the quantum system. The measured quantum expectation values          are subsequently denoised via the classical correction network prior to          the transfer layer mapping into the latent solution space.}     \label{fig:quantum-latent-map-nisq} 
\end{figure}

In this work, we intend to use the pre-trained QLSM surrogate by fixing the model weights within NISQ implementation. To bridge the gap between noiseless and NISQ readouts, a classical denoising network is introduced as a post-processing stage acting on the quantum circuit measurements. The denoiser is a compact fully-connected neural network with a single hidden layer of width 16 and $\tanh$ activation,
\begin{equation}
	\tilde{\vec{q}} = \vec{W}_2 \tanh\!\left(\vec{W}_1 \vec{q}^{\text{NISQ}} + \vec{b}_1\right) + \vec{b}_2 \, ,
	\label{eq:tanh-nisq}
\end{equation}
where $\vec{q}^{\text{NISQ}} \in \mathbb{R}^{n_l^u}$ is the vector of expectation values measured on the NISQ device and $\tilde{\vec{q}} \in \mathbb{R}^{n_l^u}$ is the corrected estimate of the corresponding noiseless readout. The hidden layer provides a nonlinear correction that compensates for distortions induced by hardware noise and finite sampling, while the output layer projects this correction back onto the readout space. The denoised readout $\tilde{\vec{q}}$ is then passed through the frozen, pre-trained transfer layer of the QLSM to recover the latent potential, as depicted in Figure~\ref{fig:quantum-latent-map-nisq}. All previously trained PQC and transfer-layer parameters remain frozen at this stage, and only the denoising weights $\{\vec{W}_1, \vec{b}_1, \vec{W}_2, \vec{b}_2\}$ are updated.

The denoiser is trained by minimising the mean squared error between its output and the noiseless simulator readout obtained for the same input,
\begin{equation}
	\mathcal{L} = \frac{1}{n_s^{\text{den}}} \sum_{i=1}^{n_s^{\text{den}}}
	\left\| \tilde{\vec{q}}_i - \vec{q}_i \right\|_2^2 \, ,
	\label{eq:loss-nisq}
\end{equation}
where $\vec{q}_i$ denotes the simulator readout, see Eq.~\eqref{eq:pauli-z}, and $\tilde{\vec{q}}_i$ is the denoised estimate produced by Eq.~\eqref{eq:tanh-nisq} from the corresponding NISQ readout $\vec{q}^{\text{NISQ}}_i$. The training dataset comprises $n_s^{\text{den}} = 100$ paired samples generated by evaluating identical inputs on both the noiseless simulator and the NISQ device, with each circuit estimated from $n^{\text{shots}} = 1000$ measurement shots. The denoiser is trained for $200$ epochs using the Adam optimiser with a fixed learning rate of $0.001$, and its performance is assessed on five held-out test samples. The underlying quantum circuit architecture follows that of Section~\ref{sec:1d-forward}, a three-layer PQC on nine qubits, and is held identical across all hardware experiments. Figure~\ref{fig:quantum-latent-map-nisq} depicts the modified quantum surrogate model, which differs from Figure~\ref{fig:quantum-latent-map} only by the insertion of the classical denoising module after the quantum measurements. The denoiser architecture is device-agnostic, but its weights are device-specific and must be retrained when the device, circuit, or calibration state changes. This retraining is a one-off calibration of $n_s^{\mathrm{den}} = 100$ samples with the PQC and transfer layer frozen.

\subsection{One-dimensional forward problem}
\label{sec:nisq-1dfwd}

\begin{figure}[]
	\centering
	\subfloat[$\mu(x)$ conductivity profile]{%
		\includegraphics[width=0.33\textwidth]{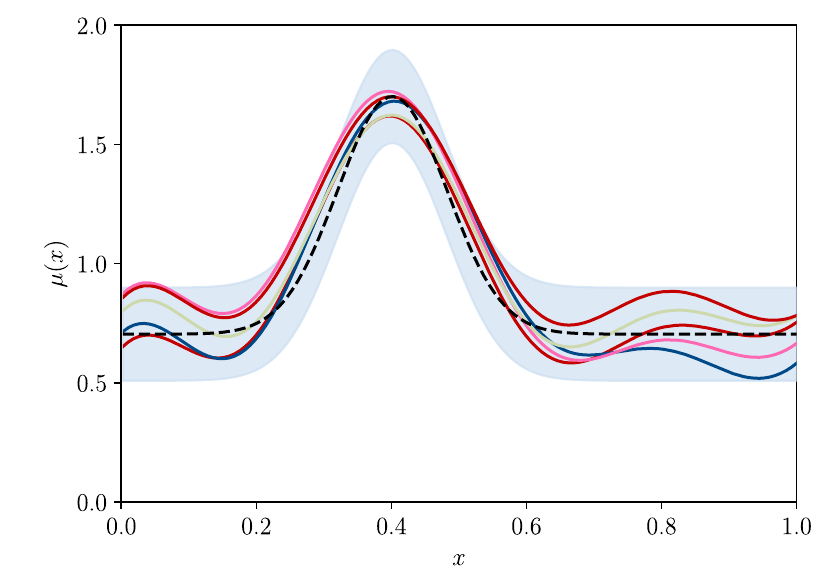}%
	}
	\hfill
	\subfloat[Noisy $u(x)$ from IonQ simulator]{%
		\includegraphics[width=0.33\textwidth]{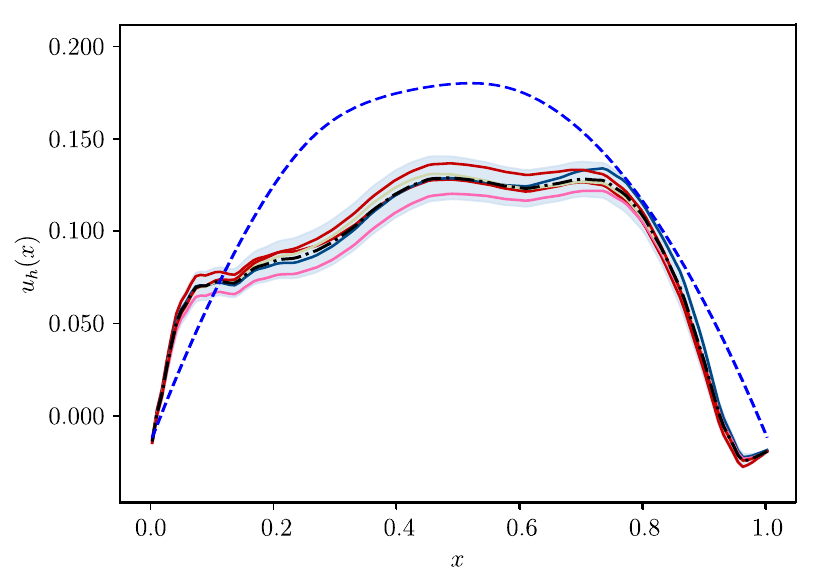}%
	}
	\hfill
	\subfloat[Denoised $u(x)$ from IonQ simulator]{%
		\includegraphics[width=0.33\textwidth]{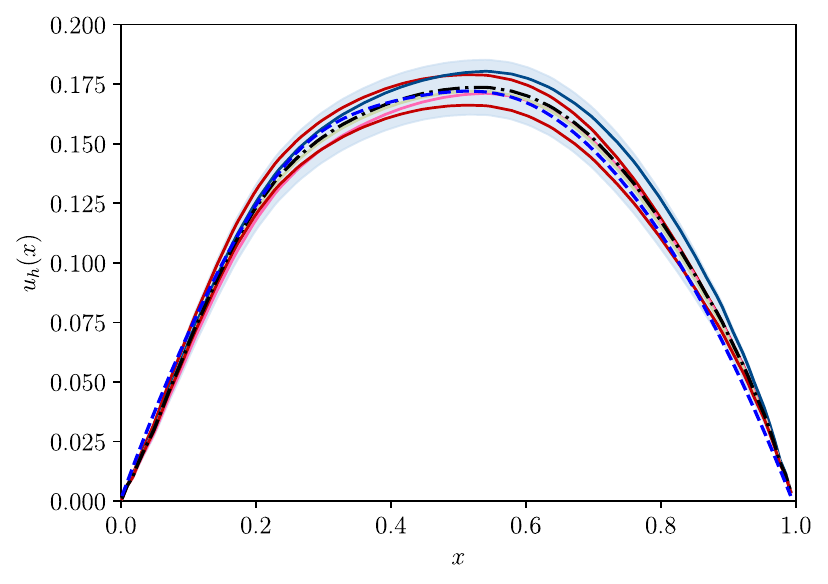}%
	}
	\\
	\subfloat[$u(x)$ from noiseless simulator]{%
		\includegraphics[width=0.33\textwidth]{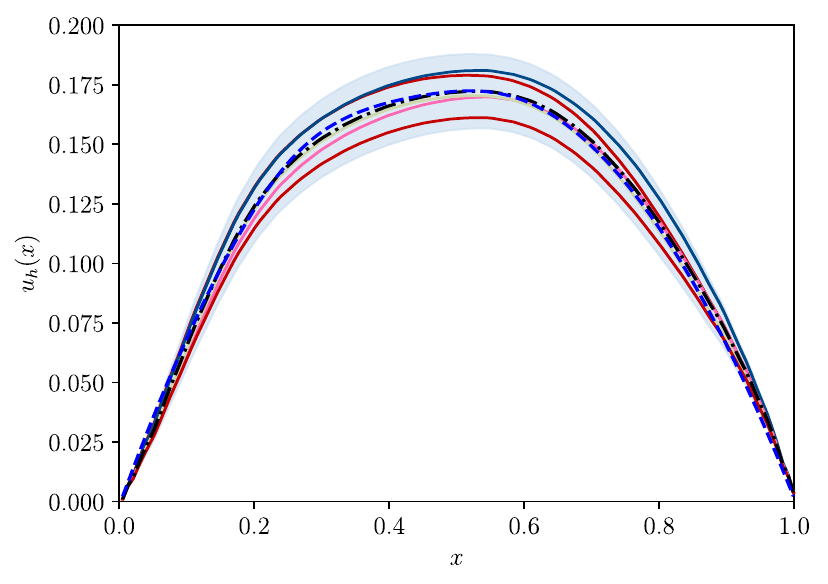}%
	}
	\hfill
	\subfloat[Noisy $u(x)$ from Rigetti Cepheus-1-108Q]{%
		\includegraphics[width=0.33\textwidth]{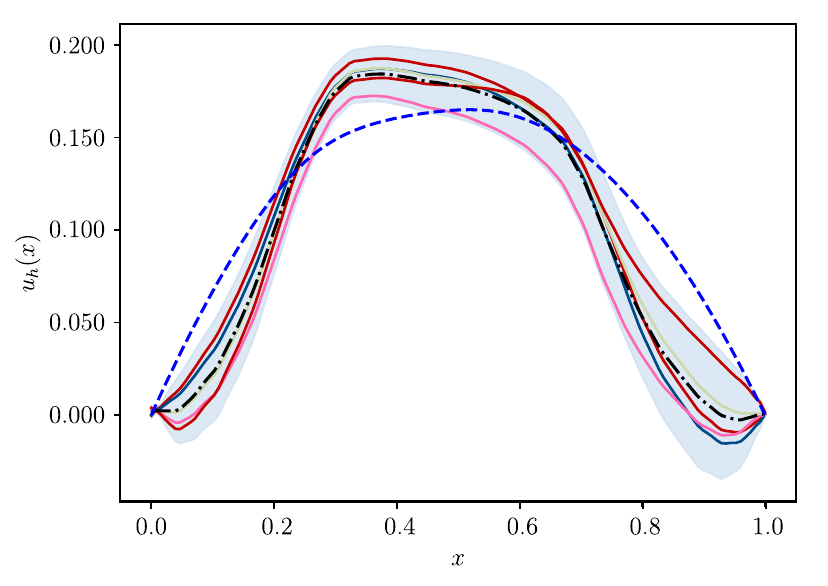}%
	}
	\hfill
	\subfloat[Denoised $u(x)$ from Rigetti Cepheus-1-108Q]{%
		\includegraphics[width=0.33\textwidth]{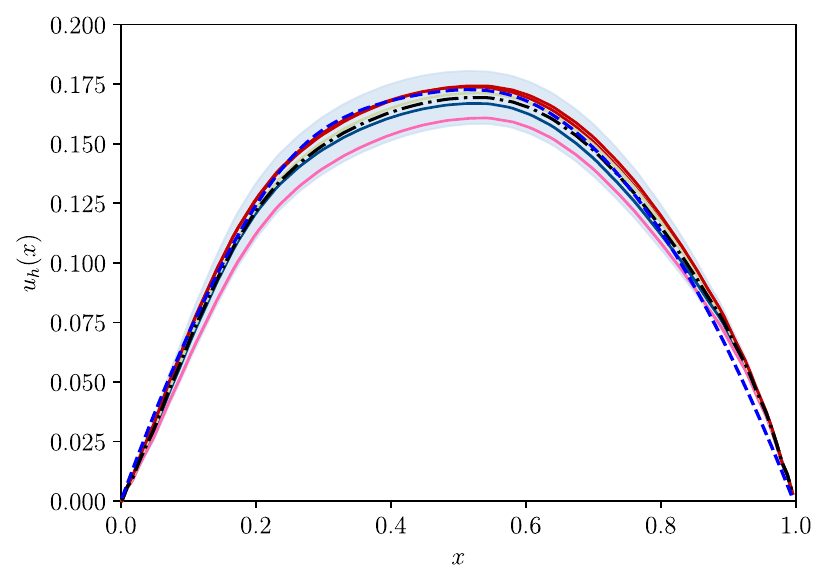}%
	}
	\caption{One-dimensional forward surrogate results under NISQ deployment, organised by column. Left column: (a) input conductivity profile $\mu(x)$ and (d) reference response $u(x)$ from the noiseless simulator. Centre column: raw noisy responses $u(x)$ from (b) the IonQ noise-aware simulator and (e) the Rigetti Cepheus-1-108Q device, prior to denoising. Right column: corresponding denoised reconstructions for (c) the IonQ simulator and (f) the Rigetti device, after applying the classical correction network. Shaded regions indicate $95\%$ confidence intervals over the test samples.}
	\label{fig:1d-nisq}
\end{figure}

We test the framework on two NISQ platforms based on contrasting quantum computing technologies: the IonQ noise-aware simulator, which emulates the noise characteristics of a trapped-ion device, and the Rigetti Cepheus-1-108Q, a physical superconducting-qubit device built from twelve interconnected nine-qubit chiplets and characterised by shorter coherence times and higher gate-error rates than the simulated environment. 
Figure~\ref{fig:1d-nisq} summarises the results, organised by column: the left column shows the input conductivity profile and the noiseless potential response used as reference; the centre column shows the raw noisy responses from the IonQ noise-aware simulator and the Rigetti Cepheus-1-108Q; and the right column shows the corresponding potential responses after denoising.

For the IonQ noise-aware simulator, the raw NISQ readout yields a relative $L^2$ error of $\epsilon_{L^2} = 0.314$ with respect to the FE reference solution. After applying the classical denoising network, the reconstructed potential field recovers the spatial structure of the noiseless solution, with the error reduced by approximately one order of magnitude to $\epsilon_{L^2} = 0.0326$. The Rigetti Cepheus-1-108Q readout exhibits a lower raw error of $\epsilon_{L^2} = 0.253$, albeit with a different spatial error profile reflecting its distinct noise channels. Following denoising, the reconstructed solution achieves $\epsilon_{L^2} = 0.0350$, marginally higher than for the IonQ case. The two technologies thus yield comparable post-correction accuracy despite their very different underlying noise mechanisms, indicating that a single denoiser architecture, retrained per device, is agnostic to the underlying noise mechanism. The residual error after denoising is approximately twice the noiseless QLSM error of $\epsilon_{L^2} = 0.0153$ reported in Section~\ref{sec:1d-forward}. In terms of computational time, a forward evaluation on the IonQ noise-aware simulator completed in about 2~s of wall-clock time. On the Rigetti Cepheus-1-108Q, a forward evaluation took about 3.3~s of device time on average. The progressive evaluation across noiseless simulation, noise-aware simulation, and physical superconducting hardware establishes a clear and reproducible pathway for the practical deployment of quantum surrogate models in computational mechanics, and demonstrates that the proposed hybrid quantum-classical framework remains robust under realistic hardware noise. 

\subsection{One-dimensional inverse problem}
\label{sec:1d-nisq-inverse}

We now proceed with the inversion of the single localised inclusion of Section~\ref{sec:1d-inverse}, with the same ground-truth coefficient vector $\vec{\mu}^{(a),\,\text{true}}$ centred on the anchor point $x^{(a)} = 0.4$, but with the forward map evaluated on the IonQ noise-aware simulator introduced in Section~\ref{sec:nisq}, which emulates the noise characteristics of a trapped-ion device. As established in Section~\ref{sec:algo-infer}, the forward map through NISQ devices admits no differentiation graph, so the pathwise gradient is unavailable and the variational parameters are instead updated using the score-function (REINFORCE) estimator of Eq.~\eqref{eq:score-function-grad}, with the baseline formed as a running average of past log-likelihoods with decay rate 0.9. At each step, the likelihood is evaluated with a single forward evaluation of the denoised noise-aware readout on the IonQ simulator, with the underlying circuit estimated from $n^{\text{shots}} = 1000$ measurement shots per evaluation; the inference is carried out for $1000$ steps within this fixed evaluation budget. To render the inversion tractable within this budget and under the higher variance inherent to the score-function estimator, the prior mean is placed close to the solution, offset from the ground truth by $0.1$ in each component,
\begin{equation}
	p(\vec{\mu}^{(a)})
	= \mathcal{N}\!\left(
		\vec{\mu}^{(a),\,\text{true}} - 0.1 \vec{1}_{11} ,\;
		0.2^{2}\, \vec{I}_{11}
	\right) \, .
	\label{eq:prior-offset-singleint}
\end{equation}
This constitutes an informative prior, chosen to demonstrate the feasibility of inversion on noise-aware hardware rather than recovery from a non-informative prior. The offset keeps the posterior search local, which is appropriate given the limited number of simulator evaluations available and the noisier descent direction of the score-function estimator. We emphasise that this demonstration establishes the feasibility of NISQ-based inversion under prescribed conditions, namely a one-dimensional domain, an informative prior centred near the ground truth, and a fixed evaluation budget. Extending it to weakly informative priors and higher-dimensional problems remains an open research question.

\begin{figure}[]
	\centering
		\includegraphics[width=0.48\textwidth]{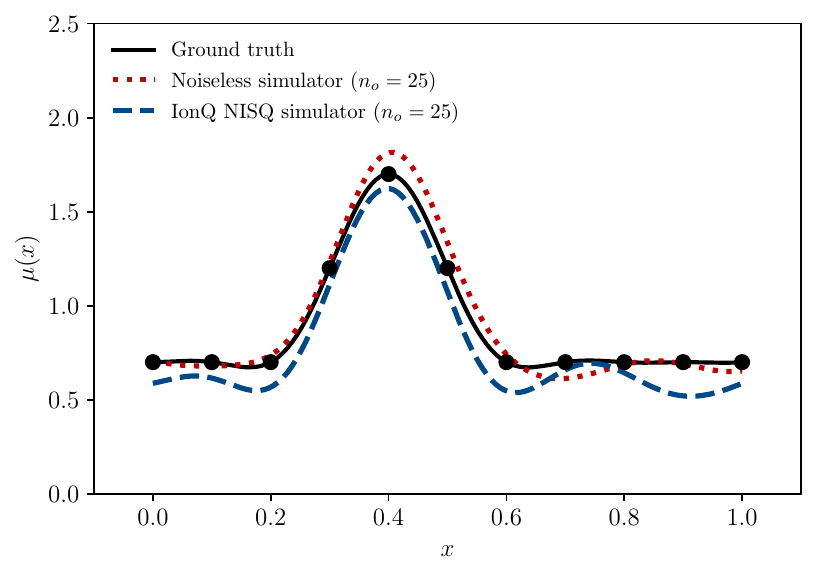}%
	\caption{Posterior mean conductivity profile $\mu(x)$ for the single inclusion: the noise-aware result, recovered on the IonQ noise-aware simulator with the score-function estimator and the prior mean offset by $0.1$ from the ground truth for $n_o = 25$ repeated observations, is shown against the posterior obtained on the noiseless simulator. The solid black line indicates the ground-truth conductivity profile, and the black dots mark the anchor locations $\{x_j^{(a)}\}$ and their true coefficient values.}
	\label{fig:1d-prediction-singleint-nisq}
\end{figure}

Figure~\ref{fig:1d-prediction-singleint-nisq} shows the posterior mean conductivity field recovered using the IonQ noise-aware simulator for $n_o = 25$ repeated measurements, alongside the corresponding profile obtained with the noiseless simulator. The values at the anchor points are again obtained as the mean of the posterior $p(\vec{\mu}^{(a)} \mid \hat{\vec{Y}})$, and the field over space is reconstructed using Eq.~\eqref{eq:gp-basis}. Both estimators correctly localise the inclusion at $x^{(a)} = 0.4$ and recover a peak amplitude close to the ground truth, with the noiseless simulator inversion achieving a relative error of $\epsilon_{L^2} = 0.0697$. The inversion through noise-aware simulator is less accurate, with the relative error increasing to $\epsilon_{L^2} = 0.1182$. This is visible in Figure~\ref{fig:1d-prediction-singleint-nisq} away from the inclusion, where the  profile departs from the ground truth background level of $0.7$, undershooting near $x \approx 0.15$--$0.2$ and producing spurious oscillations near $x \approx 0.6$--$1.0$, regions where the noiseless simulator profile remains close to background level. The near-doubling of the relative error under the noise-aware readout is consistent with the additional variance introduced both by the score-function (REINFORCE) estimator and by the noise-emulated quantum measurements.

%% file: conclusions.tex
\section{Conclusions}
\label{sec:conclusions}

We have presented QLSM, a quantum-assisted latent surrogate model for forward and inverse problems involving inhomogeneous material properties governed by elliptic partial differential equations. The proposed framework combines principal component analysis and a convolutional autoencoder to compress high-dimensional conductivity and response fields into a low-dimensional latent representation. A parameterised quantum circuit is then employed to learn the latent forward map, while stochastic variational inference is performed entirely in latent space, eliminating the need for repeated finite element solves during posterior inference. For the forward problem, QLSM accurately reproduced finite element solutions for one- and two-dimensional Poisson equations with inhomogeneous conductivity fields. The predictive accuracy was comparable to that of a classical neural-network surrogate of similar architectural depth, which we take as confirmation that the quantum-assisted route is viable rather than as evidence of any advantage over classical surrogates in accuracy or in parameter count. For the inverse problem, the learned surrogate was successfully integrated within a variational inference framework and demonstrated the ability to recover localised inclusions, structured inclusions, and unstructured conductivity fields under both informative and non-informative prior assumptions. The two research questions posed in this study were addressed directly. First, among the circuit architectures investigated, a three-layer parameterised quantum circuit provided the most favourable balance between expressivity and trainability. Second, despite the presence of hardware noise, a lightweight learned denoising stage enabled the recovery of near noiseless prediction accuracy when the model was executed on a noise-aware trapped-ion IonQ simulator and on the Rigetti Cepheus-1-108Q superconducting quantum processor. These results demonstrate that quantum latent-space surrogates can be deployed on NISQ quantum hardware while maintaining useful predictive performance. Taken together, the findings establish QLSM as a functional and reproducible framework for quantum-assisted surrogate modelling of PDE-governed forward and inverse problems. 

Several directions remain open for future research. In the present study, the Gaussian-process lengthscale governing the conductivity parameterisation was fixed a priori. Joint inference of these hyperparameters together with the anchor coefficients would relax the underlying modelling assumptions and may improve the recovery of highly localised features~\cite{Girolami2021,Rasmussen2005}. Additionally, an auxiliary random variable can be introduced to represent the model-reality mismatch~\cite{kennedy2001bayesian}, to better capture uncertainty when real data are used or when correlated or systematic discrepancies are evident between model and data. Richer variational guides that capture posterior correlations, rather than the mean-field approximation used in this work, would strengthen the uncertainty quantification~\cite{povala2022variational}. The sensitivity of the results to the autoencoder architecture and latent dimension also merits systematic study. A natural extension is the application of QLSM to vector-valued PDEs, particularly linear elasticity problems involving isotropic metals, orthotropic composites, and concrete structures. Such problems would require the simultaneous representation of multiple coupled material-property fields, appropriate nondimensionalisation prior to compression, and a reassessment of the quantum encoding strategy as the latent dimension increases~\cite{Koutsourelakis2012, Scholz2025WNVI, Jafari2025}. From a quantum algorithm perspective, extending the framework to larger and more strongly nonlinear problems will require encoding strategies that remain expressive while preserving trainability, as increasing circuit width and depth are known to exacerbate barren-plateau phenomena. Richer feature embeddings, more efficient multi-observable measurement schemes, advanced error-mitigation techniques, and ultimately fault-tolerant quantum computation represent promising directions for improving scalability and robustness~\cite{Temme2017ErrorMitigation, Cai2023QEMreview,Huang2020,Huang2026}. In particular, replacing the present angle encoding with amplitude encoding~\cite{schuld2020circuit, SchuldKilloran2019} would embed a latent dimension of up to $2^{n_q}$ in $n_q$ qubits and so relax the linear coupling between latent dimension and register width that currently limits the method to problems of low intrinsic dimension, though realising this while retaining expressivity and trainability remains an open question. We regard this as a distinct direction rather than a substitution. Finally, integrating the latent-space quantum surrogate with operator-learning approaches such as DeepONets and Fourier Neural Operators, as well as projection-based methods for sparse or incomplete observations, could substantially broaden its applicability to a wider class of PDEs, including wave propagation problems, and to realistic experimental and monitoring scenarios \cite{Lu2021,Li2021FNO}.